\RequirePackage{fix-cm}
\documentclass[twocolumn]{svjour3}          
\smartqed  
\usepackage{graphicx}
\usepackage{longtable}
\usepackage{float}
\usepackage{aeguill}
\usepackage{xcolor}
\newcommand{\stereotype}[1]{
	\guillemotleft {#1}\guillemotright
}
\usepackage{setspace} 
\usepackage{listings}
\lstdefinelanguage{OCL}
{
morekeywords={operation, body, let, in},
sensitive=false,
morecomment=[l]{//},
morecomment=[s]{/*}{*/},
morestring=[b]",
}

\newcommand{\ocl}[1]{{\fontsize{8.7}{8.7}\selectfont {\ttfamily\bfseries #1}}}

\newcommand*{\affaddr}[1]{#1} 
\newcommand*{\affmark}[1][*]{\textsuperscript{#1}}

\definecolor{codegreen}{rgb}{0,0.6,0}
\definecolor{codegray}{rgb}{0.5,0.5,0.5}
\definecolor{codepurple}{rgb}{0.58,0,0.82}
\definecolor{backcolour}{rgb}{0.95,0.95,0.92}
\definecolor{bluegreen}{RGB}{51,153,126}	
\definecolor{intenttypecolor}{RGB}{58,162,187}
\definecolor{key-color}{rgb}{0.8, 0.47, 0.196}
\definecolor{xtext-keyword}{RGB}{127, 0, 85}
\definecolor{javadocblue}{rgb}{0.25,0.35,0.75} 

\newcommand{\myOCL}[1]{\lstinputlisting[
language=MOCL, 
basicstyle=\footnotesize,
numbers=none]{#1}}

\lstdefinelanguage{MOCL}{
  keywords={ 
  (, ), *, +, ++, , -, ->, ., .., /, :, ::, ::*, ;, <, <-, <, <>, , >, >, ?, ?->, ?., @, , Bag, Boolean, Collection, Integer, Lambda, Map, OclAny, OclInvalid, OclMessage, OclState, OclVoid, OrderedSet, Real, Sequence, Set, String, Tuple, UnlimitedNatural, and, body, context, def, derive, else, elseif, endif, endpackage, extends, false, if, implies, import, in, include, init, inv, invalid, let, library, not, null, or, package, post, pre, self, static, then, true, xor, {, |, |, |?, }
  },
   keywordstyle = {\color{xtext-keyword}\bfseries},
   stringstyle=\color{javadocblue},
   commentstyle=\color{codegreen},
   numberstyle=\tiny\color{codegray},
  comment=[l]{\-\-},
  morestring=[b]',
  morestring=[b]",
  captionpos=b, 
  basicstyle=\scriptsize\ttfamily,
}

\begin{document}

\title{Model Driven Engineering for Data Protection and Privacy: Application and Experience with GDPR
}


\author{Damiano Torre \affmark[1] \and
        Mauricio Alferez \affmark[1] \and
        Ghanem Soltana \affmark[1] \and
        \\Mehrdad Sabetzadeh \affmark[2]\,\affmark[1] \and
        Lionel Briand \affmark[1]\,\affmark[2] 
}


\institute{Damiano Torre \at
           \email{damiano.torre@uni.lu}
           \and
           Mauricio Alferez \at
             \email{mauricio.alferez@uni.lu}
           \and
           Ghanem Soltana \at
             \email{ghanem.soltana@uni.lu}
           \and
           Mehrdad Sabetzadeh \at
           \email{m.sabetzadeh@uottawa.ca}
           \and
           Lionel Briand \at 
           \email{lionel.briand@uni.lu}\\
           \\
              \affaddr{\affmark[1]~SnT Centre for Security, Reliability and Trust, University of Luxembourg, Luxembourg}\\
              \\
              \affaddr{\affmark[2]~School of Electrical Engineering and Computer Science, University of Ottawa, Canada}\\
}

\date{Received: date / Accepted: date}

\maketitle

\begin{abstract}
{In Europe and indeed worldwide, the General Data Protection Regulation (GDPR) provides protection to individuals regarding their personal data in the face of new technological developments. GDPR is widely viewed as the benchmark for data protection and privacy regulations that harmonizes data privacy laws across Europe. Although the GDPR is highly beneficial to individuals, it 
presents significant challenges for organizations monitoring or storing personal information.
Since there is currently no automated solution with broad industrial applicability, organizations have no choice but to carry out expensive manual audits to ensure GDPR compliance.
In this paper, we present a complete GDPR UML model as a first step towards designing automated methods for checking GDPR compliance.
Given that the practical application of the GDPR is influenced by national laws of the EU Member States, we suggest a two-tiered description of the GDPR, generic and specialized. 
In this paper, we provide (1) the GDPR conceptual model we developed with complete traceability from its classes to the GDPR, (2) a glossary to help understand the model, (3) the plain-English description of 35 compliance rules derived from GDPR along with their encoding in OCL, and (4) the set of 20 variations points derived from GDPR to specialize the generic model.
We further present the challenges we faced in our modeling endeavor, the lessons we learned from it, and future directions for research.}
\newline
\end{abstract}

\section{Introduction}\label{introduction}
 
With the growing concerns about data protection and privacy, it is becoming increasingly important to assess compliance with the relevant regulations. In Europe and indeed worldwide, the General Data Protection Regulation (GDPR)~\cite{GDPR2018} is now widely viewed as a benchmark for data protection and privacy regulations. The GDPR came into effect in May 2018, replacing the previous Data Protection Directive, 95/46/EC. The GDPR has been designed to harmonize data privacy laws across Europe in order to provide further protection and capabilities to individuals for controlling their personal data in the face of new technological developments~\cite{EUP2019}. While undoubtedly beneficial to individuals in many ways, the reality with the GDPR is that organizations are having severe difficulties in understanding what compliance means in this new environment and how to implement the GDPR~\cite{Tankard2016}. 

In order to comply with the requirements of the GDPR, organizations need to consider the principles of personal data processing as set out in the GDPR and to make regular reviews of their measures, practices and processes regarding the collection,  use and protection of personal data. Failure to comply with the GDPR may result in fines of up to 20m or 4\% of an organization's global turnover for specific breaches \cite{GDPR2018}. In addition, organizations are liable for damages and other remedies towards individuals in case of data breaches~\cite{EU2018}. For this reason, there is now a fast-growing need for cost-effective methods that will help different business sectors achieve, demonstrate and maintain compliance with the GDPR. Given the sheer complexity of the systems and services that are subject to the GDPR, e.g., e-Government applications and cloud-based services, automated support for GDPR analysis is critically important. At the moment, there is a lack of such support on the market. This gap will become even more evident once individuals start to exercise their rights under the GDPR, likely resulting in an onslaught of new legal challenges for companies. Due to the absence of automated solutions, we have started a long-term investigation, involving both IT researchers and legal experts, into GDPR compliance automation. Our ultimate goal is to bring scalability to GDPR compliance assessment and create opportunities for developing innovative GDPR-related services.

The GDPR is considered the most far-reaching and technically demanding personal data privacy regulation ever established. The high level of rigor that ensuring GDPR compliance entails is increasingly comparable to what is required for demonstrating compliance to safety standards and regulations. GDPR compliance analysis can thus benefit from existing work where models have been employed for systematic compliance analysis in the context of safety certification, e.g.,~\cite{Panesar2013}. While highly advantageous, encoding the GDPR and its compliance mechanisms into a model-based representation is a complicated task. In particular, the level of abstraction of such a representation has to be suitable for ensuring a consistent implementation and interpretation of the regulation, national laws and case law. 

In this paper, we draw on Model-Driven Engineering (MDE)~\cite{Brambilla2016} for building a machine-analyzable representation of the GDPR as a first step towards the development of future automated methods for assessing GDPR compliance. Although MDE is primarily a paradigm for reducing the complexity of systems development \cite{France2007}, over the years, MDE has outgrown its traditional use and is now increasingly applied as a general mechanism for structuring domain knowledge. When employed in this broader sense, as we do in our work, MDE provides an effective communication bridge between IT experts and domain experts, such as legal experts, who may have little software development expertise.

What we pursue in this paper through the application of MDE is a visual and yet precise representation of the textual content of the GDPR. Since a concrete implementation of the GDPR is affected by the national laws of the EU member states, the GDPR's expanding body of case law and other contextual factors, we propose a two-tiered representation of the GDPR: a generic tier and a specialized tier. The generic tier captures the concepts and principles of the GDPR that apply to all contexts, whereas the specialized tier describes a specific tailoring of the generic tier to a given context, including the contextual variations that may impact the interpretation and application of the GDPR. We represent both the generic and specialized tiers using UML class diagrams~\cite{OMG2017} and a set of invariants expressed in the Object Constraint Language (OCL)~\cite{OMG2014}. 
In particular, as we explain in detail in Section \ref{vision}, we provide an overview of our long-term research project involving four steps: (1) building a generic tier of the GDPR, (2) tailoring the generic tier into a specialized one, (3)~developing tool support for representing technical and legal documents in a structured form, and (4) enabling checking GDPR compliance. {In this paper, we focus exclusively on conducting steps~1 and 2; the initial results of steps~3 and 4 are presented elsewhere in recently published work ~\cite{Torre2020}}.

Several strands of work employ models for expressing legal requirements and assessing whether and to what extent these requirements are met by a given system. These strands include the large body of research concerned with the application of goal models to laws and regulations, e.g., \cite{Ghanavati2014},  \cite{Ingolfo2014}, as well as a number of conceptual modeling techniques aimed at representing the semantics of legal texts, such as key legal abstractions and modalities, e.g., \cite{Zeni2015,Soltana2018,Arora2016}, and the structural representation of legal texts, e.g., \cite{Emmerich1999,Breaux2009,Sannier2017}. 

As we discuss in more detail in Section \ref{related}, existing model-based approaches for compliance verification have one of the following limitations as far as the GDPR is concerned: they (1)~have a different focus than the GDPR, e.g.,~\cite{Panesar2013}, (2)~present guidelines only for the manual application of the GDPR, e.g.,~\cite{Ayala2018}, or (3) focus exclusively on specific GDPR use cases, e.g.,~\cite{Caramujo2019,Pullonen2019}. To the best of our knowledge, there are no proposals in the literature aimed at providing a holistic model-based representation of the GDPR. 
{In order to address this gap, in a previous conference paper \cite{Torre19}, we tackled the following three research questions (RQ):}
\begin{itemize}
\item \textit{RQ1: How can we develop a generic and adaptable model-based representation of the GDPR to support automated compliance checking?}

\item \textit{RQ2: How can we tailor the generic GDPR model according to the specific needs of  a given context?}

\item \textit{RQ3: What are the challenges in modeling the GDPR?}
\end{itemize}

{
This submitted article is a major extension of our previous paper at MODEL 2019 \cite{Torre19}. In summary, the article enhances our earlier publication by providing: (1) the nine packages of the GDPR conceptual model developed in Enterprise Architect, (2) a table capturing how the classes in these packages are traceable to the GDPR (96 entries), (3) the complete glossary for our conceptual model (267 entries), (4) the plain-English description of 35 compliance rules derived from GDPR, (5) an encoding of said rules in OCL, and (6) a set of 20 variation points derived from the GDPR to specialize the generic model along with guidelines on how to apply them. None of these six complete artifacts were previously presented \cite{Torre19}.}

{The complete material regarding points 1-4 above can be found in Appendix A, point 5 is presented in Appendix B, and point 6 is discussed in Appendix A (in plain-English) and Appendix C (with the additional variability OCL constraints).}

\textbf{Contributions.} Our contributions are as follows:
{
\textit{(1)} We present the generic model of the GDPR composed of nine UML class diagrams and 35 OCL constraints. We use the term ``generic'' to imply that the model is based only on the content of the GDPR and is not encompassing any complementary information that may be necessary to contextualize the GDPR for use in a particular situation.
\newline\textit{(2)} The exact realization of the GDPR is subject to some variability depending on context. We present guidelines for tailoring the generic GDPR model into a specialized model that is suitable for application in a specific context. To this end, we describe 20 variation points that are considered acceptable by the GDPR and our strategy for handling these variations.}
\newline\textit{(3)} We reflect on the lessons learned from encoding the GDPR into a model-based representation. Our lessons, which cover model validation, traceability and  contextualization, provide a useful stepping stone for UML-based specification of other complex laws and regulations. 
\newline\textit{(4)} We present the challenges we identified during our modeling endeavor alongside a number of future directions aimed at addressing these challenges.

\textbf{Structure.} Section \ref{GDPR} introduces basic concepts related to the GDPR. Section \ref{vision} provides an overview of our approach. Section \ref{main} addresses our research questions. Section \ref{lessons} and \ref{directions} present lessons learned and future directions, respectively. Section \ref{related} compares with related work. Section~\ref{conclusion} concludes the paper.

\section{GDPR overview} \label{GDPR}
The GDPR \cite{EU2018} is a complex piece of legislation comprised of 173 recitals, and 99 articles divided into 11 chapters. 
The GDPR applies primarily to businesses established in the EU. However, the regulation may also apply to businesses outside the EU, e.g., when these businesses offer goods or services to, or monitor individuals in the EU.
If a business is subject to the GDPR, 
 it has to identify itself as either a data controller or data processor. A controller 
 determines the purpose and means of the processing, whereas a processor 
 acts on the instructions of the controller.
The responsibilities of a given business under the GDPR vary depending on whether it is a processor or a controller and depending on the kind of data processed.

Processors notably have to: (i) implement adequate technical and organizational measures to keep personal data  safe and secure, and, in cases of data breaches, notify the controllers; (ii) appoint a statutory data protection officer and conduct a formal impact assessment for certain types of high-risk processing; (iii) keep records about their data processing; and (iv) comply to the GDPR restrictions when transferring personal data outside the EU.

In comparison to processors, controllers are subject to more GDPR obligations. In particular, in addition to having to meet the obligations mentioned above, controllers have to: (i)~adhere to six core personal data processing principles, namely, fair and lawful processing, purpose limitation, data minimization, data accuracy, storage limitation, and data security; (ii)~keep identifiable individuals informed about how their personal data will be used; and (iii)~preserve 
the individual rights envisaged by the GDPR, e.g., the right to be forgotten and the right to lodge a complaint.

\section{Towards a Model-based Approach for Automated GDPR Compliance Checking} \label{vision}
\begin{figure}[!t]
\centerline{\includegraphics[width=1\linewidth]{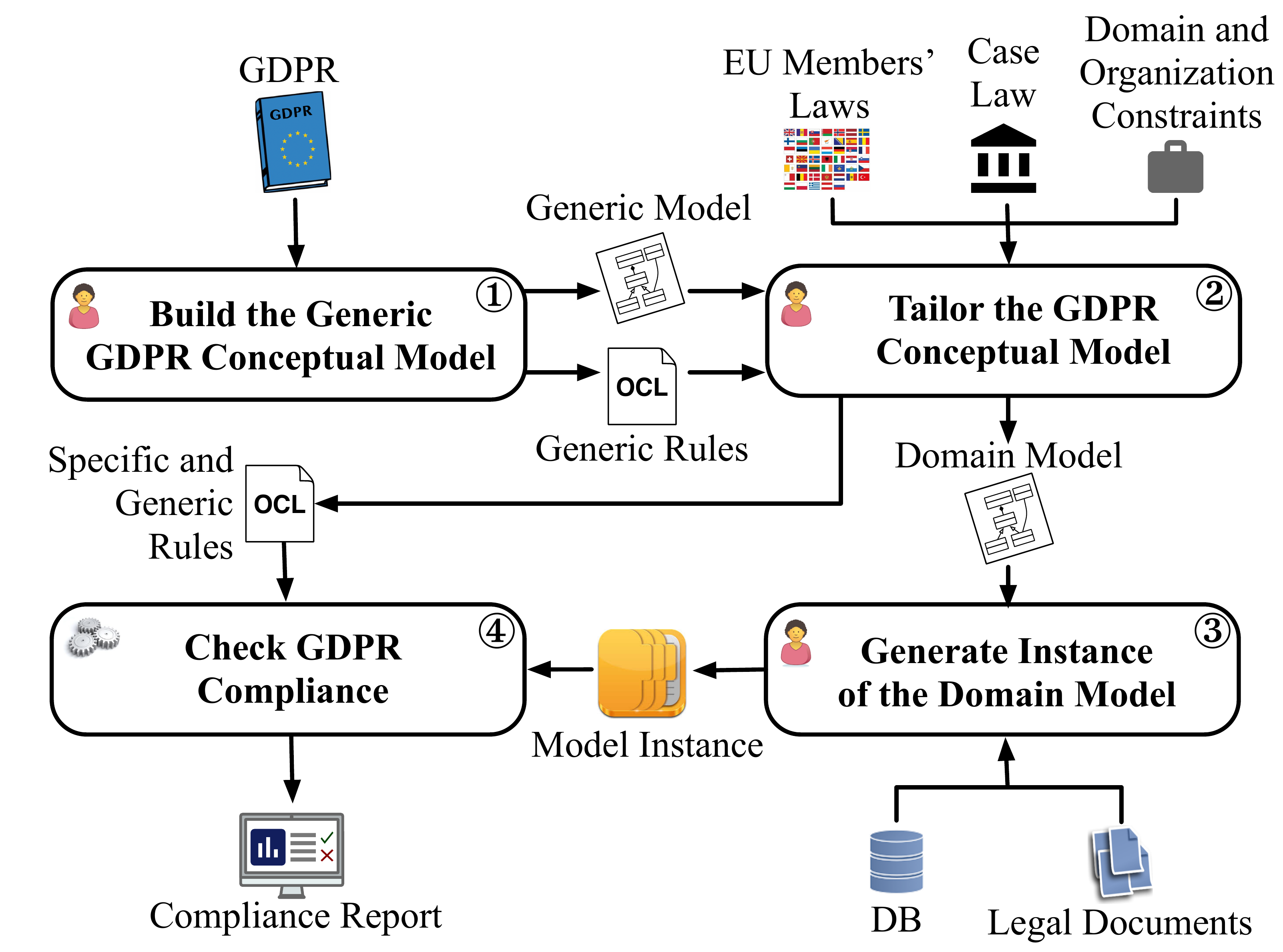}}
\caption{Approach for Automated GDPR Compliance Checking}
\label{fig:overview}
\end{figure}

Our approach for enabling automated GDPR compliance checking has four steps, as depicted in Fig. \ref{fig:overview}.

Step~1 is a manual, one-off task aimed at building a generic model of the GDPR with the help of legal experts. More specifically, the goal of this step is to build, using UML class diagrams and OCL, a context-independent representation of the GDPR that does not take into account specific situations where EU member states’ national laws, case law, or domain/organization decisions may affect the operationalization of the regulation. In this step, we develop, through a qualitative study, the following: (i)~a generic model of the GDPR's main concepts and relationships, (ii) generic OCL constraints that verify GDPR compliance, (iii)~a glossary to facilitate the understanding of the GDPR, and (iv) the variation points describing specific situations where the generic representation needs to be adapted to a given domain or organizational context. 

In step~2, we process the generic model and OCL constraints of step~1 in order to tailor them into a specialized model and a (specialized) set of OCL constraints. The goal of step~2 is to build an actionable basis for implementing the GDPR according to (i) the national laws of EU member states, (ii)~GDPR case law, and (iii) other contextual information that may complement the GDPR. 
Among other things, step~2 yields two outputs that will later enable automated compliance checking in step~3. These outputs are: (i) a specialized model that represents the model tailored according to the application context, and (ii) a set of specialized OCL constraints which contain revised versions of the generic constraints developed in step~1 and potentially new constraints.

Step~3 concerns the development of a model-instance generation tool in order to create instances of the specialized model obtained from step~2.
 This is done via a model-editing tool that allows legal experts to create representations of legal and technical documents in the form of an instance of the specialized model. An example of legal documents would be privacy policy statements, and an example of technical documents would be system requirements specifications.
Stated otherwise, step~3 generates a model instance providing a structured representation of the legal and technical documents that have a bearing on GDPR compliance.

Finally, in step~4, the model instance generated from step~3 is checked against the specialized OCL constraints obtained from step~2.
The compliance diagnostics resulting from the constraint checking process are then delivered to end-users, typically legal experts, in a user-friendly manner. 

In this paper, we describe our experience conducting steps~1 and 2. {Steps~3 and 4 are discussed in another recent piece of work~\cite{Torre2020}}. Steps 1 and 2, along with their inputs and outputs, are discussed in detail in Sections~\ref{modeling} and \ref{variability}, and in the Appendixes A-B.

\section{Modeling the GDPR} \label{main}
\subsection{Building a Generic Model for the GDPR (RQ1)} \label{modeling}
In the first step of our approach (Fig.~\ref{fig:overview}), we build a generic model representing the GDPR without accounting for the specificities of the application domain. 
This modeling activity addresses RQ1 and yields: (1) a UML Class Model (CM) that captures the GDPR's key concepts and their relationships 
{(see Section \ref{Package} of Appendix~A)}; 
(2) a set of 35 OCL constraints over the CM reflecting the GDPR's obligations {(see Section \ref{plain} of Appendix~A  for the plain-English version, and Appendix~B for their OCL version)}. Given a specific context, the applicable constraints need to be completed so that one can evaluate them in an automated and precise manner; 
(3) {a glossary of 267 terms to understand the GDPR model (see Section \ref{Glossary} of Appendix~A)};
and (4) a table that summarizes all variation points extracted from the GDPR {(see Section~\ref{variabilitytable} of Appendix~A)}. The output table mentioned above aims to facilitate the work of analysts in the subsequent tailoring step (Section~\ref{variability}). 
Below, we explain the methodology we employed to create these outputs. We then illustrate the outputs using concrete examples. 

\begin{figure}
\centering
\includegraphics[width=70mm] {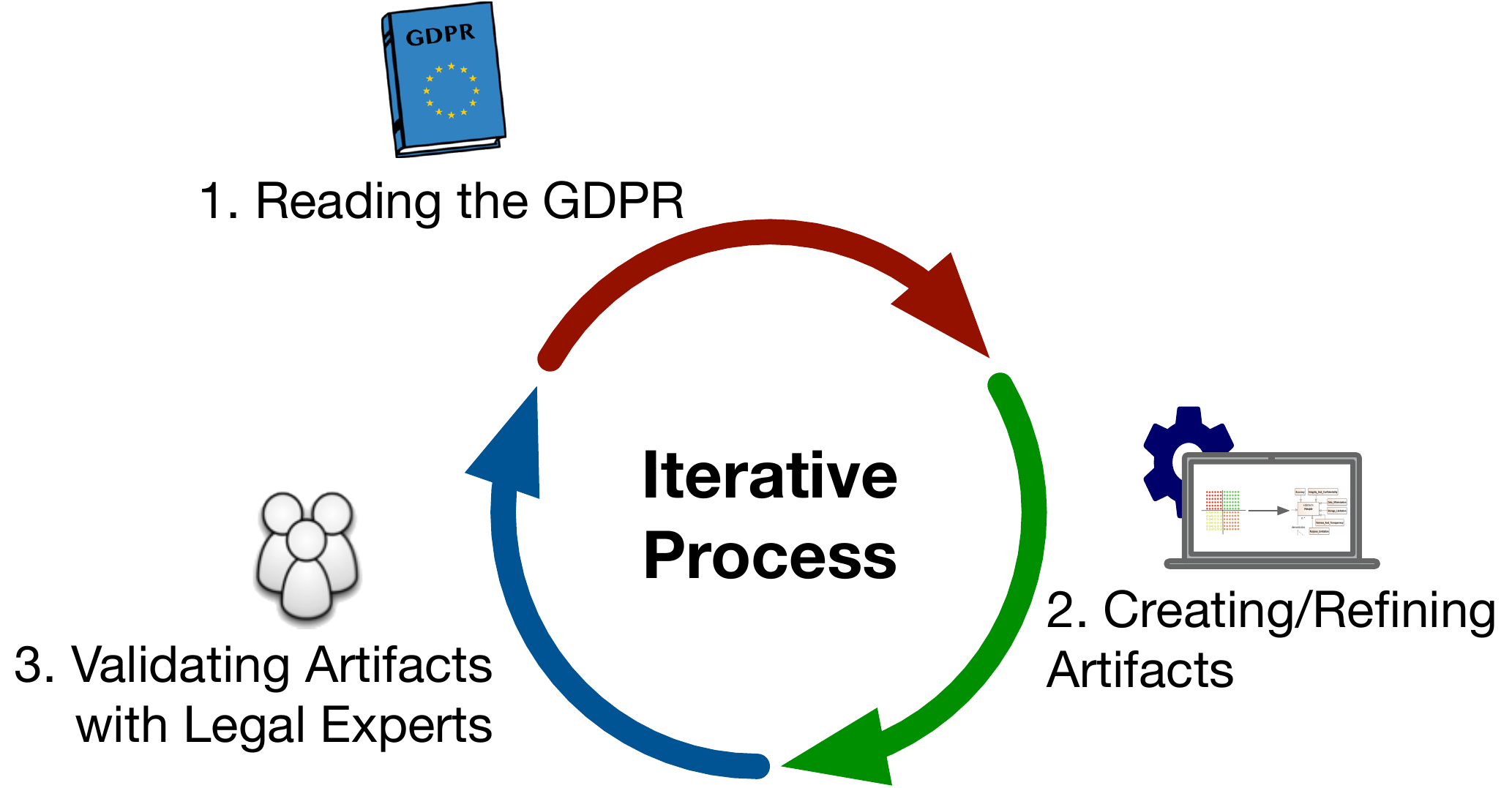}
\caption{{Iterative Process.}} \label{fig:iterativeProcess}
\end{figure}

\textbf{Modeling methodology.}
{This modeling activity was performed in an iterative and incremental manner as shown in Fig.~\ref{fig:iterativeProcess}.} Each iteration was interleaved with a thorough validation session with legal experts, noting that legal experts were already trained to understand the CM notation. 
Building the generic model for the GDPR took four iterations with each iteration requiring on average two weeks. In addition to off-line validation, we had several  face-to-face validation sessions with legal experts, with each of these sessions lasting between 2 to 3 hours. 

\begin{figure*}[t]
	\centering
	\includegraphics[width=1\linewidth]{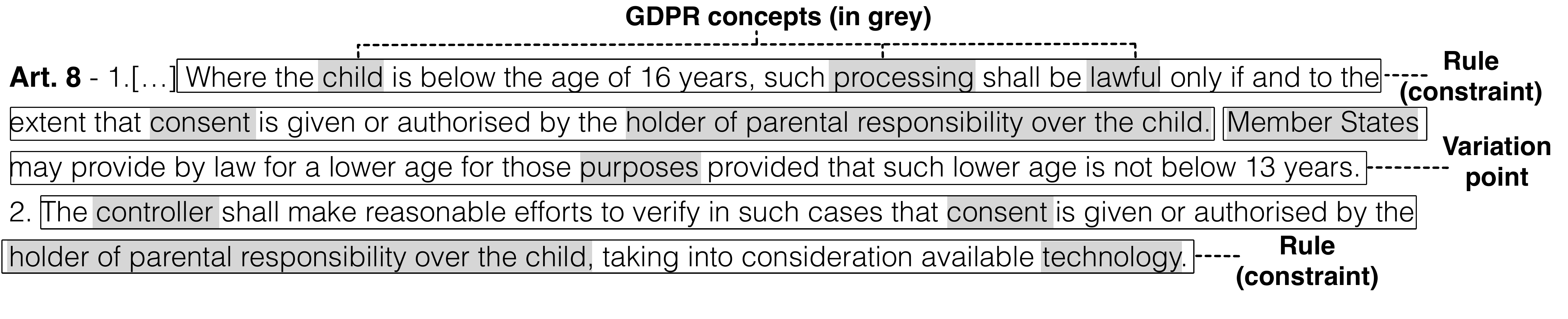}
	\caption{Example of Information Extracted from (Excerpt of) Article 8 of the GDPR \label{fig:Art8}}
	\vspace*{0.5em}
\end{figure*}

During the first iteration, we read the GDPR in its entirety and tried to extract important definitions, concepts, rules and possible variations from it. Figure~\ref{fig:Art8} illustrates the information extracted from Art.~8 -- the article regulating how a child data subject can provide consent for processing her personal data in the context of information society services.
In particular, eleven concepts (shaded gray), two rules, and one variation point were extracted from the excerpt of Art.~8 in Figure~\ref{fig:Art8}. 
Recent work uses natural language processing techniques to extract such legal information in an automated manner~\cite{DBLP:conf/re/SleimiSSBD18}. 
Nevertheless, we opted for a manual strategy to avoid overlooking any important information while deepening our understanding of the GDPR. Among other reasons, a manual strategy was essential for enabling the identification of GDPR rules in a fully precise manner.   
For example,  we have mapped the rules in Art.~8  to their corresponding OCL constraints as we illustrate later. 

Based on the extracted information, and using our understanding and interpretation, we created the modeling artifacts listed earlier.
Next, these artifacts were presented to legal experts for feedback.
{In addition to pointing out issues and omissions, our collaborating legal experts were encouraged to bring to our attention any GDPR article that they suspected might have been misinterpreted, i.e., incorrectly modeled. 
By doing so, we boosted subsequent iterations since we no longer needed to analyze the entire GDPR again.}

In practice, we observed that the corrections suggested by the legal experts were, by and large, based on conventions or articles that were not part of the GDPR itself, e.g., articles from the {Article 29 Working Party (WP)}\footnote{Art. 29 WP is the independent European working party that dealt with issues relating to the protection of privacy and personal data until 25 May 2018 (date at which the GDPR took effect). All archives from Art. 29 WP are available at: https://ec.europa.eu/newsroom/article29/news-overview.cfm. Art. WP 29 has been replaced by the European Data Protection Board; see https://edpb.europa.eu}.
For example, a data controller might need to simultaneously communicate with many supervisory authorities; such authorities are established by individual European member states to supervise compliance with the GDPR.
In such a case, the controller has to designate a unique \emph{lead} supervisory authority (Art. 56). { Subsequently, the controller should only communicate with the lead supervisory authority, which in return, will coordinate any investigation or administrative task with the other concerned authorities.} Although not explicitly stated in the GDPR, the choice of the lead supervisory authority is not arbitrary. The lead supervisory authority should be selected based on predefined rules that account, among other things, for the location of the main establishment of the controller and where the actual data processing is taking place (Working package 244 of the WP).

In the next modeling iteration, we re-read the GDPR parts and other annex documents that were noted by the legal experts in the previous iteration.
Then, we refined the outputs according to expert feedback, and so on. 
Once the conceptual model started to stabilize, we put together a general report including all the resulting outputs for off-line validation. 
The modeling step terminated when the general report, containing all the model artifacts, was approved by the legal experts.

\begin{figure*}[h]
	\centering
	\vspace*{0.5em}
	\includegraphics[width=\textwidth]{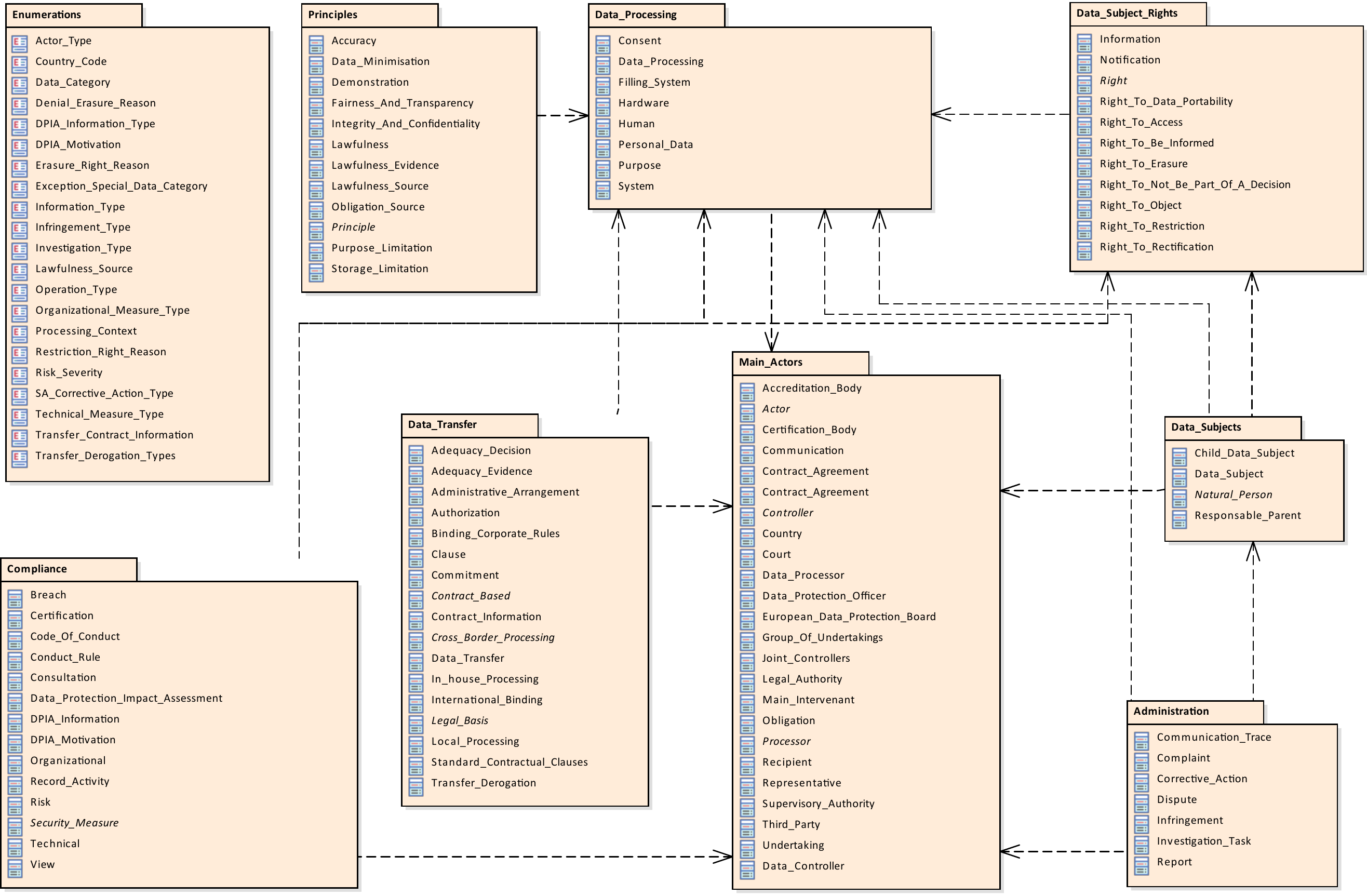}
	\vspace*{0.4em}
	\caption{Package Representation of the CM \label{fig:packages}}
	\vspace*{0.4em}
\end{figure*}

\textbf{Illustration of the modeling artifacts. }
{Fig.~\ref{fig:packages} depicts the package view of the CM. }
To keep the CM manageable and easy to grasp as it grows in size, we spread the CM classes over nine packages as follows, noting the package names are self-explanatory. Packages \emph{GDPR Principles}, \emph{Data Subject Rights}, and \emph{Data Transfer} respectively cover chapters 2, 3, and 5 of the GDPR. Concepts from chapters 1, 4, 8 and 9 were spread over the remaining packages based on their meanings and roles. For example, concepts from chapter~4, which is the longest chapter and where most GDPR compliance requirements are defined, are grouped in packages \emph{Data Processing}, \emph{Compliance Evidence}, and \emph{Actors}. Chapters 6, 7, 10, and 11 have little to no impact on compliance checking, and subsequently were excluded after the first modeling iteration. For example, chapter~6 regulates the internal functioning and composition of the public data supervisory authorities. {The nine CM packages, their traceability with GDPR, and their description are presented in Appendix A.}
{In Fig.~\ref{fig:CM}, as an example, we show an excerpt of the \emph{Data Processing} package that covers most concepts extracted from \hbox{Art.~8 in Fig.~\ref{fig:Art8}.}}

Intuitively, the CM in Fig.~\ref{fig:CM} presents the information that has to be collected when the lawfulness of data processing is based on consent.
In the CM, only data processing manipulating some personal data should be considered. Other kinds of processing are out of scope. The purposes for each processing have to be explicitly defined (see \emph{realizes} association between \emph{Data Processing} and \emph{Purpose}), noting that several instances of processing can share one or more purposes. 
A well-designed consent form should, among other things, remind data subjects of all their applicable GDPR rights.
Consent is given by data subjects, or their responsible parent in case of a child data subject, for one or more predefined processing purposes (see \emph{given for} association between \emph{Consent} and \emph{Purpose} and  \emph{gives} association between \emph{Data Subject} and \emph{Consent}). 
This is only possible when the treated personal data is sufficient for the precise identification of data subjects (see \emph{identifies} association between \emph{Personal Data} and \emph{Data Subject}). 

\begin{figure*}[t]
	\centering
	\includegraphics[width=0.85\linewidth]{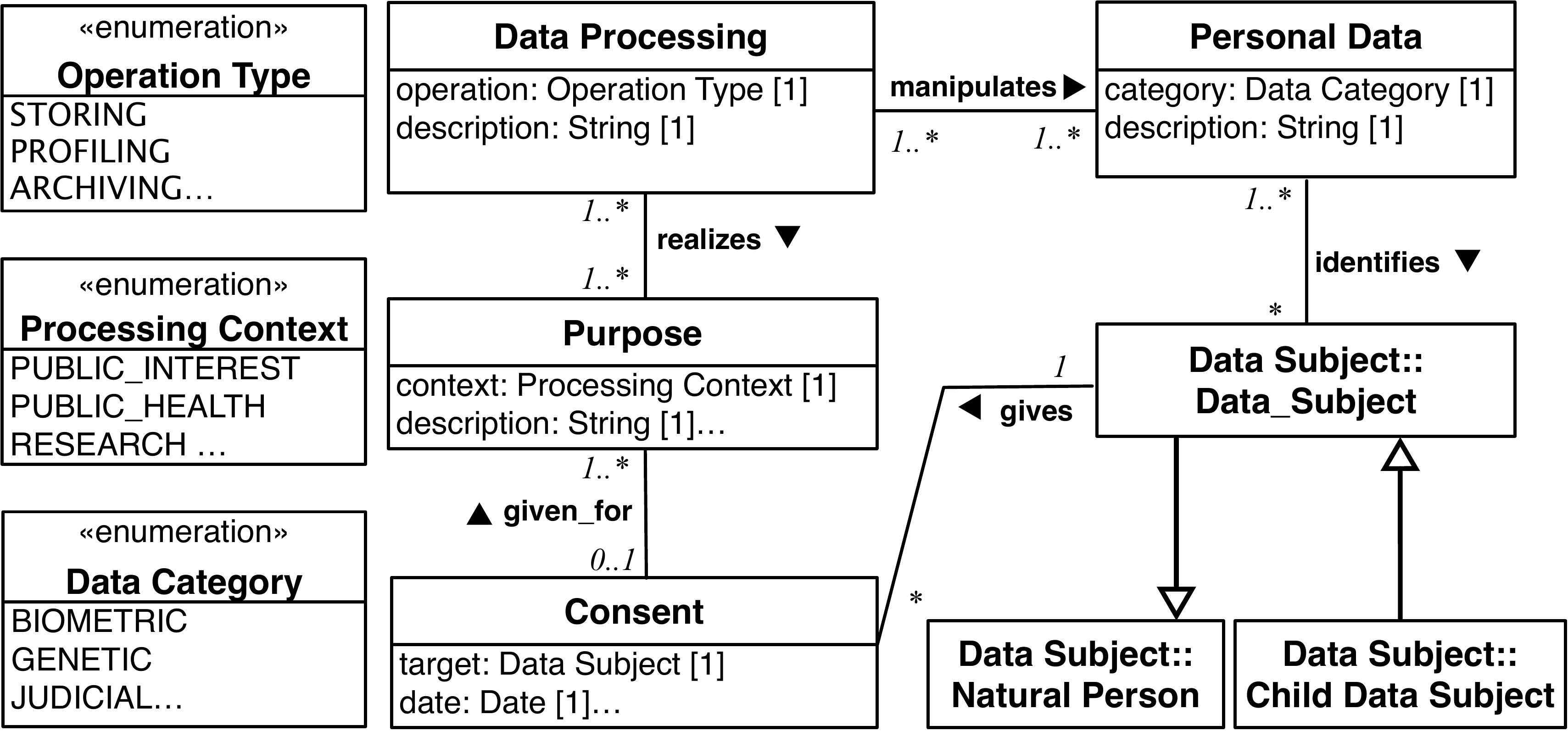}
	\caption{Excerpt of the \emph{Data Processing} Package \label{fig:CM}}
\end{figure*}

{The CM comes with 35 constraints, as presented in plain English in Section~\ref{plain} of Appendix~A, and in OCL in Appendix~B.} 
OCL constraints are expressed as invariants denoting logical conditions that must always hold over all instances of a given class.
Listing~1 presents three OCL constraints related to excerpt of the \emph{Data Processing} package in Fig.~\ref{fig:CM}. 

For example, the invariant named {\sf C5} checks that when lawfulness is based on consent (L.~2), the consent for child data subjects has to be provided by their legal responsible parent (L.~3-14).
This constraint involves no variability and does not require any additional tailoring in the subsequent step.



\vspace*{1.5em}

\begin{lstlisting}[linewidth=\columnwidth, xleftmargin=1.2em, breaklines=true, caption={Examples of OCL Constraints}]
context Data_Processing inv C5:
self.isLawfulnessOnlyByConsent() implies
let identifiableSubjects : Set(Data_Subject) = self.personal_data.data_subject->flatten()->asSet() in self.purposes->forAll(p:Purpose|
    identifiableSubjects->forAll(ds: Data_Subject|
        let eligibleToGiveConsent:Natural_Person = 
        if ds.oclIsTypeOf(Child_Data_Subject)
            then ds.getResponsibleParent()
            else ds endif in
        p.getConsents()->forAll(c:Consent|
            c.provider=eligibleToGiveConsent
            and c.target=ds
        )
    )
)
context Data_Subject inv V1:
    let minDSAge: Integer = Variability.V_getMinimumAgeForDS(self) in
    if (self.oclIsTypeOf(Child_Data_Subject))
        then self.getAge() < minDSAge
        else self.getAge() >= minDSAge endif
context Natural_Person inv V2:
    self.children - > forAll(c: Child_Data_Subject | self.V_checkParentDocuments(c))
\end{lstlisting}
\vspace*{1em}

Constraints involving variability are distinguishable by their name, which includes the ``{\sf V}" prefix, e.g., {\sf V1}. 
We handle variability in OCL constraints using partially specified operations that need to be later updated or redefined based on the context at hand. 
For example, the second constraint (L.~15-19) states that the age of data subjects should be greater than a certain dynamic threshold. 
However, when the context is known, the operation {\sf V\_getMinimumAgeForDS} should dynamically identify the value of the threshold based on the country of residence of the data subject and the locations of the involved data processing, controllers, and processors. Finally, {\sf V2} (L.~20-21), checks that a given person is indeed the holder of parental responsibility over a given child data subject. We further discuss variability in Section~\ref{variability}.

{To ease the understanding of the modeling artifacts for legal experts, we rely on a glossary of important terms. The glossary has 267 entries for the CM (see Section \ref{Glossary} of Appendix A). In addition, we include intuitive descriptions for each OCL constraint (see Section \ref{plain} of Appendix A). 
Table~\ref{tab:glossary} presents an excerpt of the glossary that supports the CM package in Fig.~\ref{fig:CM}. The plain English description of the OCL constraints in Listing~1 and their traceability to the GDPR can be found in Section \ref{plain} of Appendix A.
The first column of Table~\ref{tab:glossary}  points to the modeled concept (i.e., term in the table) such as the classes. 
The second column presents an intuitive natural-language description of the element in the first column. 
For example, the class Data Processing is described in the second row of Table~\ref{tab:glossary}.
In addition to the description, we present in Section \ref{Package} of Appendix A the GDPR source articles of the elements of the CM packages showed in Figure \ref{fig:packages}. 
Here, traceability is meant to help legal experts during the validation sessions.
In particular, it makes it easier to spot whether we have missed some important articles that might further consolidate the definition of a given concept.}

\begin{table*}[bt]
\caption{Glossary Excerpt}\label{tab:glossary}
\fontsize{7}{9}\selectfont
\begin{tabular}{|p{2.5cm}|p{14cm}|} 
\hline
\textbf{Term} & \textbf{Description} \\
\hline
Personal Data & Personal data means any information relating to an identified or identifiable natural person. 
\\
\hline
Data Processing & Is any operation performed on personal data, whether or not by automated means, including collection, recording, organization, structuring, storage, etc. 
\\
\hline

Data Subject & A natural person whose personal data is processed by a controller or processor. 
\\
\hline
Consent & It means any freely given, specific, informed and 
unambiguous indication of the data subject's wishes by which he or she, 
by a statement or by a clear affirmative action, signifies agreement to 
the processing of personal data relating to him or her.  
\\
\hline
Purpose & The purpose of data processing is said to be lawful if its legal basis matches one of the possible circumstances under which GDPR permits the processing of personal data. Example of valid legal basis for data processing are 
consent and when processing is necessary to perform or prepare for a contract with the data subject. 
 \\
\hline
\end{tabular}
\normalsize
\end{table*}

The last modeling artifact is a table including all possible variation points extracted from the GDPR. We defer the discussion and presentation of this table to Section~\ref{variability}.

\subsection{Specializing the Generic Model (RQ2)} \label{variability}

\begin{table*}[bt]
\caption{Excerpt of the Variability Table}\label{tab:variability}
\fontsize{7}{9}\selectfont
\begin{tabular}{|p{1cm}|p{1.5cm}|p{8cm}|p{5cm}|}
\hline
\textbf{ID} &\textbf{Source} & \textbf{Description} & \textbf{How to resolve}\\
\hline
V1 &  Article 8 & EMS law may provide for a lower age   (. . . )   provided   that   such lower age is not below 13 years. & Add the V1 OCL constraint to the generic model and implement {\sf V\_getMinimumAgeForDS} based on the EMS laws.   \\
\hline

V2 &  Article 8 & Checks that a given person is indeed the  holder  of  parental  responsibility  over  a  given data subject according to the EMS law. & Add the V2 OCL constraint to the generic model and implement {\sf VcheckParentDocuments} based on the EMS laws.
\\
\hline
V3 & Article 9 & The processing of sensitive personal data is prohibited unless the data subject has given explicit consent (…), except where (…) EMS law provide that the prohibition (…) may not be lifted by the DS & Implement \newline {\sf V\_prohibitionCanBeLiftedByConsent} of the constraint C6 based on the EMS laws.
\\
\hline
V4 &  Article 9 & EMS  may  maintain  or  introduce  further  conditions,including limitations, with regard to the processing of genetic data, biometric data or data concerning health & Add the V4 OCL constraint to the generic model and implement {\sf VcheckParentDocuments} based on the EMS laws.
\\
\hline
... & ... & ... & ...
\\
\hline
V10 &  Article 49 & In  the  absence  of  an  adequacy decision,  (. . . )  EMS  law  may,  for important  reasons  of  public  interest, expressly set limits to the transfer  of  specific  categories  of personal data to a third country or an international organisation.  & Add the V10 OCL constraint to the generic model and implement {\sf V\_verifyTransferLimits} based on the EMS laws. 
\\
\hline
\end{tabular}
\normalsize
\end{table*}

{In the second step of our approach (Fig.~\ref{fig:overview}), analysts tailor the generic modeling artifacts to account for the specific context and activities of the organizations seeking compliance. This step addresses RQ2. 
Generally speaking, analysts have to resolve all the variations that are relevant to the context at hand. This might introduce new constraints coming from the specific law of a European Member State (EMS), GDPR case laws, and other contextual information that may complement the GDPR. In this paper, we focus on addressing the variability that may come from the specific law of a EMS as expressed in the GDPR. 
The output of this step is a specialized and augmented version of the modeling artifacts created in the first step of our approach (Section~\ref{modeling}). 
As mentioned in Section~\ref{vision},  the variability in the GDPR comes from the fact that the interpretation or the enforcement of some provisions may be affected by additional acts and laws from the EMS. 
Table~\ref{tab:variability} presents an excerpt of the variability table that contains five variation points (two of them, namely {\sf V1} and {\sf V2}, used during the example showed in the first step of our approach). The complete set of 20 variation points is presented in Section \ref{variabilitytable} of Appendix A.
This table will guide analysts in better understanding when and how they should resolve a given variability.
} 
{The first column of the table represents the identifier of the variation point. This identifier will be later used to point to the OCL constraint or model adaptation that is needed to solve a given variation (see Appendix C). 
For example, the variation point V1 is addressed by the OCL constraint {\sf V1} shown in Fig.~\ref{fig:Art8}.} 
{The second column of Table~\ref{tab:variability} traces the variability to the GDPR. 
} 
{The third column provides an intuitive textual description of the variation. 
The third column indicates also the actor that should be consulted for resolving the variation, e.g., the EMS.
Note that the description also covers when the underlying actor is likely to influence the interpretation and enforcement  of the original GDPR rules. For example, in {\sf V10}, the EMS law may,  for important  reasons  related to  public  interest, expressly set limits to the transfer  of  specific  categories  of personal data to a third country or an international organisation. At the time this article was written, data can be transferred within the same international organization to Switzerland without additional obligations. However, unconditional data transfer to third countries is limited, for example, transfer to Canada is only limited to commercial organizations under Canadian's PIPEDA law (Personal Information Protection and Electronic Documents Act). Other sectors and domains involve additional obligations that need to be fulfilled such as the approval of the lead supervisory authority.  
} 
{The fourth column of Table~\ref{tab:variability} provides the context required by analysts to understand how they should resolve the variation. 
}

 
 \begin{table*}[bt]
\caption{Variability Resolution Table}\label{tab:actions}
\fontsize{7}{9}\selectfont
\begin{tabular}{|p{1cm}|p{3cm}|p{12cm}|}
\hline
  \centering \textbf{Ref.} &      \centering \textbf{Artifact} &  \multicolumn{1}{c|}{\textbf{Summary of actions}}  \\
  \hline
V1 & - & [Not applicable] \\
\hline
V2 & - & [Not applicable] \\
\hline
V3 &  Specialized Model &  The specialized model includes an update version of the constraint C6.  \\
\hline
V3 &  OCL constraints & Implement {\sf V\_prohibitionCanBeLiftedByConsent} of the constraint C6 based on the EMS laws.
\\
\hline
V3 &  Glossary & Add the terminology of the implementation of {\sf V\_prohibitionCanBeLiftedByConsent} to the glossary. 
\\
\hline
V4 &  Specialized Model &  The specialized model includes the new constraint V4.  \\
\hline
V4 &  OCL constraints & Add the V4 OCL constraint to the generic model and implement {\sf V\_verifyFurtherConditionsAndLimit} based on the EMS laws.
\\
\hline
V4 &  Glossary & Add the terminology of the new constraint V4 and the implementation of {\sf V\_verifyFurtherConditionsAndLimit} to the glossary. 
\\
\hline
... &... & ...\\
\hline
V10 & - & [Not applicable] \\
\hline

\end{tabular}
\normalsize
\end{table*}


{The strategy we employ for resolving most of the variation points is ``clone and own''~\cite{Clements2001}, where the generic artifacts are specialized for the organization and system(s) at hand. Examples of changes to the artifacts include updating the cloned CM, glossary, and the OCL constraints. Further, analysts can add new OCL constraints, and drop or override existing ones.
The only artifact from the first modeling step of our approach that remains unchanged is the variability table (Table~\ref{tab:variability}). 
This is because the variability table incorporates all possible variations with regard to the GDPR and is used as a checklist for guiding the analysis during the tailoring step. Concretely, analysts skim through the variability table and resolve the variations that apply to the underlying context. 
} 
{An important challenge here is keeping track of the changes made for specializing the modeling artifacts.
To do so, analysts have to record the actions they have taken to tailor the generic modeling artifacts. 
To illustrate, let us suppose that an organization~$X$ is an international commerce company located in Europe and Canada processing sensitive personal data. } 

{Table~\ref{tab:actions} presents an example of how the variability in Table~\ref{tab:variability} would be handled for $X$.
The first column of Table~\ref{tab:actions} references a particular variation ID listed in Table~\ref{tab:variability}, whereas the second column  of Table~\ref{tab:actions} lists the cloned generic artifacts that were impacted during the resolution of the variation. The final column of Table~\ref{tab:actions} describes how the artifacts in the second column were updated based on the specific context of $X$. 
$X$ must account only for {\sf V3} and {\sf V4} in Table~\ref{tab:variability}. {\sf V1} and {\sf V2} in Table~\ref{tab:variability} do not apply to $X$ since $X$  only trades with subjects aged over 18 years old (clearly stated in $X$'s privacy policy and website).  {\sf V10} does not apply to $X$ since $X$ has an adequacy decision. $X$ is also requested to conduct a DPIA (Data Privacy Impact Assessment) to be able to perform cross-border data transfer to Canada with an adequacy decision. The specific adequacy decision for Canada is called PIPEDA. Other generic constraints (i.e., {\sf C27}, {\sf C28}, {\sf C32}, {\sf C33} among other rules) handle the DPIA and the adequacy decisions.
} 

{As shown in Table~\ref{tab:actions}, only the variation points {\sf V3} and {\sf V4} are relevant for $X$. Specifically, the cloned and possibly specialized model,  OCL constraints, and glossary were altered as described in the last column of the table. 
For example, to address {\sf V3}, the OCL constraint {\sf C6} that checks that the processing of special data categories are prohibited unless the processing is based on consent needs to be updated by implementing {\sf V\_prohibitionCanBeLiftedByConsent} if the the prohibition can be lifted by consent based on the EMS laws (third to fifth row of Table~\ref{tab:actions}).  
Instead, to address {\sf V4}, the new OCL constraint {\sf V4} needs to be added to encode the variation point. In addition, the analyst has to implement {\sf V\_verifyFurtherConditionsAndLimit} in {\sf V4} as it is imposed by one of the EMS laws that are relevant to $X$ (sixth to eight row of Table~\ref{tab:actions}).
\newline The analyst that is tailoring the specialized model should take into account two main points.  First, regardless of the changes made, the OCL constraints should remain correct with respect to the cloned and possibly specialized CM. For example, if the analyst decides to drop the class \emph{Consent} and its associations, then all impacted constraints have to be either corrected or dropped. 
Second, the analyst might unintentionally introduce inconsistencies in the set of OCL constraints, e.g., two contradicting constraints.  
To avoid this, one can employ existing constraint solvers, e.g., UML2CSP~\cite{UMLtoCSP}, Alloy~\cite{UMLtoCSP} or PLEDGE~\cite{PLEDGE}, to spot UNSAT sets of constraints, and consider if any of the UML consistency rules discussed in the literature could apply to the application context~\cite{Torre2018}.}

\subsection{Challenges Encountered during the Modeling (RQ3)} \label{issues}

In this section, we address RQ3 by listing the main challenges encountered when modeling the GDPR. Later, in Section \ref{directions}, we present our vision for how we intend to address these challenges.

\textit{Specification of Compliance Rules (Challenge 1)}: In this paper, we first use OCL constraints to embed compliance rules in the generic model. We then adapt and expand these constraints to create a specialized model. We have already taken care of defining OCL constraints over the generic model; no additional effort is thus foreseen for this task. Nevertheless, additional effort, including by legal experts, will be required for defining OCL constraints over the specialized model, noting that these constraints necessarily refer to legal material (e.g., EMS laws, GDPR case law, and domain adaptations) that is more complex and fragmented than the GDPR. Due to the scarce familiarity of legal experts with OCL, the creation of the latter group of constraints may be difficult and time-consuming. 

\textit{Rationale for Model Specialization (Challenge 2)}: Although we keep track of all the actions performed during the tailoring step, we do not systematically express the rationale behind the actions; in other words, we do not document why analysts made the decisions they did~\cite{Shum1994}. In the context of our work, the rationale needs to cover the problems the analysts encountered, the options they investigated, the GDPR provisions they examined to evaluate the options, and, most importantly, the arguments that led them to make certain decisions.

\textit{Generation of the Instance Model (Challenge 3)}: The process of generating an instance of a specialized model is currently dealt with manually (recall step 3 in Fig.~\ref{fig:overview}). This would mean that a legal expert would have to create, by using a model editor, the instance. This manual process is time-consuming and tedious.




\section{Lessons Learned} \label{lessons}
In this section, we discuss the lessons we learned from modeling the GDPR.

\vspace*{.2em}\textbf{Streamline the validation process.}
We observed that modeling the GDPR, whether at a generic or specialized level, necessitates substantial legal knowledge and expertise that may go beyond the GDPR itself, e.g., knowledge of the Article 29 Working Party. Thus, putting in place an effective and efficient validation process with legal experts was paramount to ensure that the produced artifacts were as complete and precise as possible. To achieve this goal, we had to shield the legal experts from the complexity arising from the CM and its underlying OCL constraints. 

As discussed in Section~\ref{modeling}, legal experts were able to grasp the CM with relative ease. 
This was in large part thanks to the intuitiveness of UML class diagrams and the fact that non-software experts can be quickly trained to obtain a working understanding of the notation for validation purposes. In general, we observed from experience that class diagrams can be taught to non-IT experts with relative ease.
In contrast, OCL, which we use to formally express the GDPR rules, was challenging and intimidating to legal experts, despite our attempts to explain the meaning of the constraints. 
Similar communication barriers were observed when we attempted to replace OCL with other logical notations, e.g., standard first-order logic. In general, we believe such barriers are to be expected when formal logic is used directly with professionals who do not have adequate mathematical background.
We mitigated this issue by describing each OCL constraint via an intuitive but precise textual description in natural language {(see Section~\ref{plain} of Appendix A)}. 
Nevertheless, {the plain-English explanation of the OCL constraints} per se was still not enough to ensure reliable validation of the OCL constraints. 
In particular, the same rule can be often expressed over smaller and modular sub-constraints. 
For example, the excerpt of the article of Fig.~\ref{fig:Art8} was encoded over three constraints, namely {{\sf V5}, {\sf V1} and {\sf V2} in Listing 1.} The former constraint encodes the common part of the rule, whereas the latter two (variation points constraints) cover the variable part. 
With the rules getting fragmented, legal experts experienced difficulties because they could no longer relate to the original rule.  
One way to remedy this problem is by forcing one-to-one mappings, where any GDPR rule is expressed using a unique OCL constraint.
However, such a solution will further complicate the tailoring step, since variant requirements will have to be mixed with the fixed ones. This prompted us to create (1) a plain-English constraints table (see Section~\ref{plain} of Appendix A) which traces the GDPR rules to their corresponding constraints, and (2) the variability table (see Section \ref{variabilitytable} of Appendix A) that contains the variation points expressed in plain-English traced to the GDPR.
Both previous tables facilitated the validation of the OCL constraints by legal experts. 

The variability table (e.g., see Section ~\ref{variabilitytable} of Appendix A) was enough to enable the legal expert to verify that the list of extracted variation points was complete and precise. 
A simple but effective solution was to support several views for the same CM, where the level of detail to display is configured according to needs.
Although the validation of the CM was conducted package by package, legal experts still found the models to be overwhelming in terms of their information content. To this end, we found out that, in many situations, hiding class operations, attribute types, and stereotypes would be helpful.
 Further, to ensure that enough time was given for validation, we alternated on-line and off-line validation as discussed in the modeling methodology of Section~\ref{modeling}.

\vspace*{.2em}\textbf{Maintain traceability.} Another observation from our GDPR modeling experience is that both analysts and legal experts often needed to consult specific articles to refresh their memory. Being able to do so effectively required all our modeling artifacts to be traceable to their corresponding GDPR provisions.  
{Examples of traceability links can be seen in the second column of the tables in Section~\ref{Package} of Appendix A, the plain-English constraints table (Section~\ref{plain} of Appendix A) and the variability table (Section~\ref{variabilitytable} of Appendix A).} Although not shown in Fig.~\ref{fig:CM}, classes too are traceable to the specific GDPR provisions pertaining to them at the level of the CM. For example, the class \emph{Purpose} in Fig.~\ref{fig:CM} is mapped to Arts. 5, 13, 14 and 15. These links made it easy to go back and forth between the modeling artifacts and the GDPR. {We anticipate the links to be useful for other purposes as well, e.g.,  performing impact analysis when the GDPR, or the EMS laws change.}
 The only classes that were not mapped to the GDPR are those we have created to better structure the CM, e.g., the \emph{Processing Activity Record} class.

A further final observation about traceability concerns the importance of maintaining consistent relationships between the different modeling artifacts. In practice, one often needs to quickly navigate from one artifact to another, in particular during the tailoring step. For example, when resolving a given variability, it is often useful to view the list of rules whose fulfillment is likely to be impacted by the EMS laws. Similarly, analysts need to navigate to the underlying OCL constraints that need to be updated. {Examples of such links can be found in the third column of plain-English constraints tables (Section \ref{plain} of Appendix A) and the second column of variability tables (Section ~\ref{variabilitytable} of Appendix A). We received positive feedback from the legal experts about having such navigable artifacts. In particular, legal experts appreciated the intuitive and GDPR-traceable approach of our model artifacts that allows them to maintain the integrity of the artifacts by reducing arduous tasks such as going back to read a specific article of the GDPR related to a class (without traceability).
}

\vspace*{.2em}\textbf{Make the tailoring step as systematic as possible.}
During  the tailoring step,  we  observed  that  even experienced analysts  could encounter difficulties in resolving the variation points. The root cause of this was the large number and size of the modeling artifacts.   This  prompted  us  to develop simple guidelines to systematize and better organize the tailoring step. 
First, analysts have to go through the variation points and tick those that are relevant to their working context. 
 Then, analysts can focus only on the relevant variation points and apply our recommendations on how to resolve them. 
 This was facilitated by the ``How to resolve'' column of the variability table (see Section \ref{variabilitytable} in Appendix A). 
 For example, when V1 in Table~\ref{tab:variability} is relevant to the context, analysts will get to know that they have to update {\sf V\_getMinimumAgeForDS} to account for the minimum age of children as regulated by the relevant  EMS laws. 
 However, we do not recommend a sequential resolution of variation points, e.g.,  first resolving V1, then V2, then V3, and so on. 
 In particular, analysts should postpone completing the specification of the OCL constraints until all the variability for the CM has been handled. This is because some changes in the CM might break other constraints for which variability was previously resolved.  
 To help analysts follow these recommendations, we proposed to keep track of all the tailoring actions in the resolution table (see Table~\ref{tab:actions}). This facilitates resolving the variabilities in an incremental and non-sequential manner.
In line with the above, a recent work from Hajri et al. proposes a tool-supported approach that guides analysts in configuring product specific models from product line models ~\cite{hajri2018} \cite{Hajri2016}. In the future, we envisage  to operationalize our tailoring recommendations by customizing Hajri et al.'s work. 
 
Finally, we found the resolution table to be very useful when we had to deal with several similar contexts. In such cases, we started  the tailoring from specialized modeling artifacts produced for other similar contexts, rather than from the generic artifacts. This, in our experience, helps to expedite the tailoring step.

\section{Future Directions} \label{directions}
In this section, we describe the most important future directions that, we believe, are necessary for addressing the challenges identified in Section \ref{issues}.

\vspace*{.2em}
\noindent\textbf{Domain-Specific Rule Language (Challenge 1).} 
Using OCL constraints is key to achieving automation in checking GDPR compliance. 
In our approach, this is done via the specialized set of OCL constraints that encode the rules applying to a given context. 
Nevertheless, some of the specialized constraints, in particular, the new ones originating from the EMS laws, have to be validated by legal experts. As discussed in Section~\ref{lessons}, OCL impedes understandability by legal experts. To tackle this limitation and improve the tailoring of the generic model (Step 2 in Fig. \ref{fig:overview}), it would be advantageous to develop a Domain-Specific Rule Language (DSRL). 
The DSRL should, on the one hand, be expressive enough to be useful for the precise specification of GDPR compliance checking rules, and on the other hand, understandable enough to be readily used by legal experts. For example, the OCL rule presented in Listing 1, would be hardly understood by most legal experts. To ease understandability, restricted natural language (NL) could be used as the basis for the DSRL. While basing the DSRL on NL increases usability, there is still the risk that legal experts may find it difficult to articulate their rules in a proposed language. To mitigate this issue, one needs to closely interact with legal experts during the DSRL design, and  iteratively validate the language constructs with them. In addition, providing training material for the DSRL would be essential to make the language more accessible to non-software experts. Finally, to support automated compliance checking, the rules specified in the DSRL should be automatically translatable into OCL so that the rules can be checked directly over instantiations of a specialized GDPR model.

\vspace*{.2em}
\noindent\textbf{Goal Models (Challenge 2).} Using goal models can help to deal with capturing and reasoning about the rationale for model specialization. Each goal is a prescriptive statement of intent that a system should satisfy \cite{vanLamsweerde2009}. Here, the term “system” refers to a combination of IT applications, organizations, work-flows and people that together perform certain functions. A goal model is characterized by a collection of goals, the relationships (e.g., hierarchical decomposition) between the goals, and the obstacles that could hinder the satisfaction of the goals. Goal models provide a flexible instrument for arguing about model specialization. A key task related to a goal-oriented analysis of the GDPR would be to decide how the application context discussed in Section \ref{tab:variability} (Step 2 in Fig. \ref{fig:overview}) should be decomposed and analyzed in order to tailor the specialized model. This decomposition necessarily involves breaking down the GDPR's core tenets (e.g., data minimization) into more tangible sub-goals. Additionally, one may need to decompose the goals of a given system (e.g., a specific organization), and examine how the system goals map onto the goals stipulated by the GDPR. A main criterion to fulfill regarding goal decomposition would be to ensure that the decomposition process makes progress towards a set of concrete claims for which meaningful evidence about satisfaction (in term of model specialization) could be collected. Meeting this criterion necessitates that the developed goal models should provide a blueprint for the justification that is needed in order to argue about the adequacy and effectiveness of a proposed model specialization.

\vspace*{.2em}
\noindent\textbf{AI-enabled Automation Support (Challenge 3).} Legal documents typically come in the form of NL descriptions. Mining these descriptions to identify the appropriate metadata to build the instance model is a prerequisite for automated compliance checking. Metadata items relevant to GDPR are numerous. Examples of such metadata include: “purpose” to mark the purposes of the processing for which personal data is being collected, “basis” to mark the legal basis for the processing of personal data, and “right to access” to mark the clause(s) giving an individual the right to request from the controller access to their personal data. These metadata items have to be identified in legal and technical documents such as privacy policies, consent statements, records of processing activities and exemptions, and data protection impact assessments. Natural Language Processing (NLP) \cite{DBLP:books/daglib/0001548} and  Machine Learning (ML) \cite{DBLP:books/mit/026233758} provide a useful technical platform for metadata extraction \cite{DBLP:conf/re/SleimiSSBD18}. The metadata identified will be the basis for the model-based representation of the legal and technical documents to be checked. In other words, an automatic instantiation process will convert the metadata extracted with NLP and ML for a given document into a model-based representation, i.e., the instance of a specialized model. The elements of this instance model will be both fully traceable to the content of the source document as well as unambiguously mappable onto the underlying generic and specialized models. {The initial results tackling this future direction are presented elsewhere in recently published work ~\cite{Torre2020}}.

\section{Related Work} \label{related}
In  this  section,  we  present  related  work  on  (1)~modeling the GDPR, and (2)~checking compliance.

{We distinguish two categories of studies related to the work presented in this paper: proposals that report on (1) Modeling the GDPR, and (2) Checking Compliance.}

\vspace*{.2em}\textbf{Modeling the GDPR.} {There is some early work on conceptual modeling of the GDPR.} 
In particular, 
Ayala-Rivera and Pasquale \cite{Ayala2018} propose a model-based approach to help organizations understand the data protection obligations imposed by the GDPR. 
Caramujo et al. \cite{Caramujo2019} target privacy policies from the web and mobile applications, and propose a domain-specific language along with model transformations for specifying privacy-policy models. 
Pullonen and Matulevicius \cite{Pullonen2019} present a multi-level model to be used as an extension of the Business Process Model and Notation (BPMN) to enable the visualization, analysis, and communication of the privacy-policy characteristics of business processes. 
Tom et al. \cite{Tom2018} present a preliminary GDPR model aimed at providing a simple, visual overview so that process implementers can better understand the associations between different entities in the GDPR. The authors describe an approach for using their proposed model as a tool to develop an organizational privacy policy along with an illustration of compliance-rule extraction.
These existing strands of work either address narrow analytical use cases (e.g., only the compliance analysis of privacy policies) or focus on providing guidelines for the (manual) application of the GDPR.  We go beyond the existing work by modeling the GDPR in a more holistic way and providing a systematic tailoring mechanism to support GDPR compliance automation in different contexts.

\vspace*{.2em}\textbf{Checking Compliance.} To the best of our knowledge, no automated approach for checking GDPR compliance has been published so far. However, there are a few threads of work that describe methodologies for assessing system compliance. 

Chung et al. \cite{Chung2008} identify non-compliance issues in user-defined process models by matching these models against a standard model during both process specification and process execution. Panesar-Walawege et al. \cite{Panesar2013} propose a model-based approach to aid the suppliers of safety-critical systems in defining the evidence information necessary for certification according to standards and automatically detecting non-compliance issues in the collected evidence.

{Ranise and Siswantoro \cite{Ranise2017} devise an SMT-based tool for checking compliance of security policies at design time.} They introduce an implementation that uses tools for policy analysis based on efficient Satisfiability Modulo Theories (SMT) solvers. 
Guarda et al. \cite{Guarda2017} propose a logic-based framework to support the specification of information system designs, purpose-aware access control policies, and legal requirements.

While being a useful source of inspiration, none of the above approaches can be directly adapted to the GDPR due to their main focus being different than data protection and privacy.

\section{Conclusion} \label{conclusion}
{In this paper, we used UML and OCL to build a model-based representation of GDPR. The key motivation behind this research is to pave the way for the creation of automated, model-based GDPR compliance analysis solutions.
Our research resulted in the development of a generic GDPR model alongside a detailed and well-defined strategy for specializing this model according to different contexts and for satisfying the requirements of different types of GDPR-related analysis.
We presented several artifacts: (1) a GDPR conceptual model with full traceability to the GDPR, (2) a glossary to help explain the conceptual model, (3) 35 compliance rules extracted from GDPR and defined both in plain English and in OCL, and (4) a set of 20 GDPR variation points to specialize the generic model.
Building on the knowledge obtained from our modeling endeavor, we discussed several learned lessons. We also suggested potential strategies for solving the challenges we found in our work and promoting longer-term research focus on the model-based analysis of GDPR compliance.
\newline In the future, we plan to work on the directions presented in Section \ref{directions} in order to enable a full realization of the approach outlined in Section \ref{vision}. Furthermore, we will be working closely with legal experts on implementing a number of compliance analysis use cases, e.g., checking the compliance of data processing agreements with GDPR. 
Doing so will allow us to identify and address high-priority automation needs and help bridge the gap between software engineers and legal experts by developing more effective communication methods.}

\section*{Acknowledgment} \label{acknowledgment}
This paper was supported by Linklaters, Luxembourg's National Research Fund (FNR) under grant \newline BRIDGES/19/IS/13759068/ARTAGO, and NSERC of Canada under the Discovery, Discovery Accelerator and CRC programs.

\clearpage
\section*{{Appendix~A}} \label{appendixA}

In this appendix, we present (a) the packages of the CM presented in Fig. \ref{fig:packages} and the traceability of the packages to the GDPR in Section \ref{Package}, (b) the glossary (Section \ref{Glossary}), (c) the compliance rules in plain English (Section \ref{plain}), and (d) the set of variation points (Section \ref{variabilitytable}).

\subsection{Packages traceability to the GDPR}\label{Package}

\begin{figure*}[h]
	\centering
	\includegraphics[width=17cm]{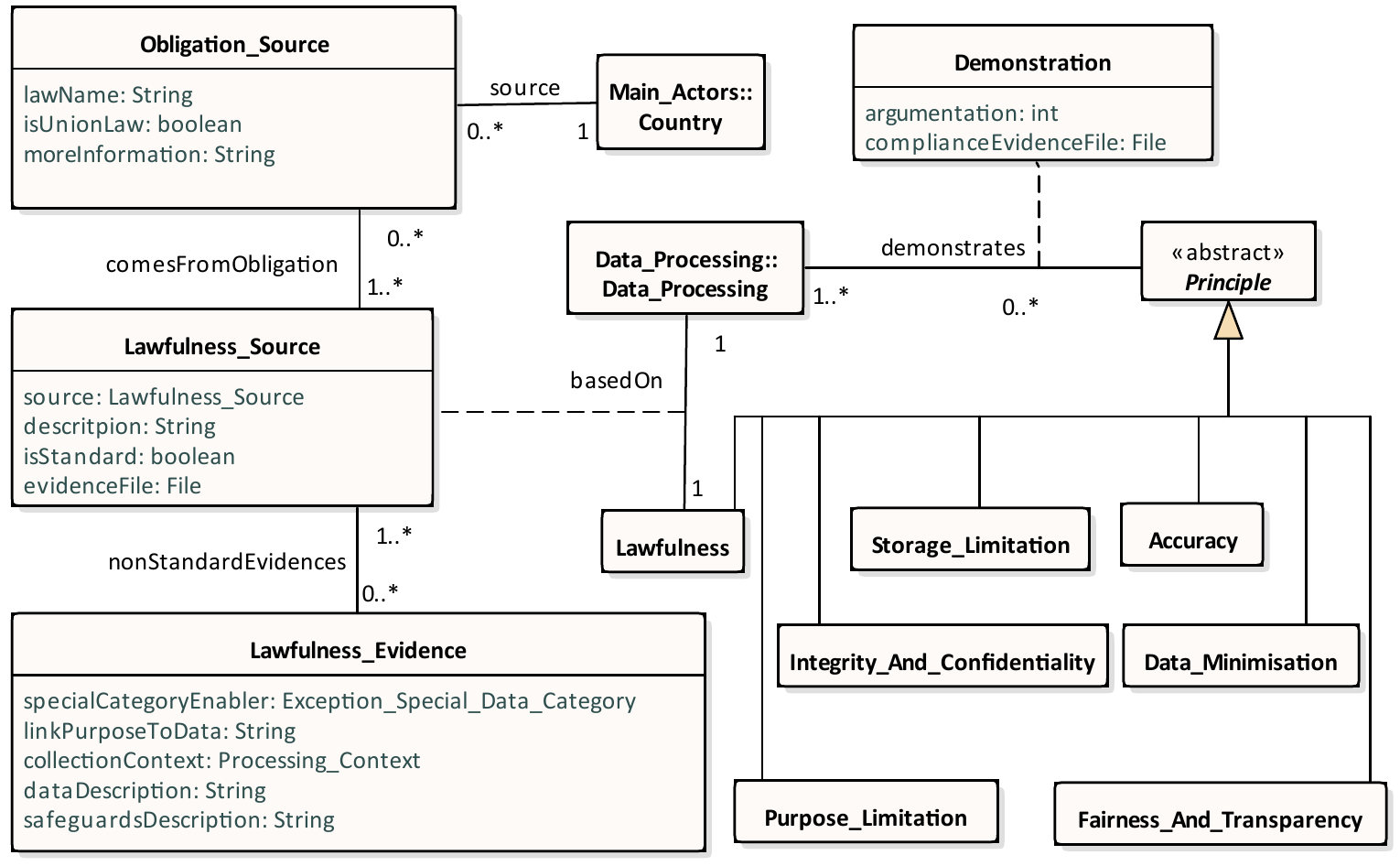}
	\caption{Principles Package \label{fig:Principles}}
\end{figure*}

\textbf{Principles package} is showed in Fig.~\ref{fig:Principles} and its traceability to the GDPR is described in Table~\ref{tab:Principles}.

\begin{table}[H]
\caption{Principles Package}\label{tab:Principles}
\centering
\begin{tabular}{|p{3.5cm}|p{3.5cm}|}
\hline
\textbf{Class} & \textbf{Traceability to GDPR} \\
\hline
{Accuracy} & Article 5 \\
\hline
{Data Minimization} & Article 5 \\
\hline
{Demonstration} & Articles 5, and 25 \\
\hline
{Fairness and 

Transparency} & Article 5 \\
\hline
{Integrity and 

Confidentiality} & Article 5 \\
\hline
{Lawfulness} & Article 5 \\
\hline
{Lawfulness Evidence} & Article 5 \\
\hline
{Lawfulness Source} & Article 6 \\
\hline
{Obligation Source} & Article 6 \\
\hline
{Principle} & Abstract concept \\
\hline
{Purpose Limitation} & Article 5 \\
\hline
{Storage Limitation} & Article 5 \\
\hline
\end{tabular}
\end{table}

\noindent\textbf{Data Processing package} is showed in Fig.~\ref{fig:DataProcessing} and its traceability to the GDPR is described in Table~\ref{tab:DataProcessing}.

\begin{table}[h]
\caption{Data Processing Package}\label{tab:DataProcessing}
\centering
\begin{tabular}{|p{3.5cm}|p{3.5cm}|}
\hline
\textbf{Class} & \textbf{Traceability to GDPR} \\
\hline
{Consent} & Articles 4, 7, and 8 \\
\hline
{Data Processing} & Article 4 \\
\hline
{Filing System} & Articles 2, and 4 \\
\hline
{Hardware} & No mapping to the law \\
\hline
{Human} & No mapping to the law \\
\hline
{Personal Data} & Articles 4, 9, and 10 \\
\hline
{Purpose} & Articles 5, and 13 to 15 \\
\hline
{System} & No mapping to the law \\
\hline
\end{tabular}
\end{table}

\begin{figure*}[h]
	\centering
	\includegraphics[width=18cm]{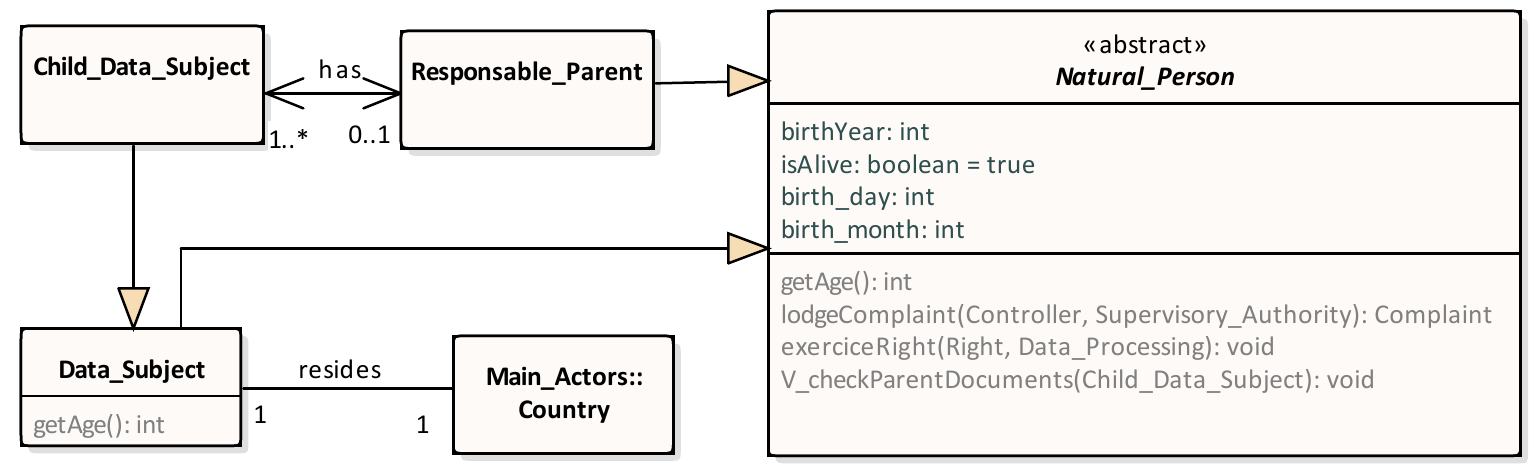}
	\caption{Data Subjects Package \label{fig:DataSubjects}}
\end{figure*}

\newpage
\noindent\textbf{Data Subjects package} is showed in Fig.~\ref{fig:DataSubjects} and its traceability to the GDPR is described in Table~\ref{tab:DataSubjects}.

\begin{table}[H]
\caption{Data Subjects Package}\label{tab:DataSubjects}
\centering
\begin{tabular}{|p{3.5cm}|p{3.5cm}|}
\hline
\textbf{Class} & \textbf{Traceability to GDPR} \\
\hline
{Child Data Subject} & Article 8 \\
\hline
{Data Subject} & Article 4 \\
\hline
{Natural Person} & Abstract concept \\
\hline
{Responsible Parent} & Article 8 \\
\hline
\end{tabular}
\end{table}

\begin{figure*}[t]
	\centering
	\includegraphics[width=24cm, angle=90]{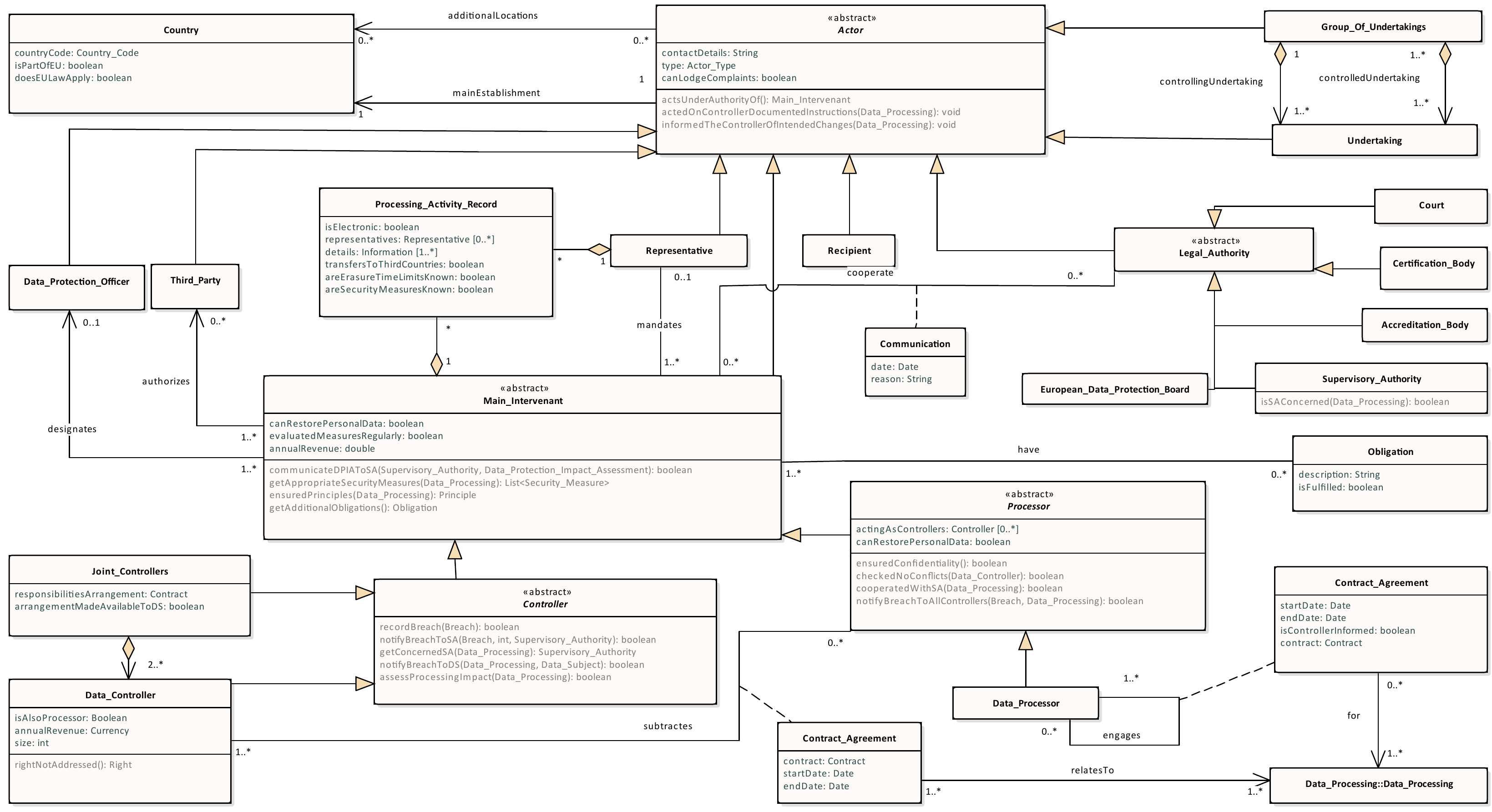}
	\caption{Main Actors Package \label{fig:MainActors}}
\end{figure*}

\noindent\textbf{Main Actors package} is showed in Fig.~\ref{fig:MainActors} and its traceability to the GDPR is described in Table~\ref{tab:MainActors}.

\begin{table}[h]
\caption{Main Actors Package}\label{tab:MainActors}
\centering
\begin{tabular}{|p{3.5cm}|p{3.5cm}|}
\hline
\textbf{Class} & \textbf{Traceability to GDPR} \\
\hline
{Accreditaion Body} & Article 43 \\
\hline
{Processing Activity Record} & No mapping the law \\
\hline
{Actor} & Abstract concept \\
\hline
{Certification Body} & Article 43 \\
\hline
{Communication} & Articles 31, 57, 58, and 33 \\
\hline
{Contract Agreement} & Article 28 \\
\hline
{Controller} & Abstract concept \\
\hline
{Country} & Article 3 \\
\hline
{Court} & Articles 78, 79, and 81 \\
\hline
{Data Processor} & Articles 4, 28, and 29 \\
\hline
{Data Protection 

Officer} & Articles 37 to 39 \\
\hline
EU Data Protection & Articles 68 to 76 \\
\hline
Group Of 

Undertakings & Articles 4, and 47 \\
\hline
{Joint Controllers} & Articles 4, and 26 \\
\hline
{Legal Authority} & Abstract concept \\
\hline
{Main Intervenant} & Abstract concept \\
\hline
{Obligation} & Articles 24 to 31 \\
\hline
{Processor} & Abstract concept \\
\hline
{Recipient} & Chapter 5\\
\hline
{Representative} & Articles 4, 27, and 80 \\
\hline
{Supervisory Authority} & Articles 4, and 51 to 59 \\
\hline
{Third Party} & Articles 4, and 44 \\
\hline
{Undertaking} & Articles 4, and 47 \\
\hline
{Data Controller} & Articles 4, and 24 \\
\hline
\end{tabular}
\end{table}

\begin{figure*}[t]
	\centering
	\includegraphics [width=24cm, angle=90] {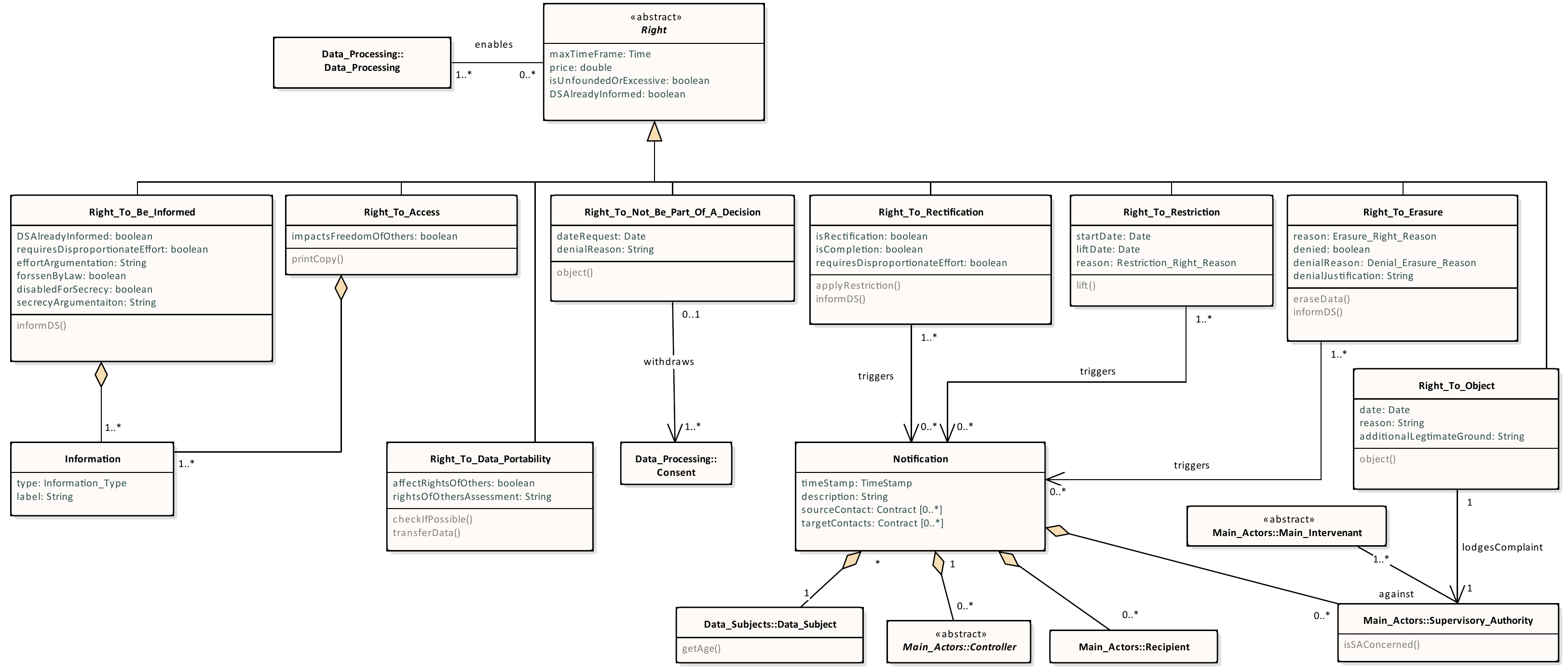}
	\caption{Data Subjects Rights Package \label{fig:DSR}}
\end{figure*}

\newpage
\noindent\textbf{Data Subject Rights package} is showed in Fig.~\ref{fig:DSR} and its traceability to the GDPR is described in Table~\ref{tab:DSR}.

\begin{table}[H]
\caption{Data Subjects Rights Package}\label{tab:DSR}
\centering
\begin{tabular}{|p{3.5cm}|p{3.5cm}|}
\hline
\textbf{Class} & \textbf{Traceability to GDPR} \\
\hline
{Notification} & Article 19 \\
\hline
{Right} & Abstract concept \\
\hline
Right To  
Portability & Article 20 \\
\hline
{Right To Access} & Article 15 \\
\hline
{Right To Be Informed} & Articles 13 to 14 \\
\hline
{Right To Erasure} & Article 17 \\
\hline
Right To Not Be Part Of A Decision & Article 22 \\
\hline
{Right To Object} & Article 21 \\
\hline
{Right To Restriction} & Article 18 \\
\hline
{Right To Rectification} & Article 16 \\
\hline
{Information} & Articles 13 to 15 \\
\hline
\end{tabular}
\end{table}

\begin{figure*}[t]
	\centering
	\includegraphics [width=24cm, angle=90] {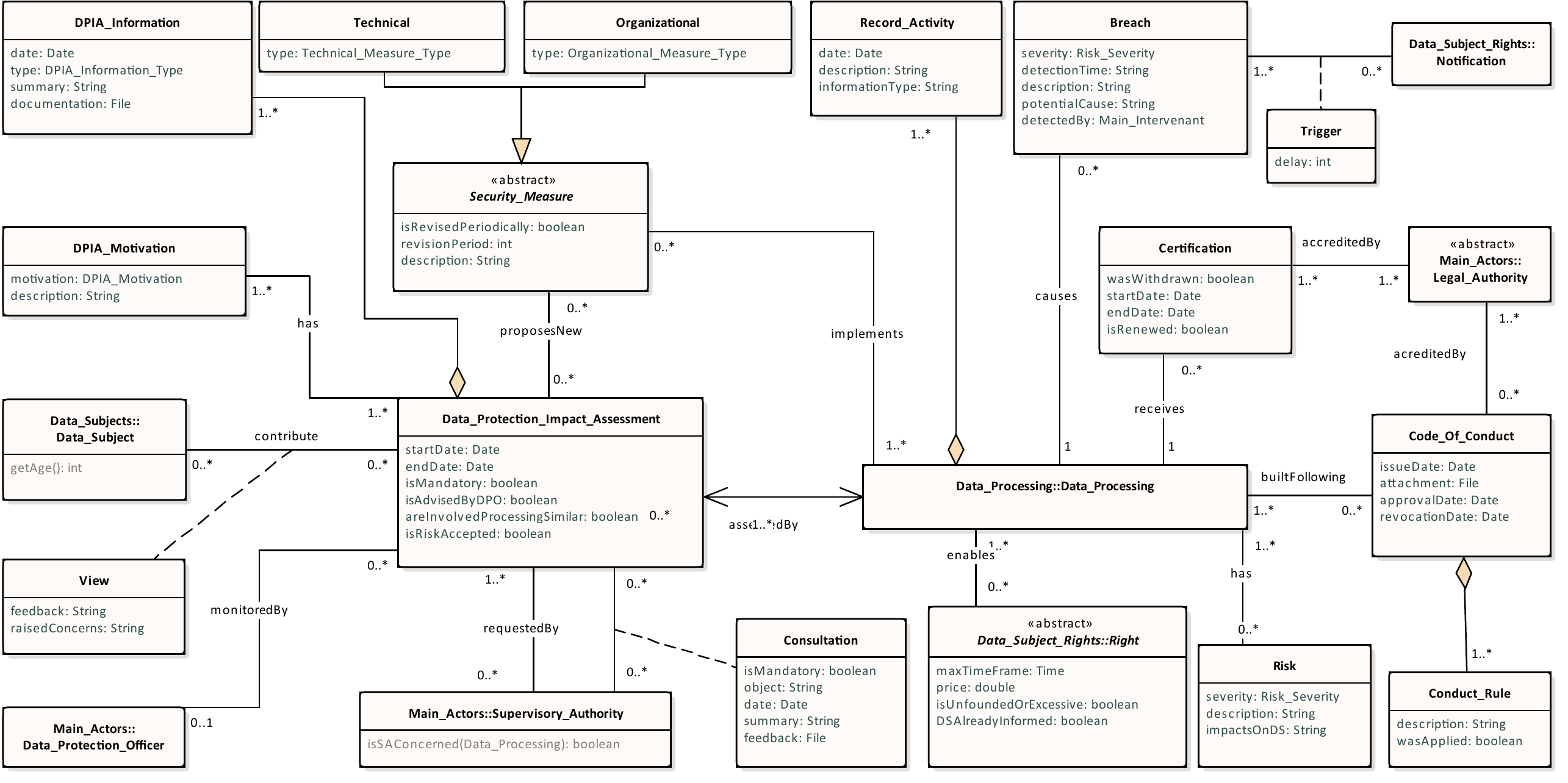}
	\caption{Compliance Package \label{fig:Compliance}}
\end{figure*}

\noindent\textbf{Compliance package} is showed in Fig.~\ref{fig:Compliance} and its traceability to the GDPR is described in Table~\ref{tab:Compliance}.

\begin{table}[H]
\caption{Compliance Package}\label{tab:Compliance}
\centering
\begin{tabular}{|p{3.5cm}|p{3.5cm}|}
\hline
\textbf{Class} & \textbf{Traceability to GDPR} \\
\hline
{Breach} & Articles 4, 33, and34 \\
\hline
{Certtification} & Article 42 \\
\hline
{Code Of Conduct} & Article 40 \\
\hline
{Conduct Rule} & Article 40 \\
\hline
{Consultation} & Article 36 \\
\hline
{Data Protection Impact Assessment} & Article 35 \\
\hline
{DPIA Information} & Article 35 \\
\hline
{DPIA Motivation} & Article 35 \\
\hline
{Organizational} & Articles 24, 28, and 32 \\
\hline
{Record Activity} & Article 30 \\
\hline
{Risk} & Article 32 \\
\hline
{Security Measure} & Abstract concept \\
\hline
{Technical} & Articles 24, 28, and 32 \\
\hline
{View} & Article 35 \\
\hline
\end{tabular}
\end{table}

\begin{figure*}[ht]
	\centering
	\includegraphics[width=24cm, angle=90]{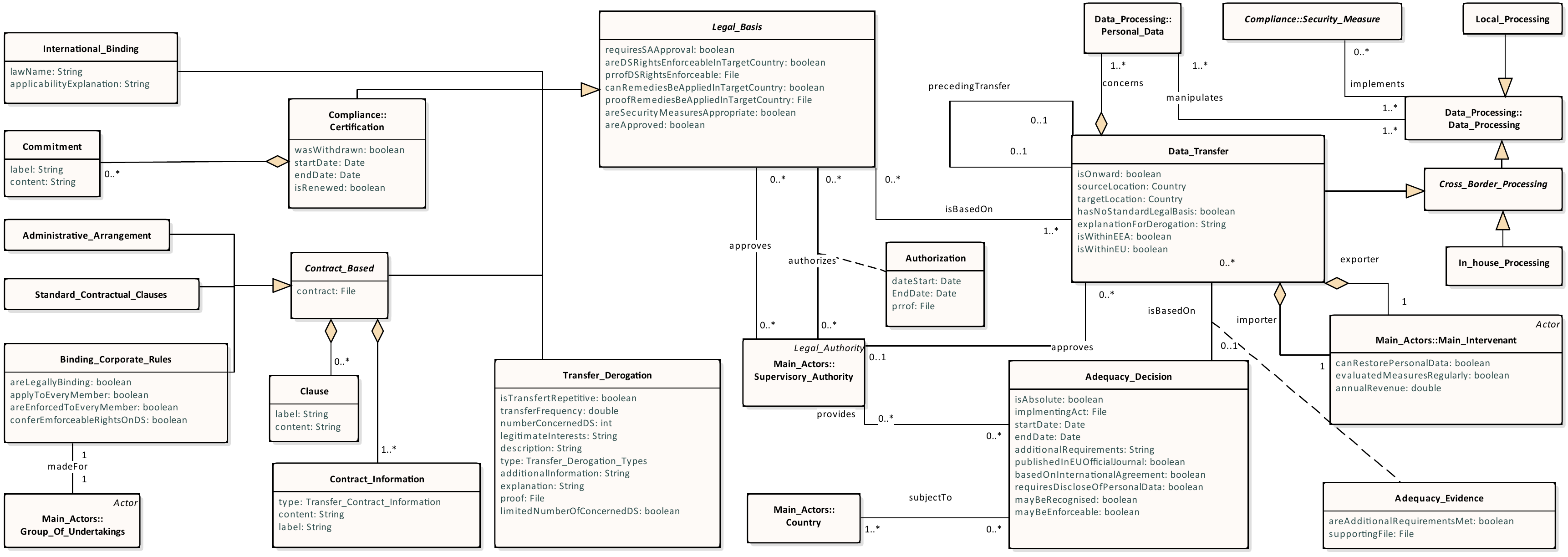}
	\caption{Data Transfer Package \label{fig:DataTransfer}}
\end{figure*}

\newpage
\noindent\textbf{Data Transfer package} is showed in Fig.~\ref{fig:DataTransfer} and its traceability to the GDPR is described in Table~\ref{tab:DataTransfer}.

\begin{table}[H]
\caption{Data Transfer Package}\label{tab:DataTransfer}
\begin{tabular}{|p{3.5cm}|p{3.5cm}|}
\hline
{Class} & \textbf{Traceability to GDPR} \\
\hline
{Adequacy Decision} & Article 45 \\
\hline
{Adequacy Evidence} & Article 45 \\
\hline
Admnistrative 

Arrangment & Article 46 \\
\hline
{Authorization} & Article 46 \\
\hline
{Binding Corporate Rules} & Article 47 \\
\hline
{Clause} & Articles 46 to 47 \\
\hline
{Commitment} & Articles 46, and 49 \\
\hline
{Contact Based} & Abstract concept \\
\hline
{Contact Information} & Articles 46 to 47 \\
\hline
{Cross Boarder Processing} & Abstract concept \\
\hline
{Data Transfer} & Article 44 \\
\hline
{International Binding} & Article 49 \\
\hline
{Legal Basis} & Abstract concept \\
\hline
{Local Processing} & No mapping to the law \\
\hline
{Standard Contractual Clauses} & Article 46 \\
\hline
{Transfer Derogation} & Article 49 \\
\hline
{In house Processing} & WP 244, and Article 29 \\
\hline
\end{tabular}
\end{table}

\begin{figure*}[h]
	\centering
	\includegraphics[width=1\linewidth]{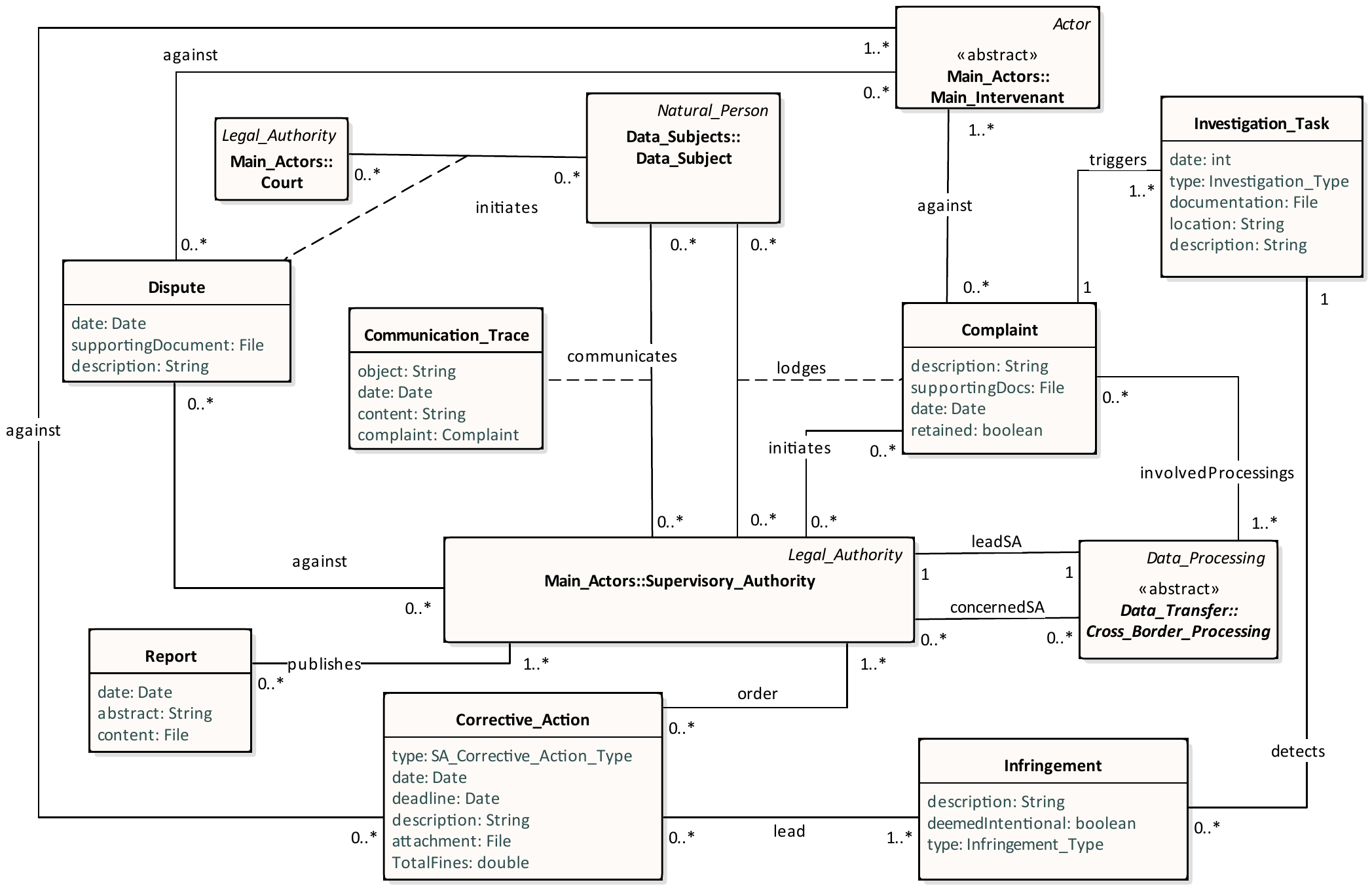}
	\caption{Administration Package \label{fig:Administration}}
\end{figure*}

\noindent\textbf{Administration package} is showed in Fig.~\ref{fig:Administration} and its traceability to the GDPR is described in Table~\ref{tab:Administration}.

\begin{table}[H]
\caption{Administration Package}\label{tab:Administration}
\begin{tabular}{|p{3.5cm}|p{3.5cm}|}
\hline
\textbf{Class} & \textbf{Traceability to GDPR} \\
\hline
{Dispute} & Articles 78, 79, and 81 \\
\hline
{Complaint} & Articles 4, and 77 \\
\hline
{Communication\_Trace} & Article 77 \\
\hline
{Report} & Article 59 \\
\hline
{Corrective\_Action} & Article 58 \\
\hline
{Infringment} & Article 83 \\
\hline
{Investigation\_Task} & Article 57 \\
\hline
\end{tabular}
\end{table}

\begin{figure*}[h]
	\centering
	\includegraphics[width=24cm, angle=90]{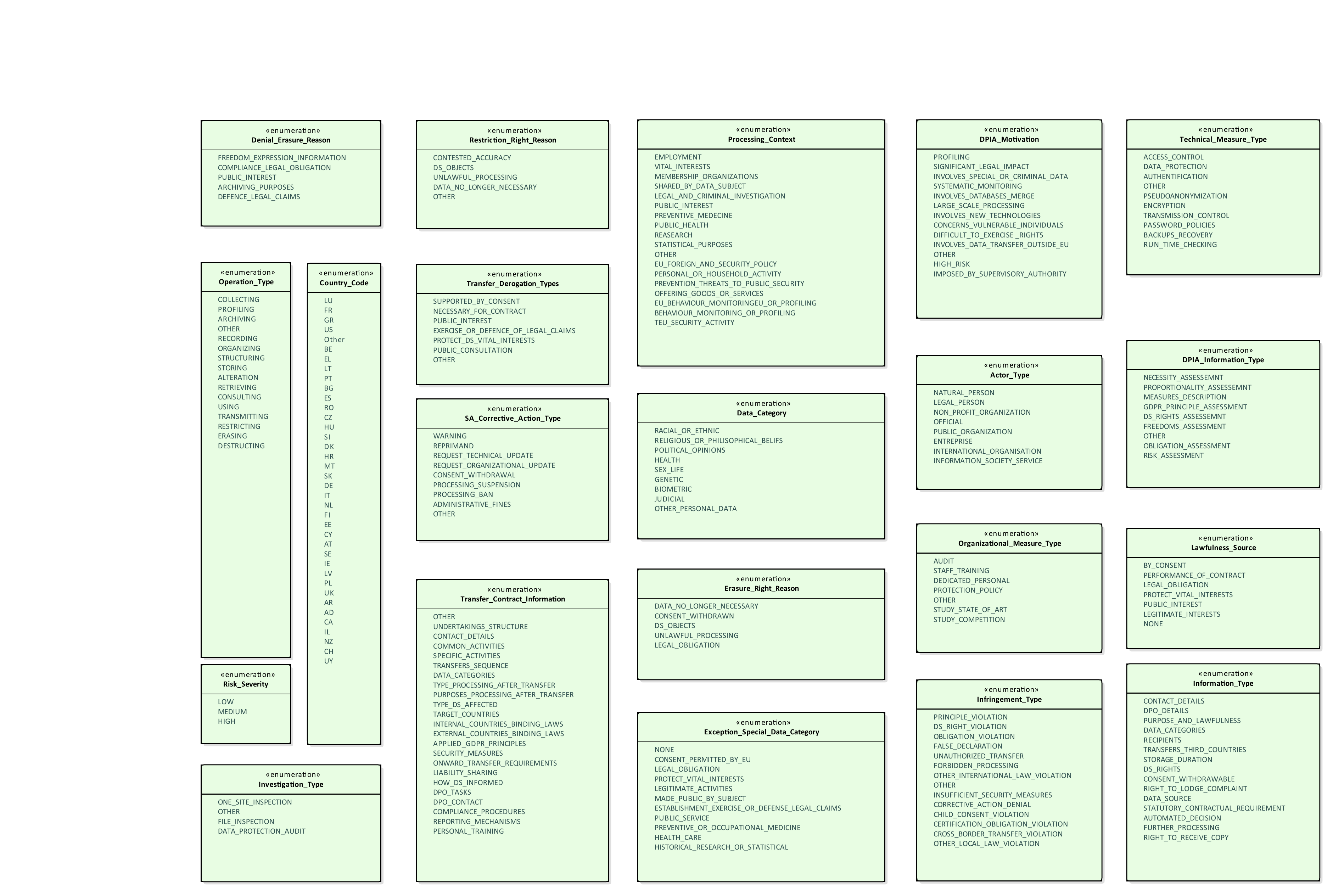}
	\caption{Enumeration Package \label{fig:Enemuration}}
\end{figure*}

\noindent\textbf{Enumeration package.} We describe in detail the Enumeration package showed in Fig.~\ref{fig:Enemuration}. 
\\
\\

{
\begin{figure*}[h]
	\centering
	\includegraphics[width=16cm]{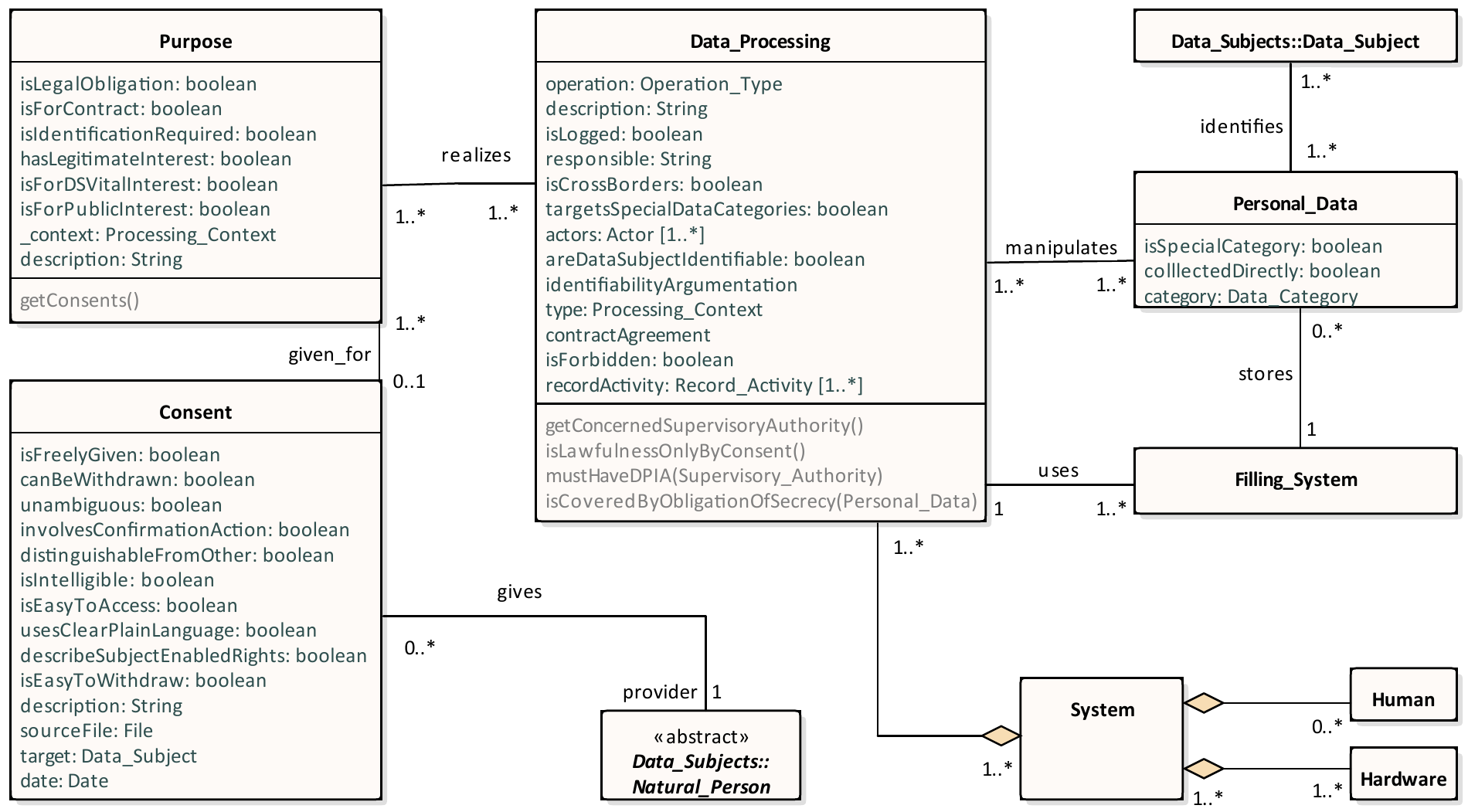}
	\caption{Data Processing package \label{fig:DataProcessing}}
\end{figure*}
}
\

\clearpage

\onecolumn
\subsection{Glossary}\label{Glossary}

As follows, we present the glossary related to the CM's packages previously presented.

\bigbreak

\noindent \textbf{Country Code class:} Lists the abbreviation codes for countries. The ISO Country Codes standard where used for the literals, e.g., LU denotes Luxembourg.

\noindent \textbf{Actor Type class:} Lists the possible types for a real actor involved in the data processing.

\begin{table}[H]
\caption{Actor Type class}\label{tab:ActorType}
\centering
\begin{tabular}{|p{2cm}|p{14.5cm}|}
\hline
\textbf{Literal} & \textbf{Description} \\
\hline
{Natural 

Person} & A natural person is a person that is an individual human being. \\
\hline
{Legal Person} & A legal person is any individual, company, or 
other entity which has legal rights and is subject to privileges and 
obligation such as having the ability to enter into contracts, to sue, 
and to be sued. \\
\hline
{Non Profit 

Organization} & When they are formed to further 
cultural, educational, religious, professional, or public service 
objectives. Their start-up funding is provided by their members, 
trustees, or others who do not expect repayment, and who do not share in 
the organization's profits or losses which are retained or absorbed. \\
\hline
{Official} & Any establishment recognized as an Authority 
entity or a public body. \\
\hline
{Public 

Organization} & When they are owned and 
operated by the government and exist to provide services for citizens. 
Often, organizations in the public sector do not seek to generate a 
profit. \\
\hline
{Enterprise} & It is a business or a company. \\
\hline
International Organization & An international organization 
is an organization with an international membership, scope, or presence. 
\\
\hline
{Information Society 

Service} & An information society 
service is any society/person that provides service provided, via 
electronic means upon an individual request from a user. \\
\hline
\end{tabular}
\end{table}
\noindent\textbf{Restriction Right Reason class:}Lists 
the situations where data processing can be restricted.

\begin{table}[H]
\caption{Restriction Right Reason class}\label{tab:RestrictionRight}
\centering
\begin{tabular}{|p{2cm}|p{14.5cm}|}
\hline
\textbf{Literal} & \textbf{Description} \\
\hline
{Contested 

Accuracy} & If the data subject (DS) contest the accuracy of the 
data concerning him or her. \\
\hline
{DS Objects} & If the DS has lodged a complaint against the 
processing. \\
\hline
{Unlawful 

Processing} & If the processing has no legal basis. 
\\
\hline
{Data no 

Necessary} & If the personal data of a data 
subject is no longer needed for a given processing. \\
\hline
{Other} & Groups the other exceptional situations where the 
processing can be restricted. \\
\hline
\end{tabular}
\end{table}

\newpage
\noindent \textbf{Processing Context class:} Provides standard general contexts for which personal data is processed.

\begin{table}[H]
\caption{Processing Context class}\label{tab:ProcessingContext}
\centering
\begin{tabular}{|p{2cm}|p{14.5cm}|}
\hline
\textbf{Literal} & \textbf{Description} \\
\hline
{Employment} & A relationship between two parties, usually based on a contract where work is paid for, where one party is the employer and the other is the employee. \\
\hline
{Vital 

Interests} & It is necessary to protect life 
and death of individuals. \\
\hline
{Membership Organization} & Any organization that allows people to subscribe, and often requires them to pay a membership fee or "subscription". \\
\hline
{Legal and Criminal Investigation} & The study of facts, used to identify and prove the guilt of an accused person according to the law. \\
\hline
{Public 

Interest} & Any processing conducted in the 
interest is "the welfare or well-being of the general public". \\
\hline
{Preventive Medicine} & It consists of 
measures taken for disease prevention, as opposed to disease treatment. 
\\
\hline
{Public Health } & The science of 
protecting the safety and improving the health of communities through education, policy making and research for disease and injury prevention. 
\\
\hline
{Research} & The systematic investigation into and 
study of materials and sources in order to establish facts and reach new 
conclusions. \\
\hline
{Statistical Purposes} & Any 
operation of collection and the processing of personal data necessary 
for statistical surveys or for the production of statistical results. \\
\hline
{EU Foreign and Security Policy} & EU foreign and security 
policy, which has developed gradually over many years, enables the EU to 
speak and act as one in world affairs. \\
\hline
{Persona or Household Activity} & Activities 
which are exclusively personal or household. \\
\hline
{Prevention Threats to Public Security} & The 
activity of identification of possible threats in order to prevent 
threats for the public security. \\
\hline
{Offering Goods or Service} & The process 
(carried out by an actor) of offering goods or services. \\
\hline
EU 

Behaviour Monitoring or Profiling & The 
process of monitoring or profiling the behaviour of individuals that
reside within the EU. \\
\hline
{Behaviour Monitoring or Profiling} & The 
process of monitoring or profiling the behaviour of data subject. \\
\hline
{EU Security Activity} & The EU activities aimed 
at keeping secure the EU. \\
\hline
{OTHER} & Any additional processing context not in the enumeration. \\
\hline
\end{tabular}
\end{table}

\newpage
\noindent\textbf{Data Category class:}Lists the 
different categories of personal data envisaged by GDPR.

\begin{table}[H]
\caption{Data Category class}\label{tab:DataCategory}
\centering
\begin{tabular}{|p{2cm}|p{14.5cm}|}
\hline
\textbf{Literal} & \textbf{Description} \\
\hline
{Racial or Ethnic} & Any data that might reveal 
the race or ethnicity of an individual. \\
\hline
{Religious or Philosop. 

Believes} & Any 
data that might reveal the religious or philosophical believes of an 
individual. \\
\hline
{Political Opinion} & Any data that might reveal 
the political opinions of an individual. \\
\hline
{Health} & Any data that reveals information about 
the health status of a given person. \\
\hline
{Sex Life} & Any data that reveals information 
about the sexual orientations and life a given person. \\
\hline
{Genetic} & Data relating to the inherited or 
acquired genetic characteristics of a natural person. 
\\
\hline
{Biometric} & Physical, physiological or behavioural characteristics of a natural person. \\
\hline
{Judicial} & Data relating to criminal convictions 
and offenses \\
\hline
{Other 

Personal Data} & Other types of personal data that 
might make the DS identifiable. \\
\hline
\end{tabular}
\end{table}

\noindent\textbf{Operation Type class:}Provides some 
standard types of automated data processing.

\begin{table}[H]
\caption{Operation Type class}\label{tab:Operation Type}
\centering
\begin{tabular}{|p{2cm}|p{14.5cm}|}
\hline
\textbf{Literal} & \textbf{Description} \\
\hline
{Collecting} & The activity of bringing or gather 
together personal data. \\
\hline
{Profiling} & The activity of recording and 
analysis of a person's psychological and behavioural characteristics, so 
as to assess or predict their capabilities in a certain sphere or to 
assist in identifying categories of people. \\
\hline
{Archiving} & The activity of transferring 
(personal data) to a less frequently consulted storage. \\
\hline
{Recording} & The action or process of recording 
personal data for subsequent use. \\
\hline
{Organizing} & The activity of arranging/ordering 
personal data systematically. \\
\hline
{Structuring} & The activity constructing or 
arranging personal data according to a give a pattern. \\
\hline
{Storing} & The activity of keeping or 
accumulating personal data for future use. \\
\hline
{Altering} & The activity of changing in character 
or composition the personal data, typically in a comparatively small but 
significant way. \\
\hline
{Retrieving} & The activity of getting or bringing 
personal data back from somewhere. \\
\hline
{Consulting} & The activity of accessing personal 
data. \\
\hline
{Using} & The activity of taking, holding, or 
deploying personal data as a means of processing the data. \\
\hline
{Transmitting} & The activity of passing personal 
data on from one person, organization, etc. to another. \\
\hline
{Restricting} & The activity of limiting personal 
data somehow. \\
\hline
{Erasing} & The activity of delegating recorded 
personal data. \\
\hline
{Destroying } & The activity of completely 
eliminating personal data from more than one support and from one or
more tiers. \\
\hline
{Other} & Additional specific processing types can be 
specified. \\
\hline
\end{tabular}
\end{table}

\newpage
\noindent\textbf{Technical Measure Type class:}
Provides some standard types for technical measure that can be 
implemented as safeguards to protect personal data and comply with GDPR.

\begin{table}[H]
\caption{Technical Measure Type class}\label{tab:TechnicalMeasuresType}
\centering
\begin{tabular}{|p{2cm}|p{14.5cm}|}
\hline
\textbf{Literal} & \textbf{Description} \\
\hline
{Access  

Control} & Policy to limit access to personal data for 
an authorized employee to ensure that employees have access only to 
information or systems applicable to their job. \\
\hline
{Data 

Protection} & It is the branch of information security dealing with the proper handling of data concerning consent, notice, sensitivity, and regulatory 
concerns. It specifies the relationship between the collection and 
dissemination of data, technology, and the legal and political issues 
surrounding them. \\
\hline
{Authentifica-tion} & The process or action of 
proving or showing some credential to be valid. \\
\hline
{Pseudonymiza-tion} & The activity of processing 
personal data in such a manner that the personal data can no longer be 
attributed to a specific data subject without the use of additional 
information, provided that such additional information is kept 
separately and is subject to technical and organizational measures to 
ensure that the personal data are not attributed to an identified or 
identifiable natural person. \\
\hline
{Encryption} & Personal data that is protected through 
technological measures to ensure that the data is only 
accessible/readable by those with specified access. \\
\hline
{Transmission Control} & It is responsible for the 
transmission of personal data. \\
\hline
{Password 

Polices} & A password policy is a set of rules 
designed to enhance computer security by encouraging users to employ 
strong passwords and use them properly. \\
\hline
{Backups 

Recovery} & Backups refer to make a copy of the data 
which should be replicated to another device or location. That data may 
be used to restore the original after a data loss event. \\
\hline
{Run the checking} & It is an analysis and 
execution approach based on extracting information from a running system 
and using it to detect and possibly react to observed behaviours 
satisfying or violating certain properties such as unauthorized access. 
\\
\hline
{Other} & Other data protection technical measure can be 
specified. \\
\hline
\end{tabular}
\end{table}

\noindent\textbf{Organizational Measure Type class:}
Provides some standard types for organizational measures that can be
implemented as safeguards to protect personal data and comply with GDPR.

\begin{table}[H]
\caption{Organizational Measure Type class}\label{tab:OrganizationalMeasureType}
\centering
\begin{tabular}{|p{2cm}|p{14.5cm}|}
\hline
\textbf{Literal} & \textbf{Description} \\
\hline
{Audit} & Inspection of an organization and system functioning, 
often by an independent body. \\
\hline
{Staff 

Training} & Training employees and raising awareness 
about GDPR and best practices to protect personal data. \\
\hline
{Dedicated Personal} & Dedicate some employee to the specific 
task of ensuring and monitoring compliance with GDPR. \\
\hline
{Protection Policy} & Protection policy is a document which 
provides guidelines to users on the processing, storage, and 
transmission of sensitive information. \\
\hline
{Study State of Art} & Be up-to-date and informed about the 
latest technology and organizational advances in terms of data 
protection. \\
\hline
{Study 

Competition} & Study and analyze the technical and 
organizational measures applied by similar companies (competition) to 
protect personal data. \\
\hline
{Other} & Other protection organizational measures can be 
specified. \\
\hline
\end{tabular}
\end{table}

\newpage
\noindent\textbf{Denial Erasure Reason class:} Specify 
the situation where, without further authorizations, a controller can deny or request for data erasure made by a given DS.

\begin{table}[H]
\caption{Denial Erasure Reason class}\label{tab:DenialErasureReason}
\centering
\begin{tabular}{|p{2cm}|p{14.5cm}|}
\hline
\textbf{Literal} & \textbf{Description} \\
\hline
{Freedom  Expression  Information} & If the controller deems 
that the processing is done in under the freedom of expression (and does 
not violate other rights). \\
\hline
{Compliance Legal 

Obligation} & If the controller must keep 
the data to comply to legal obligation. \\
\hline
{Public 

Interest} & If the controller must keep the data to 
for public interest related reasons. \\
\hline
{Archiving Purposes} & If the controller must keep the data 
for archive purposes. \\
\hline
{Defense Legal Claims} & If the controller must keep the data 
for the defense of legal claims (data is kind of a legal proof). \\
\hline
\end{tabular}
\end{table}

\noindent\textbf{Lawfulness Sources class:} Lists the 
possible legal basis, as envisaged by the GDPR, for a given processing of personal data.

\begin{table}[H]
\caption{Lawfulness Sources class}\label{tab:AwfulnessSources}
\centering
\begin{tabular}{|p{2cm}|p{14.5cm}|}
\hline
\textbf{Literal} & \textbf{Description} \\
\hline
{By Consent} & The processing is enabled by a consent. Consent 
means any freely given, specific, informed and unambiguous indication of 
the data subject's wishes by which he or she, by a statement or by a 
clear affirmative action, signifies agreement to the processing of 
personal data relating to him or her. \\
\hline
{Performance of Contract} & The processing is necessary for 
the performance of a contract to which the data subject is party or in 
order to take steps at the request of the data subject prior to entering 
into a contract. \\
\hline
{Legal 

Obligation} & It covers processing that is necessary to 
fulfil a legal obligation to which the data controller is subject. 
Examples include obligations in the fields of employment and social 
security, including those that require processing sensitive (now known 
as "special category") personal data. \\
\hline
{Protect Vital Interests} & The processing is necessary to 
protect the vital interests of the DS. This is relevant for emergency 
medical care, when you need to process personal data for medical 
purposes but the individual is incapable of giving consent to the 
processing. \\
\hline
{Public 

Interest} & The processing is necessary for Public 
interest reasons. Examples include taxation, reporting crimes, 
humanitarian purposes, preventive or occupational medicine, public 
health, social care, quality and safety of products, devices and 
services, and election campaigns. \\
\hline
{Legitimate Interest} & A legitimate interest is a clearly 
articulated benefit to a single company, or to society as a whole, that 
can be derived from processing personal data in a lawful way. \\
\hline
{None} & None of the standard legal basis is possible. This can 
happen when the processing is not the main reason for which the data was 
initially collected or when the legal basis for processing is based on a 
derogation as stated in Article 6. \\
\hline
\end{tabular}
\end{table}

\newpage
\noindent\textbf{Information Type class:}Summarizes the 
type of information that controller might provide when exercising the information and access rights for a given individual.

\begin{table}[H]
\caption{Information Type class}\label{tab:InformationType}
\centering
\begin{tabular}{|p{2cm}|p{14.5cm}|}
\hline
\textbf{Literal} & \textbf{Description} \\
\hline
{Contact 

Details} & An individual's private or 
personal information by which another person, business, or entity can 
use to reach the individual. \\
\hline
{DPO Details} & The Data Protection Officer (DPO) 
information by which another person, business, or entity can use to 
reach that DPO. \\
\hline
{Purpose and Lawfulness} & The reason for which 
some personal data are somehow processed according to the law. \\
\hline
{Data 

Categories} & The different categories of 
personal data that are involved in the data processing. \\
\hline
{Recipients} & The natural or legal person, public 
authority, agency or another body, to which the personal data are 
disclosed, whether a third party or not. \\
\hline
{Transfer Third 

Countries} & Explicitly states whether the 
processing involves data transfers to third countries. \\
\hline
{Storage 

Duration} & The storage duration. \\
\hline
{DS Right} & Summarizes the enables DS rights for a given 
processing. \\
\hline
{Consent Withdrawal} & States whether the consent is 
withdrawable at any time. \\
\hline
{Right to Lodge 

Complaint} & Recalls DS that they have the 
right to lodge a complaint. \\
\hline
{Data Source} & The sources used to gather personal 
data. \\
\hline
{Statutory Contractual Requirement} & The 
regulatory obligations that are required by a statutory contractual 
requirement. (if applicable) \\
\hline
{Automated Decision} & The involved automated 
decisions that might be linked to the underlying processing. \\
\hline
{Further 

Processing} & States whether additional (to what 
already exists) processing of personal data are planned or possible. \\
\hline
{Right to 

Receive Copy} & Recall DS about their right to 
receive the personal data concerning him or her in a structured, 
commonly used and machine-readable format. \\
\hline
\end{tabular}
\end{table}

\newpage
\noindent\textbf{DPIA Information Type class:}
Summarizes the standard type of information that a Data Processing 
Impact Assessment might contain.

\begin{table}[H]
\caption{DPIA Information Type class}\label{tab:DPIA}
\centering
\begin{tabular}{|p{2cm}|p{14.5cm}|}
\hline
\textbf{Literal} & \textbf{Description} \\
\hline
{Necessity 

Assessment} & A systematic description of the 
envisaged processing operations and the purposes of the processing, 
including, where applicable, the legitimate interest pursued by the 
controller. \\
\hline
{Proportionali-ty 

Assessment} & An assessment of the necessity 
and proportionality of the processing operations in relation to the 
purposes. \\
\hline
{Measures 

Description} & The measures envisaged to address the 
risks, including safeguards, security measures, and mechanisms to ensure 
the protection of personal data and to demonstrate compliance with this 
Regulation taking into account the rights and legitimate interests of 
data subjects and other persons concerned. \\
\hline
{GDPR 

Principle 

Assessment} & An assessment that checks that 
the processing respects all applicable GDPR principles. \\
\hline
{DS Rights

Assessment} & An assessment that checks that the 
processing respects all applicable DS rights. \\
\hline
{Freedoms 

Assessment} & An assessment of the risks to the 
rights and freedoms of data subjects and others. \\
\hline
{Obligation

Assessment} & An assessment that verifies that the 
involved controllers and processors have fulfilled all their GDPR 
imposed obligations. \\
\hline
{Risk

Assessment} & Summary of the potential risks related to 
the processing and how they were mitigated. \\
\hline
{Other} & Any additional assessment that supports DPIA. \\
\hline
\end{tabular}
\end{table}

\newpage
\noindent\textbf{Exception Special Data Category class:} Lists the different possible scenarios where processing of 
special data categories can be lawful.

\begin{table}[H]
\caption{Exception Special Data Category class}\label{tab:Exception1}
\centering
\begin{tabular}{|p{2cm}|p{14.5cm}|}
\hline
Literal & Description \\
\hline
{Consent Permitted by EU} & The special data can 
be processed when the processing is clearly permitted by the EU law. \\
\hline
{Legal 

Obligation} & The special data can be 
processed when the processing is necessary for the purposes of carrying 
out the obligations and exercising specific rights of the controller or 
of the data subject in the field of employment and social security and protection law. \\
\hline
{Protect Vital Interests} & The special data can 
be processed when it is necessary to protect the vital interests of the 
data subject or of another natural person where the data subject is 
physically or legally incapable of giving consent. \\
\hline
{Legitimate Activities} & The special data can be 
processed when in the context of legitimate activities with appropriate 
safeguards by a foundation, association or any other not-for-profit body 
with a political, philosophical, religious or trade union aim and on 
condition that the processing relates solely to the members or to former 
members of the body or to persons who have regular contact with it in 
connection with its purposes and that the personal data are not 
disclosed outside that body without the consent of the data subjects. \\
\hline
{Made Public by Subject} & The special data can 
be processed when the data are manifestly made public by the data 
subject. \\
\hline
{Establishment Exercise or Defense Legal Claims} & The 
special data can be processed for the establishment, exercise or defense of legal claims or 
whenever courts are acting in their judicial capacity. \\
\hline

{Public 

Service} & The special data can be 
processed when the processing is necessary for reasons of substantial 
public interest on the basis of Union or Member State law which shall be 
proportionate to the aim pursued. \\
\hline
{Preventive or Occupational Medicine} & The special data can be processed when the processing is necessary for the 
purposes of preventive or occupational medicine, for the assessment of 
the working capacity of the employee, medical diagnosis, the provision 
of health or social care or treatment or the management of health or 
social care systems and services on the basis of Union or Member State 
law or pursuant to contract with a health professional. \\
\hline
{Health Care} & The special data can be processed 
when the processing is necessary for reasons of public interest in the 
area of public health, such as protecting against serious cross-border 
threats to health or ensuring high standards of quality and safety of 
health care and of medicinal products or medical devices. \\
\hline
{Historical 

Research or 

Statistical} & The special data 
categories can be processed when the processing is necessary for 
archiving purposes in the public interest, scientific or historical 
research purposes or statistical purposes. \\
\hline
{None} & No valid reason for processing sensitive data found 
(non-compliant state). \\
\hline
\end{tabular}
\end{table}

\noindent\textbf{Erasure right reason class:}
Summarizes the situations where a DS can request the erasures of his or her personal data.

\begin{table}[h]
\caption{Erasure right reason class}\label{tab:ErasureRight}
\centering
\begin{tabular}{|p{2cm}|p{14.5cm}|}
\hline
\textbf{Literal} & \textbf{Description} \\
\hline
{Data no Longer 

Necessary} & It refers to the situation when 
some specific personal data of a data subject is no longer needed for a 
given processing. \\
\hline
{Consent Withdrawal} & Once the consent is withdrawn, DS might 
ask for the erasures of the data concerning him or her. \\
\hline
{DS Objects} & In case DS objects the processing, DS might ask 
for the erasures of the data concerning him or her. \\
\hline
{Unlawful 

Processing} & DS might ask for the erasures of the 
data concerning him or her if she deems that the processing personal has 
no legal basis. \\
\hline
{Legal 

Obligation} & DS might ask for the erasures of the data 
concerning him or her in order to meet some legal or judicial 
obligation. \\
\hline
\end{tabular}
\end{table}

\noindent\textbf{Risk Severity class:}Defines the level 
of risk severity related to a given data breach.

\begin{table}[H]
\caption{Risk Severity class}\label{tab:RiskSeverity}
\centering
\begin{tabular}{|p{2cm}|p{14.5cm}|}
\hline
\textbf{Literal} & \textbf{Description} \\
\hline
{Low} & The risk is low. \\
\hline
{Medium} & The risk is deemed to be average. \\
\hline
{High} & The risk is deemed to be extremely high. \\
\hline
\end{tabular}
\end{table}

\noindent\textbf{DPIA Motivation class:}Lists the 
standard situations where a Data Privacy Impact Assessment (DPIA) is required or recommended.

\begin{table}[H]
\caption{DPIA Motivation class}\label{tab:DPIAMotivation1}
\centering
\begin{tabular}{|p{2cm}|p{14.5cm}|}
\hline
\textbf{Literal} & \textbf{Description} \\
\hline
{Profiling} & If the processing involves behavior profiling. \\
\hline
{Significant 

Legal Impact} & A data protection impact 
assessment shall, in particular, be required in case of a systematic and 
extensive evaluation of personal aspects relating to natural persons 
which is based on automated processing and on which decisions are based 
that produce legal effects concerning the natural person or similarly 
significantly affect the natural person. \\
\hline

{Involves 

Special or 

Criminal Data} & A data protection 
impact assessment shall, in particular, be required in case of personal 
data relating to criminal convictions and offenses. \\
\hline
{Systematic Monitoring} & A data protection impact assessment 
shall, in particular, be required in case of systematic monitoring of a 
publicly accessible area on a large scale. \\
\hline
{Involves Database Merge} & A DPIA is welcomed when the data 
processing involves database merging. \\
\hline
{Large Scale Processing} & A data protection impact 
assessment shall, in particular, be required in case of processing on a 
large scale of special categories of data. \\
\hline
{Involves New Technologies} & A DPIA is welcomed when it 
involves the use of new (non-mature) technologies. \\
\hline
{Concerns 

Vulnerable 

Individuals} & A DPIA is welcomed when 
it concerns vulnerable individuals. \\
\hline
{Difficult to Exercise Rights} & A DPIA is welcomed when the 
controller cannot enable DS rights for technical or disproportionate 
effort reasons \\
\hline
{Involves Data 

Transfer 

Outside EU} & A DPIA is welcomed 
when cross-border transfers are involved. \\
\hline
{High Risk} & A DPIA is a must when the processing has a high 
residual risk. \\
\hline
{Imposed by Supervisory Authority} & A DPIA can be 
requested by a supervisory authority. \\
\hline
{Other} & Other viable reason for conduction a thorough DPIA. 
\\
\hline
\end{tabular}
\end{table}

\newpage
\noindent\textbf{Transfer Contract Information class:} Summarizes the list of information to be provided when a cross-border 
data transfer is based on standard contractual clauses, agreements, or binding corporate rules.

\begin{table}[H]
\caption{Transfer Contract Information class} \label{tab:TransferContract2}
\centering
\begin{tabular}{|p{2cm}|p{14.5cm}|}
\hline
\textbf{Literal} & \textbf{Description} \\
\hline
{DPO Tasks} & A description of the DPO(s) tasks and 
responsibilities with regard to the data transfer and processing. \\
\hline
{DPO 

Contact} & The contact details of the involved DPS(s). \\
\hline
{Compliance Procedures} & A description of the technical and 
organizational measures and the procedures employed for ensuring and 
monitoring compliance. \\
\hline
{Reporting Mechanisms} & A description of the mechanisms for 
reporting and recording changes to the data transfer rules and reporting 
those changes to the supervisory authority. \\
\hline
{Personal Training} & A description of the training that the 
employee involved in data transfer and processing have taken. \\
\hline
{Other} & Any other information that is relevant to 
cross-border transfer rules. \\
\hline
{Undertaking Structure} & The structure and relationship of 
the undertakings that are involved in some common activities 
(cooperation). \\
\hline
{Contact 

Details} & The contact details of the groups of 
undertakings and each of its constituent members. \\
\hline
{Common 

Activities} & A detailed description of the common 
activities carried out within the groups of undertakings. \\
\hline
{Specific 

Activities} & A detailed description of the 
(non-common) activities of each undertaking involved in a group of 
undertakings. \\
\hline
{Transfers 

Sequence} & A complete description of data flow 
transfers, including all onward transfers and processing. \\
\hline
{D.
Categories} & A description of the concerned personal 
data categories. \\
\hline

{Type 
Proces. 
After Trans.} & A description of the type 
processing that will be applied on the personal data after transfer. \\
\hline
{Purposes 

Processing 

After Trans.} & A description of the 
purpose of each processing that will be applied to the personal data after transfer. \\
\hline
{Type DS 

Affected} & A description of the DS segment that is 
likely to be impacted by the transfer and subsequent processing. \\
\hline
{Target 

Countries} & A list of all countries and territories 
through which the data will flow. \\
\hline
{I. Countries Binding Laws} & A description of all 
internal law bindings of the controllers and processors to which the 
data is disclosed. \\
\hline
{E. Countries Bindings Laws} & A description of all 
relevant and enforceable laws of the countries and territories through 
which the data will flow. \\
\hline
{Applied GDPR 

Principles} & A demonstration that all GDPR 
principles are being satisfied when transferring the data and when 
processing it afterward. \\
\hline
{Security 

Measures} & A description of the security measures 
to safeguard data during and after the transfer. \\
\hline
{Onward Transfer 

Requirements} & A description of onward 
requirements for the transfer to bodies not bound by binding corporate 
rules. \\
\hline
{Liability 

Sharing} & A description of the responsibility and 
liability sharing between all actors that are involved in the transfer 
and processing. \\
\hline
{How DS 

Informed} & A clear description of how and when DS 
will be informed about the data transfer and subsequent processing. \\
\hline
\end{tabular}
\end{table}

\newpage
\noindent\textbf{Transfer Derogation Types class:}
Lists the standard situation where a cross-border data transfer is 
possible although it is not based on an adequacy decision, contractual clauses, agreements, nor binding corporate rules.

\begin{table}[H]
\caption{Transfer Derogation Types class}\label{tab:TransferDerogation}
\centering
\begin{tabular}{|p{2cm}|p{14.5cm}|}
\hline
\textbf{Literal} & \textbf{Description} \\
\hline
{Supported by Consent} & The DS has explicitly consented to 
the proposed transfer, after having been informed of the possible risks 
of such transfers for the data subject due to the absence of an adequacy 
decision and appropriate safeguards. \\
\hline
{Necessary for Contact} & The transfer is necessary for the 
performance of a contract or the implementation of pre-contractual 
measures where the DS is involved. \\
\hline
{Public 

Interest} & The transfer is necessary for important 
reasons of public interest. \\
\hline
{Exercise or Defence of Legal Claims} & The transfer is 
necessary for the establishment, exercise or defense of legal claims. \\
\hline
{Protect DS Vital Interests} & The transfer is necessary to 
protect the vital interests of the DS or of other persons, where the DS 
is physically or legally incapable of giving consent. \\
\hline
{Public 

Consultation} & The transfer is made from a register 
which according to Union or Member State law is intended to provide 
information to the public and which is open to consultation either by 
the public in general or by any person who can demonstrate a legitimate 
interest. \\
\hline
{Other} & Other derogations that rely on the authorization of a 
supervisory authority are also possible. \\
\hline
\end{tabular}
\end{table}

\noindent\textbf{SA Corrective Action Type class:} Standard actions that a supervisory authority (SA) can apply on controller/processor.

\begin{table}[H]
\caption{SA Corrective Action Type class}\label{tab:SACorrective}
\centering
\begin{tabular}{|p{2cm}|p{14.5cm}|}
\hline
\textbf{Literal} & \textbf{Description} \\
\hline
{Warning} & SA can issue a warning. \\
\hline
{Reprimand} & SA reprimand a controller. \\
\hline
{Request

Technical

Update} & SA can order a technical update 
based on state of the art. \\
\hline
{Request 

Organizational 

Update} & SA can order a change in 
the organizational measures and procedures of the establishment. \\
\hline
{Certification

Withdrawal} & SA can withdraw a certification 
(involving administrative fines). \\
\hline
{Processing

Suspension} & SA can order the temporary 
suspension of some processing. \\
\hline
{Processing Ban} & SA can ban processing and order the 
destruction of data. \\
\hline
{Administrative

Fines} & SA impose administrative fines 
based on the infringement. \\
\hline
{Other} & Other corrective actions. \\
\hline
\end{tabular}
\end{table}

\newpage
\noindent\textbf{Investigation Type class:}Lists some
common whys for supervisory authorities to conduct their investigations.

\begin{table}[H]
\caption{Investigation Type class}\label{tab:InvestigationType}
\centering
\begin{tabular}{|p{2cm}|p{14.5cm}|}
\hline
\textbf{Literal} & \textbf{Description} \\
\hline
{One Site 

Inspection} & Involves the inspection at the 
premises of the controller/processor, for example, to check the 
installation of a video surveillance system and the data retention period. \\
\hline
{File 

Inspection} & Supervisory authorities' subject 
controllers/processors to questionnaire and document request and review 
of answers and relevant documents. \\
\hline
{Data 

Protection 

Audit} & A more general, complete and 
in-depth review of the relevant documentation. \\
\hline
{Other} & Other type of investigations. \\
\hline
\end{tabular}
\end{table}

\noindent\textbf{Infringement Type class:}Some standard 
GDPR infringement that might give rise to administrative fines.

\begin{table}[H]
\caption{Infringement Type class }\label{tab:Infringement}
\centering
\begin{tabular}{|p{2cm}|p{14.5cm}|}
\hline
\textbf{Literal} & \textbf{Description} \\
\hline
{Principle 

Violation} & One or more GDPR principles were 
violated. \\
\hline
{DS Right 

Violation} & One or more GDPR principles were 
violated or not supported. \\
\hline
{Obligation 

Violation} & Violation of the controller/processor 
obligations. \\
\hline
{False 

Declaration} & Any false information transmitted to a 
supervisory authority. \\
\hline
{Unauthorized 

Transfer} & Cross-border data that has no legal 
justification according to GDPR. \\
\hline
{Forbidden

Processing} & Processing of categories of 
personal data not included into the derogations that enable 
such processing. \\
\hline
{Other

International

Law 

Violation} & Violation of 
international laws and treaties other than GDPR. \\
\hline
{Insufficient

Security

Measures} & Not enough safeguards and 
logistics were put to protect data and its free flow. \\
\hline

{Corrective Action

Violation} & Non-compliance with order or 
request from the SA. \\
\hline
{Child 

Consent

Violation} & It lays down for the establishing consent for a child DS. \\
\hline
{Certification 

Obligation

Violation} & Non-compliance to the 
commitments and obligations imposed by a certification body. \\
\hline
{Cross-Boarder 

Transfer

Violation} & Non-compliance with the 
contractual rules that govern cross-border data transfers. \\
\hline
{Other Local 

Law 

Violation} & Violation of member state 
local laws. \\
\hline
{Other} & Other non-common infringement. \\
\hline
\end{tabular}
\end{table}

\newpage
\noindent\textbf{Glossary:} As follows, we present the GDPR-related terminology we use across the different model presented in this paper.

\begin{table*}[hbt]
\caption{Glossary (1 of 4)}\label{tab:glossary_1}
\begin{tabular}{|p{2cm}|p{14.5cm}|}
\hline
\textbf{Term} & \textbf{Description} \\
\hline
{Personal Data} & Personal data means any information relating 
to an identified or identifiable natural person. \\
\hline
{Data 

subject} & A natural person whose personal data is 
processed by a controller or processor. \\
\hline
{Union Law} & Regulation (i.e. laws) of the European Union. \\
\\
\hline
{Directive} & A legislative act that sets out a goal that all 
EU countries must achieve through their own national laws. \\
\hline
{Controller} & A natural or legal person, public authority, 
agency or any other body which, alone or jointly with others, determines 
the purposes and means of the processing of personal data where the 
purposes and means of such processing are determined by national or EU 
laws or regulations, the controller or the specific criteria for its 
nomination may be provided by national or EU law. \\
\hline
{Processor} & A natural or legal person, public authority, 
agency or other body which processes personal data on behalf of the 
controller. Moreover, data processors need to assist controllers in 
various circumstances where relevant, for example in a potential 
personal data breach notification or in considering a Data Protection 
Impact Assessment. \\
\hline
{Processing} & Any operation performed on personal data, 
whether or not by automated means, including collection, recording, 
organization, structuring, storage, adaptation or alteration, retrieval, 
consultation, use, disclosure by transmission, dissemination or 
otherwise making available, alignment or combination, restriction, 
erasure or destruction \\
\hline
{Identifiable natural person} & An identifiable natural person 
is one who can be identified, directly or indirectly, in particular by 
reference to an identifier such as a name, an identification number, 
location data, an online identifier or to one or more factors specific 
to the physical, physiological, genetic, mental, economic, cultural or 
social identity of that natural person. \\
\hline
{Restriction of processing} & It means the marking of stored 
personal data with the aim of limiting their processing in the future. \\
\hline
{Profiling} & It means to analyze or predict aspects concerning 
that natural person's performance at work, economic situation, health, 
personal preferences, interests, reliability, behavior, location or 
movements. \\
\hline
{Pseudo-nymisation} & It means the processing of personal data 
in such a manner that the personal data can no longer be attributed to a 
specific data subject without the use of additional information, 
provided that such additional information is kept separately and is 
subject to technical and organizational measures to ensure that the 
personal data are not attributed to an identified or identifiable 
natural person \\
\hline
{Filing 

system} & It means any structured set of personal data 
which are accessible according to specific criteria, whether 
centralized, decentralized or dispersed on a functional or geographical 
basis. \\
\hline
{Recipient} & It means that a natural or legal person, public 
authority, agency or another body, to which the personal data are 
disclosed, whether a third party or not. However, public authorities 
which may receive personal data in the framework of a particular inquiry 
in accordance with Union or Member State law shall not be regarded as 
recipients. \\
\hline
{Third party} & It means a natural or legal person, public 
authority, agency or body other than the data subject, controller, 
processor and persons who, under the direct authority of the controller 
or processor, are authorized to process personal data \\
\hline
{Consent} & It means any freely given, specific, informed and 
unambiguous indication of the data subject's wishes by which he or she, 
by a statement or by a clear affirmative action, signifies agreement to 
the processing of personal data relating to him or her. \\
\hline
{Personal data breach} & It means a breach of security leading 
to the accidental or unlawful destruction, loss, alteration, 
unauthorized disclosure of, or access to, personal data transmitted, 
stored or otherwise processed. \\
\hline

\end{tabular}
\end{table*}

\begin{table*}[hbt]
\caption{Glossary (2 of 4)}\label{tab:glossary_2}
\begin{tabular}{|p{2cm}|p{14.5cm}|}
\hline
\textbf{Term} & \textbf{Description} \\ \hline

{Biometric data} & It means personal data resulting from 
specific technical processing relating to the physical, physio logical 
or behavioral characteristics of a natural person, which allow or 
confirm the unique identification of that natural person, such as facial 
images or dactyloscopic data. \\
\hline
{Data 

concerning health} & It means personal data related to 
the physical or mental health of a natural person, including the 
provision of health care services, which reveal information about his or 
her health status. \\
\hline
{Representa-tive} & It means a natural or legal person 
established in the Union who, designated by the controller or processor. 
\\
\hline
{Enterprise} & It means a natural or legal person engaged in an 
economic activity, irrespective of its legal form, including 
partnerships or associations regularly engaged in an economic activity. \\
\hline

{Group of 

undertakings} & It means a controlling undertaking 
and its controlled undertakings. \\
\hline
{Supervisory authority} & It means an independent public 
authority which is established by a Member State. \\
\hline
{Relevant and 

reasoned 

objection} & It means an objection to a 
draft decision as to whether there is an infringement of this 
Regulation, or whether envisaged action in relation to the controller or 
processor complies with this Regulation, which clearly demonstrates the 
significance of the risks posed by the draft decision as regards the 
fundamental rights and freedoms of data subjects and, where applicable, 
the free flow of personal data within the Union. \\
\hline
{International organization} & It means an organization and its 
subordinate bodies governed by public international law, or any other 
body which is set up by, or on the basis of, an agreement between two or 
more countries. \\
\hline
{Purpose} & The purpose of data processing is said to be lawful if its legal basis matches one of the possible circumstances under which GDPR permits the processing of personal data. Example of valid legal basis for data processing are 
consent and when processing is necessary to perform or prepare for a contract with the data subject. \\
\hline
{Member State Law} & Specific regulation (i.e. laws) from a 
European member state. \\
\hline
{Processing personal data 

collected for 

different 

purpose} & 
Where the processing for a purpose other than that for which the 
personal data have been collected is not based on the data subject's 
consent or on a Union or Member State law. \\
\hline
{Encryption} & Personal data that is protected through 
technological measures to ensure that the data is only 
accessible/readable by those with specified access. \\
\hline
{Child} & Any identifiable natural person that is below of 16 
years. \\
\hline
{Special 

categories of personal data} & Special categories of 
personal data include: racial or ethnic origin, political opinions, 
religious or philosophical beliefs, or trade union membership, genetic 
data, biometric data for the purpose of uniquely identifying a natural 
person, data concerning health or data concerning a natural person's sex 
life or sexual orientation shall be prohibited. \\
\hline
{Criminal convictions and offenses data} & Criminal convictions 
and offenses data represent personal data relating to criminal 
convictions and offenses or related security measures. \\
\hline
{Transparent information} & It means processing to the data 
subject in a concise, transparent, intelligible and easily accessible 
form, using clear and plain language, in particular for any information 
addressed specifically to a child. \\
\hline
{Right to 

rectification} & The data subject shall have the 
right to obtain from the controller without undue delay the 
rectification of inaccurate personal data concerning him or her. \\
\hline

\end{tabular}
\end{table*}

\begin{table*}[hbt]
\caption{Glossary (3 of 4)}\label{tab:glossary_3}
\begin{tabular}{|p{2cm}|p{14.5cm}|}
\hline
\textbf{Term} & \textbf{Description} \\
\hline
{Right to be Forgotten} & Also known as Data Erasure, it 
entitles the data subject to have the data controller erase his/her 
personal data, cease further dissemination of the data, and potentially 
have third parties cease processing of the data. \\
\hline
{Right to 

restriction} & The data subject shall have the right 
to obtain from the controller restriction of processing the data. \\
\hline
{Right to data portability} & The data subject shall have the 
right to receive the personal data concerning him or her, which he or 
she has provided to a controller, in a structured, commonly used and 
machine-readable format and have the right to transmit those data to 
another controller without hindrance from the controller to which the 
personal data have been provided. \\
\hline
{Right to 

object} & The data subject shall have the right to 
object, on grounds relating to his or her particular situation, at any 
time to processing of personal data concerning him or her. \\
\hline
{Right not to be 
subject to a decision} & The data subject 
shall have the right not to be subject to a decision based solely on 
automated processing, including profiling, which produces legal effects 
concerning him or her or similarly significantly affects him or her. \\
\hline
{Data 

Protection by Design} & A principle that calls for the 
inclusion of data protection from the onset of the designing of systems, 
rather than an addition. \\
\hline
{Joint 

controllers} & Joint controllers shall in a transparent 
manner determine their respective responsibilities for compliance with 
the obligations under the GDPR regulation. \\
\hline
{Personal data breach } & It is a security incident in which 
information is accessed without authorization. \\
\hline
{Data 

protection Impact Assessment} & A tool used to identify 
and reduce the privacy risks of entities by analyzing the personal data 
that are processed and the policies in place to protect the data. \\
\hline
{Legislative measures} & They are laws, treaties and 
regulations on the protection of traditional knowledge and traditional 
cultural expressions, and legislative texts relevant to personal data. 
\\
\hline
{Data 

protection 

officer} & A data protection officer (DPO) is 
an enterprise security leadership role required by the General Data 
Protection Regulation (GDPR). Data protection officers are responsible 
for overseeing data protection strategy and implementation to ensure 
compliance with GDPR requirements. \\

\hline
{Right to 

Access} & Also known as Subject Access Right, it 
entitles the data subject to have access to and information about the 
personal data that a controller has concerning them. \\
\hline
{Code of 

conduct} & It is intended to contribute to the proper 
application of the GDPR taking account of the specific features of the 
various processing sectors and the specific needs of micro, small and 
medium-sized enterprises. \\
\hline
{Data 

protection certification} & It is meant to demonstrate 
compliance with the GDPR of processing operations by controllers and 
processors. \\
\hline
{Infringeme-nt} & It is meant to the action of breaking the 
terms of the GDPR regulation. \\
\hline
{Cross-border data transfer} & The GDPR permits personal data 
transfers to a third country or international organization subject to 
compliance with set conditions, including conditions for onward 
transfer. \\
\hline
{Standard Contractual Clauses} & They are a common, 
standardized method for transferring personal data to controllers and 
processors located in non-adequate countries outside of the EEA. These 
act as a contract between two legal entities and they do not require a 
license. \\
\hline
{Binding 

Corporate 

Rules} & They are internal rules which 
define the international policy in a multinational group of companies 
and international organizations regarding intra-organizational personal 
data cross-border transfers. \\
\hline

\end{tabular}
\end{table*}

\begin{table*}[hbt]
\caption{Glossary (4 of 4)}\label{tab:glossary_4}
\begin{tabular}{|p{2cm}|p{14.5cm}|}
\hline
\textbf{Term} & \textbf{Description} \\

\hline
{Genetic data} & It means personal data relating to the 
inherited or acquired genetic characteristics of a natural person which 
give unique information about the physiology or the health of that 
natural person and which result, in particular, from an analysis of a 
biological sample from the natural person in question. \\
\hline
{Adequacy Decision} & An adequacy decision is a decision taken 
by the European Commission establishing that a third country provides a 
comparable level of protection of personal data to that in the European 
Union, through its domestic law or its international commitments. \\
\hline
Data 

Controller & A natural or legal person, public authority, agency or any other body which, alone or jointly with others, determines the purposes and means of the processing of personal data. \\
\hline
{Cross-border processing} & This type of processing often 
involves a unique multi-international controller/processor. Cross-border 
processing might take place in more than one Member State where the 
controller or processor is established. Cross-border processing might 
take place in the context of the activities of a single establishment of 
a controller or processor in the Union, but which substantially affects 
or is likely to substantially affect data subjects in more than one 
EU State. \\
\hline
{Data 

Processing} & Is any operation performed on personal data, whether or not by automated means, including collection, recording, organization, structuring, storage, etc. \\
\hline
{Delegated Acts} & Non-legislative acts enacted in order to 
supplement existing legislation and provide criteria or clarity. \\
\hline
\end{tabular}
\end{table*}
\clearpage

\subsection{Constraints expressed in natural language.} \label{plain}

As follows, we present the 35 compliance rules, in plain-English, we extracted from the GDPR.

\begin{table*}[hbt]
\caption{{Chapter I}: General provisions}\label{tab:Chapter_I}
\begin{tabular}{|p{1cm}|p{1.5cm}|p{13.5cm}|}
\hline
\textbf{ID} &\textbf{Source} & \textbf{Description} \\
\hline
C1 & Articles 2, and 3 & GDPR applies if some personal data is being processed or transferred AND The involved organizations are either operating within the EU
OR operating outside the EU, but offering goods or services to individuals within the EU NEVERTHELESS GDPR does not apply for processing concerned with national security purposes OR processing covered by law enforcement directives OR processing carried out by individuals for personal/household 
activities \\
\hline
\end{tabular}
\end{table*}

\begin{table*}[hbt]
\caption{{Chapter II}: Principles}\label{tab:Chapter_II}
\begin{tabular}{|p{1cm}|p{1.5cm}|p{13.5cm}|}
\hline
\textbf{ID} &\textbf{Source} & \textbf{Description} \\
\hline
C2 & Article 5 & Each processing of personal data must 
demonstrate that it meets GDPR principles (Principles::Demonstration). 
\\
\hline
C3 & Article 6 & IF lawfulness is based 
on a legal obligation, THEN further information about the obligation 
need to be provided (Principles::Obligation Source) AND IF the processing is for a purpose other than that for which the 
personal data have been collected AND processing is not based on the 
data subject's consent OR on a Union or Member State law, THEN additional information to demonstrate lawfulness are required 
(Principles::Lawfulness\_Evidence). \\
\hline
C4 & Article 7 & All the Boolean attributes of 
Data\_Processing:: Consent, which represent the requirement that a 
consent agreement must meet, should evaluate to true \\
\hline
C5 & Article 8 & When lawfulness is based on consent the DS must be at least 16 years old. OTHERWISE, the DS must be a child DS (Data Subjects::Child Data Subject) AND

Consent for data child subject has to be provided by their legal 
responsible parent (Data Subjects::Responsible Parent) \\
\hline
C6 & Article 9 & Processing of Special data categories (all the enumeration 
in Enumerations::Data Category, except OTHER PERSONAL DATA) are prohibited unless the processing: is based on consent (if prohibition can be lifted by consent)
OR is necessary for employment or social security
OR is necessary to protect vital subject interest (consent is not 
possible) OR is based on legitimate activities carried by any legal not-for-profit body OR concerns data that are manifestly made public by the data subject OR is necessary for the establishment, exercise or defense of legal claims OR is necessary for reasons of substantial public interest OR is necessary for the purposes of preventive or occupational medicine OR is necessary for reasons of public interest in the area of public health OR is necessary for scientific, historical, research or statistical purposes OR allowed by Member state law\\
\hline
C7 & Article 10 & Processing based on consent that deals with criminal convictions AND offenses or related security measures shall be carried out only under the control of official 
authority or When the processing is authorized by Union or Member State 
law providing for appropriate safeguards for the rights AND freedoms of 
data subjects AND

Any comprehensive register of criminal convictions shall be kept only 
under the control of official authority\\
\hline
C8 & Article 11 & IF 
processing deals with data that cannot make DS identifiable, THEN 
processing does not have to enable DS personal rights such as the right 
of access \\
\hline
\end{tabular}
\end{table*}

\begin{table*}[hbt]
\caption{{Chapter III}: Rights of the data subject (1 of 2)}\label{tab:Chapter_III_1}
\begin{tabular}{|p{1cm}|p{1.5cm}|p{13.5cm}|}
\hline
\textbf{ID} &\textbf{Source} & \textbf{Description} \\
\hline
C9 & Article 12 & Each processing 
of personal data must support DS rights when they are applicable
AND
Controllers have to react for a request to exercise a DS right within 
one month which can be extended by two further months where necessary, 
taking into account the complexity and number of the requests (DS must 
be informed of extension)
AND
IF the controller is not in a position to address a given requested DS 
right, THEN DS should be informed about the reasons related to this 
denial AND remind DS of his right to lodge a complaint
AND 
DS right should be free of charge, UNLESS request unfounded or 
excessive (e.g., repetitive requests)
AND
A controller must verify the DS identity (might ask for further 
information for that purpose) \\
\hline
C10 & Article 13 & 
Where personal data relating is collected directly from DS, the 
controller shall, at the time when personal data are obtained, 
provide: the identity and contact details of the controller AND DPO contact details (if exists) AND purpose and lawfulness sources AND data recipients details (if exists) AND transfer to a third party and safeguards details (if applicable) AND storage duration AND DS enabled rights AND the right to withdraw consent (if exists) AND the right to lodge a complaint AND contractual requirements, e.g., mandatory and possible consequences of non-collaboration (if applicable) AND automated decision-making details (if applicable) AND details about any further processing (if applicable) UNLESS the DS is already informed \\
\hline
C11 & Article 14 & Where personal data is not directly collected from DS, the controller shall provide DS with the following information: The identity and contact details of the controller AND DPO contact details (if exists) AND Purpose and lawfulness sources AND Data categories and details about the data concerning the DS AND Data recipients details (if exists) AND Transfer to a third party and safeguards details (if applicable) AND Storage duration AND DS enabled rights AND The right to withdraw consent (if exists) AND The right to lodge a complaint AND Automated decision-making details (if applicable) AND Details about any further processing (if applicable) UNLESS the DS is already informed OR Providing information requires dispassionate effort OR collection is foreseen by law OR professional secrecy obligations \\
\hline
C12 & Article 15 & The data subject shall have the right to obtain from the controller confirmation as to whether or not personal data concerning him or her are being processed, and, where that is the case, access to the personal data and the following information: Purpose and lawfulness sources AND Data recipients details AND Storage duration AND DS enabled rights AND The right to lodge a complaint AND Source of data (if collected indirectly) AND Automated decision-making details (if applicable) AND The right to receive a copy (free if first time) of personal data (UNLESS request affects rights and freedom of others) AND Appropriate safeguards in case of a cross-border transfer (if applicable) \\
\hline
C13 & Article 16 & Obligation to notify the 
rectification to each recipient to whom the data has been disclosed 
(unless disproportionate effort) AND Obligation to inform the DS of other recipients (if exists) \\
\hline
C14 & Article 17 & The controller must satisfy the DS right to be forgotten when: the data are is no longer necessary OR the data subject withdraws consent OR DS objects OR data have to be erased for compliance with a legal obligation in Union or Member State law to which the controller is subject OR consent or processing unlawful UNLESS the processing is necessary for: exercising the right of freedom of expression and information OR compliance with a legal obligation which requires processing by Union or Member State law to which the controller is subject OR the performance of a task carried out in the public interest OR the exercise of official authority vested in the controller OR public interest in the area of public health OR for archiving purposes in the public interest, scientific or historical research purposes or statistical purposes. IF triggered, other recipient and DS have to be informed \\

\hline
C15 & Article 18 & The data subject shall have the right to obtain restriction of processing when: the accuracy of the personal data is contested OR the processing is unlawful and the DS opposes the erasure of the personal data and requests the restriction of their use instead OR the controller no longer needs the personal data for the purposes of the processing, but they are required by the DS for the establishment, exercise or defense of legal claims OR the data subject has objected the processing and Notification obligation when the restriction is posed or is lifted 
\\
\hline
\end{tabular}
\end{table*}

\begin{table*}[hbt]
\caption{{Chapter III}: Rights of the data subject (2 of 2)}\label{tab:Chapter_III_2}
\begin{tabular}{|p{1cm}|p{1.5cm}|p{13.5cm}|}
\hline
\textbf{ID} &\textbf{Source} & \textbf{Description} \\
\hline

C16 & Article 20 & Data portability should be 
realized (upon request) when: the processing is not necessary for the performance of a task carried out in the public interest AND the processing is not necessary for the exercise of official authority 
vested in the controller AND DS identity verified AND technically feasible AND related lawfulness is based on consent or contract AND relevant data was directly collected from DS AND data portability does not affect the rights of all parties and laws OTHERWISE, the controller may deny the request (inform DS for reasons) 
\\
\hline
C17 & Article 21 & The DS shall have the right to object when processing purpose is for marketing OR The DS shall have the right to object when lawfulness is based on which is based on Consent OR legitimate interests OR public interest UNLESS controller demonstrates legitimate grounds which override the rights of DS, e.g., defense of legal claims. In any case, personal data shall no longer be processed for such purposes if objection accepted \\
\hline
C18 & Article 22 & The DS shall have the right not to be subject to a decision based solely on automated processing, including profiling, which produces legal effects concerning him or her or similarly significantly affects him or her UNLESS the automated decision is: is necessary for entering into, or performance of, a contract between the DS and a controller (but no special categories) OR is authorized by Union or Member State law to which the controller is 
subject OR is based on the DS's explicit consent. \\
\hline
\end{tabular}
\end{table*}

\begin{table*}[hbt]
\caption{{Chapter IV}: Controller and processor (1 of 2)}\label{tab:Chapter_IV_1}
\begin{tabular}{|p{1cm}|p{1.5cm}|p{13.5cm}|}
\hline
\textbf{ID} &\textbf{Source} & \textbf{Description} \\
\hline
C19 & Articles 24 and 25 & The  controller shall implement appropriate technical and organizational measures, including data protection policies, to ensure and to be able to demonstrate that processing is performed in accordance with GDPR AND Implemented measures shall be reviewed and updated where necessary AND The controller shall adhere to Data protection by design and by default principle \\
\hline
C20 & Article 26 & Joint controllers shall in a 
transparent manner determine their respective responsibilities for 
compliance with the obligations under the GDPR AND The responsibilities arrangement shall be made available to the data 
subject \\
\hline
C21 & Article 27 & When the controller or the processor is 
outside EU, it shall designate a representative in the Union UNLESS processing is occasional AND does not include processing of 
special categories of data and is unlikely to result in a risk to the rights and freedom of natural persons OR is consulted by a public authority or body AND IF a representative is designated, THEN it shall be established in one Member States\\
\hline
C22 & Articles 28 and 29 & A processor shall only act on the controller's documented instructions or the processor shall inform the controller of any intended changes AND A processor shall check that there are no conflicts between the controller's instructions and applicable EU law AND A processor shall ensure confidentiality AND A processor shall cooperate with supervisory authority AND A processor shall instore appropriate technical and organizational security measures AND A processor shall designate and collaborate with a DPA (when needed) \\
\hline
C23 & Article 30 & Controllers and their representatives shall record and maintain (in electronic form) processing activities, in particular: the name and contact details of the controller (or joint controllers), representatives and DPO (if applicable) AND the purposes of the processing AND a description of the categories of data subjects and of the categories of personal data AND the recipients to whom the personal data have been or will be disclosed AND where applicable, transfers of personal data to a third country or an international organization (+ the documentation of suitable safeguards) AND where possible, the envisaged time limits for erasure of the different categories of data AND where possible, a general description of the technical and organizational security measures Similarly, processors have to record: the name and contact details of the processor or processors and of each controller on behalf of which the processor is acting, and, where applicable, of the controller's or the processor's representative, and the data protection officer AND the categories of processing carried out on behalf of each controller AND where applicable, transfers of personal data to a third country or an international organization (+ the documentation of suitable safeguards) AND where possible, a general description of the technical and  organizational security measures. \\
\hline
C24 & Article 31 & The controller and the processor and, where applicable, their representatives, shall cooperate, on request, with the supervisory authority in the performance of its tasks. \\
\hline

\end{tabular}
\end{table*}

\begin{table*}[hbt]
\caption{{Chapter IV}: Controller and processor (2 of 2)}\label{tab:Chapter_IV_2}
\begin{tabular}{|p{1cm}|p{1.5cm}|p{13.5cm}|}
\hline
\textbf{ID} &\textbf{Source} & \textbf{Description} \\
\hline
C25 & Article 32 & The controller and the processor shall implement appropriate technical and organizational measures to ensure a level of security appropriate to the risk, including: the pseudonymisation and encryption of personal data AND the ability to ensure confidentiality, integrity, availability and resilience of processing systems and services AND the ability to restore the availability and access to personal data in the event of a physical or technical incident AND a process for regularly testing, assessing and evaluating the effectiveness of technical and organizational measures \\
\hline
C26 & Articles 33 and 34 & When data breaches are detected by a controller, the controller should act according to the associated risk: 
"Case low": record breaches
"Case Medium": Case low + notify the concerned SA within 72 hours (any delay requires justification and might lead to administrative fines)
"Case High": "Case Medium" + communication to DS without undue delay
When data breaches are detected by a processor, the processor informs all involved controllers without undue delay after becoming aware of a personal data breach \\
\hline
C27 & Article 35 & Where a type of 
processing is likely to result in a high risk to the rights and freedoms of natural persons, the controller shall, prior to the processing, carry out an assessment of the impact of the envisaged processing operations on the protection of personal data.

Is mandatory when : the processing is a systematic and extensive evaluation of personal aspects (e.g., profiling) or processing is used for automated decisions that might affect the natural person OR the processing is processing on a large scale of special categories of data OR the processing is a systematic monitoring of a publicly accessible area on a large scale OR type of processing is present in the list of processing types that must have a DPIA as published by SA OR Member state law requires processing without DPIA UNLESS the type of processing is present in the list of processing types that should no longer have a DPIA as published by SA OR similar DPIA exist (re-use measures) OR Member state law allows processing without DPIA AND The controller shall seek the advice of the data protection officer AND IF DPIA exists THEN, it should contain at least: A description of the processing and its purposes AND An assessment of the necessity and proportionality AND An assessment of the risks to individuals AND The measures in place to address risks, including security measures and compliance demonstration \\
\hline
C28 & Article 36 & When a data protection impact 
assessment indicates that the processing would result in a high risk, the controller shall consult the supervisory authority prior to processing AND IF a consultation exists, THEN, The supervisory authority shall, within a period of up to eight weeks of 
receipt of the request for consultation, provide written advice to the controller (period extensible to six weeks + notification) AND A controller shall provide the all relevant information for 
communication \\
\hline
C29 & Articles 37 and 38 & The designation of a DPA is mandatory when: The controller or processor is a public authority or body (except for courts acting in their judicial capacity) OR If processing concerns large scale systematic monitoring of individuals (e.g., online behavior tracking) OR large scale processing of special categories of data or data relating to criminal convictions and offenses \\
\hline
C30 & Article 42 & The certification shall be voluntary and available via a process that is transparent AND A certification must be provided by an accredited body (accredited by an SA) AND A certification last three years AND A certification does not make controllers/processors less accountable \\
\hline
\end{tabular}
\end{table*}

\begin{table*}[hbt]
\caption{{Chapter V}: Transfers of personal data to third countries or international organizations}\label{tab:Chapter_V}
\begin{tabular}{|p{1cm}|p{1.5cm}|p{13.5cm}|}
\hline
\textbf{ID} &\textbf{Source} & \textbf{Description} \\
\hline
C31 & Articles 44, 45, 46, 49 and 50 & 
An across-border transfer, including onward ones, shall only be 
permitted IF the transfer: remains within EU OR is supported by an adequacy decision OR is supported by one of the following: (Approved binding corporate rules in case of a corporation OR Approved standard contractual clauses OR Authorized (by SA) provisions in administrative arrangements OR Approved code of conduct or certifications OR Authorized (by SA) Binding and enforceable instrument between public bodies) UNLESS, transfer is part of one of the flowing derogations: DS has explicitly consented to the proposed transfer, (DS must be informed of possible risks + does not apply to public authorities) OR Transfer is necessary for the performance of a contract (requested by DS or concluded in the interest of DS + does not apply to public authorities) OR Transfer is necessary to protect the DS vital interests (when DS is incapable of giving consent) OR Transfer is necessary for reasons of public interest (recognized in Union law or in the law of the Member State) OR Transfer is necessary for the establishment, exercise or defense of 
legal claims OR Transfer made from a register open to consultation by the public OR If the transfer is not repetitive, concerns only a limited number of DS, is necessary for the purposes of legitimate interests pursued by the controller (SA and DSs must be informed) \\
\hline
C32 & Article 45 & IF the adequacy decision involves some additional requirements (e.g., partial  adequacy requirement that covers a specific sector within a third country), THEN the importer of the data must provide evidence of the satisfaction of these \\
\hline
C33 & Article 48 & Any 
legal decision from a third country requiring the disclose personal data 
may only be recognised or enforceable if decision based on an 
international agreement, such as a mutual legal assistance treaty \\
\hline
C34 & Article 47 & To be approved, a binding 
corporate rules: are legally binding and apply to and are enforced by every member concerned involved in a joint economic activity, including their employees; AND confer enforceable rights on DS AND included the following information: The structure and contact details of the group and of each of its members AND The categories of personal data, the type of processing and its purposes, the type of data subjects affected AND The third countries and their legally binding nature, both internally and externally AND The application of the general data protection principles AND The rights of data subjects in regard to processing and the means to exercise those rights AND The liability sharing in case of a breach or infringement AND The used means to communicate with data subjects AND The tasks of any DPO or any other person or entity in charge of data 
protection AND The mechanisms for ensuring the verification of compliance with the BCRs (e.g., audits and methods for ensuring corrective actions) AND The mechanisms for reporting and recording changes to the BCRs and reporting those changes to the SA AND The appropriate data protection training to personnel having permanent 
or regular access to personal data AND The cooperation and reporting mechanisms to be used with the SA (e.g., 
eBay and ArcelorMittal) \\
\hline
C35 & Article 83 & The 
total maximum administrative fine can cumulate up to Max(10.000.000 EUR, 
2 \% total world-wide annual turnover of preceding financial year) when:  controllers or processors obligations are not fulfilled, e.g., non-compliance to the requirements for child consent OR the criteria for the certification are no longer met OR the monitoring body reports an infringement due to miss-application of code of conducts. The total maximum administrative fine can add up to Max(20.000.000 EUR, 4 \% total world-wide annual turnover of preceding financial year) when the infringements concern the violation of: the basic principles for processing OR DS rights OR cross-border transfers of personal data rules OR any additional obligation imposed by Member states OR orders from an SA, for example, non-compliance to a temporary or definitive limitation on processing \\
\hline
\end{tabular}
\end{table*}

\clearpage

\subsection{Variability Table} \label{variabilitytable}

As follows, we present the 20 variation points we extracted from the GDPR.

\begin{table*}[hbt]
\caption{Variability Table (1 of 2)}\label{tab:VarTable_1}
\begin{tabular}{|p{1cm}|p{1.5cm}|p{8cm}|p{5cm}|}
\hline
\textbf{ID} &\textbf{Source} & \textbf{Description} & \textbf{How to resolve}\\
\hline
	V1 & Article 8 & EMS law may provide for a lower age (…) provided that such lower age is not below 13 years & Add the V1 OCL constraint to the generic model and implement {\sf V\_getMinimumAgeForDS} based on the EMS laws.  \\ \hline
	V2 & Article 8 & Checks that a given person is indeed the  holder  of  parental  responsibility  over  a  given data subject according to the EMS law & Add the V2 OCL constraint to the generic model and implement {\sf V\_checkParentDocuments} based on the EMS laws.  \\ \hline
	V3 & Article 9 & The processing of sensitive personal data is prohibited unless the data subject has given explicit consent (…), except where (…) EMS law provide that the prohibition (…) may not be lifted by the DS & Implement {\sf V\_prohibitionCanBeLiftedByConsent} of the constraint C6 based on the EMS laws.\\ \hline
	V4 & Article 9 & EMS may maintain or introduce further conditions, including limitations, with regard to the processing of genetic data, biometric data or data concerning health  & Add the V4 OCL constraint to the generic model and implement {\sf V\_verifyFurtherConditionsAndLimit} based on the EMS laws. \\\hline
	V5 & Article 23 & EMS law (…) may restrict by way of a legislative measure the scope of the obligations and rights provided for in Articles 12 to 22, as well as Article 5 (…) when such a restriction respects the essence of the fundamental rights and freedoms and is a necessary and proportionate measure in a democratic society &   Adapt the OCL constraints C2,   and  from C9   to   C22, that are already in the generic model, based on the restrictions of EMS laws.\\ \hline
	V6 & Article 29 & The processor and any person acting under the authority of the controller or of the processor, who has access to personal data, shall not process those data except on instructions from the controller, unless required to do so by (…) EMS law & Implement {\sf V\_processWithoutControllerInstruct-ions} used in the constraint C22 based on the EMS laws.\\ \hline
	V7 & Article 32 & The controller and processor shall take steps to ensure that any natural person acting under the authority of the controller or the processor who has access to personal data does not process them except on instructions from the controller, unless he or she is required to do so by (…) EMS law & Add the V7 OCL constraint to the generic model and implement {\sf V\_processWithoutControllerInstruc-tions} based on the EMS laws. \\ \hline
	V8 & Article 36 & EMS shall by law reconcile the right to the protection of personal data pursuant to this Regulation with the right to freedom of expression and information & Add the V8 OCL constraint to the generic model, RECONCILIATION\_ASSESSMENT enumeration literal in {\sf DPIA\_Information\_Type}, and implement {\sf V\_ReconcileByLaw} based on the EMS laws.\\ \hline
	V9 & Article 37 & In cases other than those referred to in paragraph 1, the controller or processor or associations and other bodies representing categories of controllers or processors may or, where required by (…) EMS law shall, designate a data protection officer. The data protection officer may act for such associations and other bodies representing controllers or processors & Implement {\sf V\_bodiesMustDesignateDPO} used in the constraint C29 based on the EMS laws.\\ \hline

\end{tabular}
\end{table*}

\begin{table*}[hbt]
\caption{Variability Table (2 of 2)}\label{tab:VarTable_2}
\begin{tabular}{|p{1cm}|p{1.5cm}|p{8cm}|p{5cm}|}
\hline
\textbf{ID} &\textbf{Source} & \textbf{Description} & \textbf{How to resolve}\\
\hline
	V10 & Article 49 & In the absence of an adequacy decision, (…) EMS law may, for important reasons of public interest, expressly set limits to the transfer of specific categories of personal data to a third country or an international organisation & Add the V10 OCL constraint to the generic model and implement {\sf V\_verifyTransferLimits} based on the EMS laws. \\ \hline
	V11 & Article 80 & EMS may provide that any body, organisation or association (…) has the right to lodge, in that MS, a complaint with the supervisory authority (…) and to exercise the rights referred to in Articles 78 and 79 if it considers that the rights of a data subject under this Regulation have been infringed & Add the V11 OCL constraint to the generic model.\\ \hline
	V12 & Article 83 & Without prejudice to the corrective powers of supervisory authorities pursuant to Article 58(2), each EMS may lay down the rules on whether and to what extent administrative fines may be imposed on public authorities and bodies established in that EMS& Replace the C35 OCL constraint with the V12\_1 OCL constraint, and add the V12\_2 OCL constraint to the generic model\\ \hline
	V13 & Article 84 & MS shall lay down the rules on other penalties applicable to infringements of this Regulation & Add the V13 OCL constraint to the generic model.\\ \hline
V14 & Article 85 &  Notwithstanding paragraph 1, EMS law may require controllers to consult with, and obtain prior authorisation from, the supervisory authority in relation to processing by a controller for the performance of a task carried out by the controller in the public interest, including processing in relation to social protection and public health & Add the V14 OCL constraint to the generic model.\\ \hline
V15 & Article 87 & EMS may further determine the specific conditions for the processing of a national identification number or any other identifier of general application & Add the V15 OCL constraint to the generic model, implement {\sf V\_checkedIDProcessing} based on the EMS laws, and add the IDENTIFICATION literal to the {\sf Data\_Category} enumeration. \\ \hline
	V16 & Article 88 & EMS may, by law or by collective agreements, provide for more specific rules to ensure the protection of the rights and freedoms in respect of the processing of employees' personal data in the employment context & Add the EMPLOYMENT\_ASSESSMENT literal to the {\sf DPIA\_Information\_Type} enumeration.\\ \hline
	V17 & Article 89 & Where personal data are processed for scientific or historical research purposes or statistical purposes, (…) EMS law may provide for derogations from the rights referred to in Articles 15, 16, 18 and 21  & Add the V17 OCL constraint to the generic model, and implement {\sf V\_checkDerrogationsFromRights} based on the EMS laws.\\ \hline
	V18 & Article 89 & Where personal data are processed for archiving purposes in the public interest, (…) EMS law may provide for derogations from the rights referred to in Articles 15, 16, 18, 19, 20 and 21 & Add the V18 OCL constraint to the generic model, and implement {\sf V\_checkDerrogationsFromRights} based on the EMS laws.\\ \hline
	V19 & Article 90 & EMS may adopt specific rules to set out the powers of the supervisory authorities (…) in relation to controllers or processors that are subject, under (…) EMS law or rules established by national competent bodies, to an obligation of professional secrecy & Add the V19 OCL constraint to the generic model, and implement {\sf V\_checkDerrogationsFromRights} based on the EMS laws. \\ \hline
	V20 & Article 91 & Where in a EMS, churches and religious associations or communities apply, at the time of entry into force of this Regulation, comprehensive rules relating to the protection of natural persons with regard to processing, such rules may continue to apply, provided that they are brought into line with this Regulation & Add the V20 OCL constraint to the generic model, and the CHURCH\_OR\_RELIGIOUS\_OR-GANIZATION literal to the {\sf Actor\_Type} enumeration  \\ \hline
\end{tabular}
\end{table*}

\clearpage

\section*{{Appendix~B}} \label{appendixB}
In this section, we present the 35 compliance rules, presented in plain English in Section~\ref{plain} of Appendix~A, expressed in OCL.

\begin{lstlisting}[linewidth=\columnwidth,xleftmargin=1em, breaklines=true]
context Data_Processing
def: getIdentifiableDS(dp:Data_Processing):Set(Data_Subject) = dp.personal_data.data_subject->flatten()->asSet()
inv C1: 
    processing->notEmpty() 
    and (
         processing.actors->exists(a|a.countries->exists(c|c.isEUMemberState or c.EULawApplies))
        or processing->select(dp:Data_Processing|dp.type=OFFERING_GOODS_OR_SERVICES or 
        dp.type=EU_BEHAVIOUR_MONITORING_OR_PROFILING)
        ->exists(dp|getIdentifiableDS(dp).residence->exists(country|country.isEUMemberState or 
        country.EULawApplies)
        )
    )
    and 
    not processing->forAll(p|p.type <> TEU_SECURITY_ACTIVITY and p.type <> PERSONAL_OR_HOUSEHOLD_ACTIVITY and p.type <> LEGAL_AND_CRIMINAL_INVESTIGATION
    )
\end{lstlisting}
\bigbreak
\begin{lstlisting}[linewidth=\columnwidth,xleftmargin=1em, breaklines=true]
inv C2:  
    processing->forAll(p|p->select(demostrates->notEmpty()))
\end{lstlisting}
\bigbreak
\begin{lstlisting}[linewidth=\columnwidth,xleftmargin=1em, breaklines=true]
context Lawfulness
inv C3:
    let processing: Set(Data_Processing)=Data_Processing.allInstances()->select(personal_data->notEmpty()),
    legal_obligation_lawfulness:Set(Lawfulness)=Lawfulness.allInstances()->select(l|l.Lawfulness_Source.source=LEGAL_OBLIGATION) in
    if (legal_obligation_lawfulness->notEmpty())
    then  
        legal_obligation_lawfulness->
        forAll(l:Lawfulness|l.Lawfulness_Source->
            forAll(ls|ls.obligation_source.moreInformation<>)
            )
    else false
    endif
    and
    if (processing->select(dp|dp.purposes._context <> dp.new_purpose._context)->
        forAll(dp:Data_Processing|dp.new_purpose.consent->size()=1 
            or not dp.actors->exists(a|a.countries->exists(c|c.isEUMemberState or c.EULawApplies))
        )
    )
\end{lstlisting}
\bigbreak
\begin{lstlisting}[linewidth=\columnwidth,xleftmargin=1em, breaklines=true]
inv C4:
Consent.allInstances()->forAll(c| c.isFreelyGiven 
    and c.canBeWithdrawn 
    and c.unambiguous 
    and c.involvesConfirmationAction 
    and c.distinguishableFromOther 
    and c.isIntelligible 
    and c.isEasyToAccess 
    and c.usesClearPlainLanguage 
    and c.describeSubjectEnabledRights 
    and c.isEasyToWithdraw)
\end{lstlisting}
\bigbreak
\begin{lstlisting}[linewidth=\columnwidth,xleftmargin=1em, breaklines=true]
inv C5:
self.isLawfulnessOnlyByConsent() implies
let identifiableSubjects : Set(Data_Subject) = self.personal_data.data_subject->flatten()->asSet() in
self.purposes->forAll
(p:Purpose|
    identifiableSubjects->forAll(ds:Data_Subject|
        let eligibleToGiveConsent : Natural_Person =
        if ds.oclIsTypeOf(Data_Subject)
        then ds
        else ds.getResponsibleParent()
        endif in p.getConsents()->exists(c:Consent|
            c.provider=eligibleToGiveConsent
            and c.target=ds
        )
    )
)
\end{lstlisting}
\bigbreak
\begin{lstlisting}[linewidth=\columnwidth,xleftmargin=1em, breaklines=true]
inv C6:
processing->forAll(dp:Data_Processing|dp.personal_data->forAll(pd|
if (Bag{Data_Category::RACIAL_OR_ETHNIC, Data_Category::RELIGIOUS_OR_PHILISOPHICAL_BELIFS, Data_Category::POLITICAL_OPINIONS, Data_Category::HEALTH, Data_Category::SEX_LIFE, Data_Category::GENETIC, Data_Category::BIOMETRIC, Data_Category::JUDICIAL}->includes(pd.category)
    and not dp.purposes->forAll(p|
        if Variability.V_prohibitionCanBeLiftedByConsent(dp) 
        then p.consent->notEmpty() 
        else false 
        endif
        or Bag{Processing_Context::EMPLOYMENT, Processing_Context::VITAL_INTERESTS, 
            Processing_Context::SHARED_BY_DATA_SUBJECT, Processing_Context::STATISTICAL_PURPOSES, 
            Processing_Context::RESEARCH, Processing_Context::LEGAL_AND_CRIMINAL_INVESTIGATION, 
            Processing_Context::PUBLIC_INTEREST, Processing_Context::PREVENTIVE_MEDICINE, 
            Processing_Context::PUBLIC_HEALTH}->includes(p._context)
            or p.hasLegitimateInterest))
then isForbidden=true
else isForbidden=false
endif))
\end{lstlisting}
\bigbreak
\begin{lstlisting}[linewidth=\columnwidth,xleftmargin=1em, breaklines=true]
inv C7:
processing->forAll(dp |
    if dp.Lawfulness_Source.source = BY_CONSENT and dp.purposes -> exists(p | p._context =
        LEGAL_AND_CRIMINAL_INVESTIGATION) and(dp.security_measure -> notEmpty()) then 
        dp.getConcernedSupervisoryAuthority().type = Actor_Type::OFFICIAL
    else false
    endif)
\end{lstlisting}
\bigbreak
\begin{lstlisting}[linewidth=\columnwidth,xleftmargin=1em, breaklines=true]
inv C8:
processing->forAll(dp |
    if dp.areDataSubjectIdentifiable then dp.right - > isEmpty()
    else false
    endif)
\end{lstlisting}
\bigbreak
\begin{lstlisting}[linewidth=\columnwidth,xleftmargin=1em, breaklines=true]
inv C9: 
    processing->forAll(dp|dp.right->notEmpty())
    and
    Data_Controller.allInstances()->forAll(dc|
        if dc.Contract_Agreement.duration() < 1 then 
            true
        else if (dc.Contract_Agreement.duration() >= 1 and dc.Contract_Agreement.duration() <=3) then
                dc.Contract_Agreement.data_processing->forAll(dp|dp.right->
                    select(oclIsKindOf(Right_To_Be_Informed))->forAll(r|r.oclAsType(Right_To_Be_Informed).
                    informDS())
                )
                else false
            endif  
        endif
    )
    and
    Data_Controller.allInstances()->forAll(dc|
        if dc.rightNotAddressed()<>null then 
            dc.Contract_Agreement.data_processing->forAll(dp|dp.right
                ->select(oclIsKindOf(Right_To_Be_Informed))
                    ->forAll(r|
                        r.oclAsType(Right_To_Be_Informed).information
                        ->exists(i|i.type=Information_Type::RIGHT_TO_LODGE_COMPLAINT
                        )
                        and r.oclAsType(Right_To_Be_Informed).informDS()))
        else false
        endif
    )
    and 
    Right.allInstances()->forAll(r|
        if r.isUnfoundedOrExcessive 
        then r.price>0 else r.price=0 endif
    )
\end{lstlisting}
\bigbreak
\begin{lstlisting}[linewidth=\columnwidth,xleftmargin=1em, breaklines=true]
def:data_controllers(dp : Data_Processing) : Set(Actor) = dp.actors->select(a|a.oclIsTypeOf(Data_Controller))
def:processing_related_to_controller (actor:Actor) : Set(Data_Processing) = actor.oclAsType(Data_Controller).Contract_Agreement.data_processing
inv C10: 
processing->forAll(dp|
    if (dp.operation=COLLECTING) then (
        dp.personal_data->forAll(pd|
        if pd.colllectedDirectly then (
            data_controllers(dp)->forAll(actor|
                processing_related_to_controller(actor)->forAll(dp|dp.right->select(oclIsKindOf(
                Right_To_Be_Informed))
                ->forAll(r| 
                let right:Right_To_Be_Informed=r.oclAsType(Right_To_Be_Informed) in
                if not right.DSAlreadyInformed then
                let actual_info:Bag(Information_Type)=right.information->collect(i|i.type) in
                actual_info->includes(Bag{
                    Information_Type::CONTROLLER_DETAILS,   
                    Information_Type::DPO_DETAILS,
                    Information_Type::PURPOSE_AND_LAWFULNESS,  
                    Information_Type::RECIPIENTS, 
                    Information_Type::TRANSFERS_THIRD_COUNTRIES,
                    Information_Type::STORAGE_DURATION,
                    Information_Type::DS_RIGHTS,
                    Information_Type::CONSENT_WITHDRAWABLE,
                    Information_Type::RIGHT_TO_LODGE_COMPLAINT,
                    Information_Type::STATUTORY_CONTRACTUAL_REQUIREMENT,
                    Information_Type::AUTOMATED_DECISION,
                    Information_Type::FURTHER_PROCESSING
                })
                else false
                endif
                )
            )
        )
    )
    else false
    endif
    )
    )
    else false
    endif
)
\end{lstlisting}
\bigbreak
\begin{lstlisting}[linewidth=\columnwidth,xleftmargin=1em, breaklines=true]
inv C11:
processing->forAll(dp|
    if (dp.operation=COLLECTING) then (
        dp.personal_data->forAll(pd|
        if not pd.colllectedDirectly then (
            data_controllers(dp)->forAll(actor|
                processing_related_to_controller(actor)->forAll(dp|dp.right->select(oclIsKindOf(
                Right_To_Be_Informed))
                ->forAll(r| 
                let right:Right_To_Be_Informed=r.oclAsType(Right_To_Be_Informed) in
                if (not right.DSAlreadyInformed 
                    or right.requiresDisproportionateEffort
                    or right.forssenByLaw
                    or right.disabledForSecrecy)
                then
                let actual_info:Bag(Information_Type)=right.information->collect(i|i.type) in
                actual_info->includes(Bag{
                    Information_Type::CONTROLLER_DETAILS,  
                    Information_Type::DPO_DETAILS,
                    Information_Type::PURPOSE_AND_LAWFULNESS,
                    Information_Type::DATA_CATEGORIES,
                    Information_Type::CONTACT_DETAILS,
                    Information_Type::RECIPIENTS, 
                    Information_Type::TRANSFERS_THIRD_COUNTRIES,
                    Information_Type::STORAGE_DURATION,
                    Information_Type::DS_RIGHTS,
                    Information_Type::CONSENT_WITHDRAWABLE,
                    Information_Type::RIGHT_TO_LODGE_COMPLAINT,
                    Information_Type::AUTOMATED_DECISION, 
                    Information_Type::FURTHER_PROCESSING 
                    })
                else false
                endif
                )
            )
        )
    )
    else false
    endif
    )
    )
    else false
    endif
)
\end{lstlisting}
\bigbreak
\begin{lstlisting}[linewidth=\columnwidth,xleftmargin=1em, breaklines=true]
inv C12: 
processing->forAll(dp|
    dp.personal_data->forAll(pd|
        data_controllers(dp)->forAll(actor|
            processing_related_to_controller(actor)->forAll(dp|
                dp.right->select(oclIsKindOf(Right_To_Access))
                ->forAll(r|
                    let actual_info:Bag(Information_Type)=r.oclAsType(Right_To_Access).information->
                    collect(i|i.type) in
                    actual_info->includes(Bag{
                    Information_Type::PURPOSE_AND_LAWFULNESS,
                    Information_Type::RECIPIENTS, 
                    Information_Type::STORAGE_DURATION,
                    Information_Type::DS_RIGHTS,
                    Information_Type::RIGHT_TO_LODGE_COMPLAINT,
                    Information_Type::DATA_SOURCE,
                    Information_Type::AUTOMATED_DECISION
                    })
                    and r.oclAsType(Right_To_Access).impactsFreedomOfOthers implies
                        actual_info->includes(Information_Type::RIGHT_TO_RECEIVE_COPY)
                    and dp.isCrossBorders implies 
                         actual_info->includes(Information_Type::TRANSFERS_THIRD_COUNTRIES)
                )
            )
        )
    )
)
\end{lstlisting}
\bigbreak
\begin{lstlisting}[linewidth=\columnwidth,xleftmargin=1em, breaklines=true]
inv C13:
Right_To_Rectification.allInstances() ->
    forAll(r | r.isRectification implies
        if not r.requiresDisproportionateEffort then r.notification -> forAll(n | n.recipient -> 
        notEmpty() and r.informDS())
        else false
        endif
    )
\end{lstlisting}
\bigbreak
\begin{lstlisting}[linewidth=\columnwidth,xleftmargin=1em, breaklines=true]
inv C14:
    Right_To_Erasure.allInstances()->forAll(
    rte| let processing_is_not_necessary: Boolean=rte->collect(r|r.reason)
    ->select(r1|r1=Denial_Erasure_Reason::FREEDOM_EXPRESSION_INFORMATION or
        r1=Denial_Erasure_Reason::COMPLIANCE_LEGAL_OBLIGATION or
        r1=Denial_Erasure_Reason::PUBLIC_INTEREST or
        r1=Denial_Erasure_Reason::ARCHIVING_PURPOSES or
        r1=Denial_Erasure_Reason::DEFENCE_LEGAL_CLAIMS)->isEmpty() in
        processing_is_not_necessary 
        implies 
            rte.denied=true 
            and 
                rte->collect(r|r.denialReason)->includes (
                Erasure_Right_Reason::DATA_NO_LONGER_NECESSARY, 
                Erasure_Right_Reason::CONSENT_WITHDRAWN,
                Erasure_Right_Reason::DS_OBJECTS,
                Erasure_Right_Reason::LEGAL_OBLIGATION,
                Erasure_Right_Reason::UNLAWFUL_PROCESSING})
            and
                rte.notification->forAll(n|n.recipient->notEmpty() and rte.informDS())
    )
\end{lstlisting}
\bigbreak
\begin{lstlisting}[linewidth=\columnwidth,xleftmargin=1em, breaklines=true]
inv C15:
    processing->forAll(dp|
    dp.personal_data->forAll(pd|
        data_controllers(dp)->forAll(actor|
            processing_related_to_controller(actor)->forAll(dp|
                dp.right->select(oclIsKindOf(Right_To_Restriction))
                ->forAll(r|
                    Set{
                        Restriction_Right_Reason::CONTESTED_ACCURACY,
                        Restriction_Right_Reason::UNLAWFUL_PROCESSING, 
                         Restriction_Right_Reason::DATA_NO_LONGER_NECESSARY,
                        Restriction_Right_Reason::DS_OBJECTS
                    }->includes(r.oclAsType(Right_To_Restriction).reason)
                )
            )
        )
    )
)
\end{lstlisting}
\bigbreak
\begin{lstlisting}[linewidth=\columnwidth,xleftmargin=1em, breaklines=true]
inv C16:
Right_To_Data_Portability.allInstances() -> forAll(rtdp |
    if (rtdp.data_processing ->
        forAll(dp | dp.purposes._context <> Processing_Context::PUBLIC_INTEREST and 
        dp.areDataSubjectIdentifiable and(dp.Lawfulness_Source.source = 
        Lawfulness_Source::BY_CONSENT or dp.Lawfulness_Source.source = 
        Lawfulness_Source::PERFORMANCE_OF_CONTRACT)) and rtdp.checkIfPossible() and 
        rtdp.data_processing -> forAll(dp | dp.personal_data.colllectedDirectly = true) and not
        rtdp.affectRightsOfOthers)
        then rtdp.transferData() = true
    else rtdp.transferData() = false
    endif)
\end{lstlisting}
\bigbreak
\begin{lstlisting}[linewidth=\columnwidth,xleftmargin=1em, breaklines=true]
inv C17:
Right_To_Object.allInstances()->forAll(rto| 
if (
    rto.data_processing->forAll(dp|
        dp.Lawfulness_Source.lawfulness_evidence->select(le|le.specialCategoryEnabler=
        ESTABLISHMENT_EXERCISE_OR_DEFENSE_LEGAL_CLAIMS)->size()=0 
        and (dp.purposes._context=Processing_Context::MARKETING
        or Bag{
            Lawfulness_Source::BY_CONSENT,
            Lawfulness_Source::LEGITIMATE_INTERESTS,
            Lawfulness_Source::PUBLIC_INTEREST
            }->exists(dp.Lawfulness_Source.source)
        )
    )
)
then rto.object() = true
else rto.object() = false
endif
)
\end{lstlisting}
\bigbreak
\begin{lstlisting}[linewidth=\columnwidth,xleftmargin=1em, breaklines=true]
inv C18:
Right_To_Not_Be_Part_Of_A_Decision.allInstances()->forAll(right| 
right.data_processing->forAll(dp|
let nAutomatedPurposes = dp.purposes->collect(p:Purpose|p._context=Information_Type::AUTOMATED_DECISION)->size() in 
if (
    dp.purposes->size()=nAutomatedPurposes
    and not dp.purposes->forAll(p|
        if (
            p._context=Information_Type::AUTOMATED_DECISION 
            and dp.contractAgreement->notEmpty()
            and not dp.targetsSpecialDataCategories
        )
        then true
        else false
        endif
    )
)
then right.object()=true
else right.object()=false
endif
)
)
\end{lstlisting}
\bigbreak
\begin{lstlisting}[linewidth=\columnwidth,xleftmargin=1em, breaklines=true]
inv C19:
processing->forAll(dp:Data_Processing|
    let measures=dp.security_measure, 
    technicalMeasures=dp.security_measure->collect (sm|sm.oclIsTypeOf(Organizational)),
    organizationalMeasures=dp.security_measure->collect (sm|sm.oclIsTypeOf(Technical)) in  
    technicalMeasures->notEmpty()
    and organizationalMeasures->notEmpty()
    and organizationalMeasures->exists(om|om.oclAsType(Organizational).type=Technical_Measure_Type::DATA_PROTECTION)
    and measures->forAll(isRevisedPeriodically)
)
\end{lstlisting}
\bigbreak
\begin{lstlisting}[linewidth=\columnwidth,xleftmargin=1em, breaklines=true]
inv C20:
    Joint_Controllers.allInstances() -> forAll(jc | jc.responsibilitiesArrangement -> notEmpty() and jc.arrangementMadeAvailableToDS)
\end{lstlisting}
\bigbreak
\begin{lstlisting}[linewidth=\columnwidth,xleftmargin=1em, breaklines=true]
inv C21:
processing->forAll(dp| 
    let controller_or_processor_is_outside_EU:Set(Actor) = dp.actors-> 
    select(a| a.oclIsKindOf(Controller) or a.oclIsKindOf(Processor) and a.countries->exists(c:Country|c.isEUMemberState)) in
    if (controller_or_processor_is_outside_EU->notEmpty() 
        and not dp.isOccasional
        )
        and (
            not (dp.targetsSpecialDataCategories)  
             or dp.actors->exists(oclIsKindOf(Legal_Authority))
        )
    then controller_or_processor_is_outside_EU->forAll(a|a.oclAsType(Representative)<>null)
    else false
    endif
)
\end{lstlisting}
\bigbreak
\begin{lstlisting}[linewidth=\columnwidth,xleftmargin=1em, breaklines=true]
inv C22: 
processing->forAll(dp:Data_Processing|
    processors(dp)->forAll(a|
        let processor = a.oclAsType(Data_Processor) in
        not Variability.V_processWithoutControllerInstructions(dp,processor)
        implies (if processor.actedOnControllerDocumentedInstructions(dp) = true
            then true
            else processor.informedTheControllerOfIntendedChanges(dp) = true
            endif)
        
        and dp.actors->select(a|a.oclIsKindOf(Data_Controller))
            ->forAll(c|processor.checkedNoConflicts(c.oclAsType(Data_Controller)))
        and 
        processor.ensuredConfidentiality()
        and processor.cooperatedWithSA(dp)
        and (let measures=dp.security_measure, 
        technicalMeasures=dp.security_measure->collect (sm|sm.oclIsTypeOf(Organizational)),
        organizationalMeasures=dp.security_measure->collect (sm|sm.oclIsTypeOf(Technical)) in  
        technicalMeasures->notEmpty()
        and organizationalMeasures->notEmpty())
        and processor.data_protection_officer->notEmpty()
        )
)
\end{lstlisting}
\bigbreak
\begin{lstlisting}[linewidth=\columnwidth,xleftmargin=1em, breaklines=true]
def:controllers(dp:Data_Processing) : Set(Actor) = dp.actors->select(a|a.oclIsKindOf(Controller))
inv C23:
processing->forAll(dp:Data_Processing|
    controllers(dp)->forAll(a|
        let controller = a.oclAsType(Controller),
        activity = controller.processing_activity,
        details = activity->collect(details) in
        activity->forAll(isElectronic)
        and details->includes 
        (Bag{
            Information_Type::CONTROLLER_NAME,
            Information_Type::CONTROLLER_DETAILS,
            Information_Type::DPO_DETAILS,
            Information_Type::PROCESSING_PURPOSES,
            Information_Type::DS_CATEGORIES,
            Information_Type::PERSONAL_DATA_CATEGORIES,
            Information_Type::RECIPIENTS
        })
        and activity->forAll(a|
            a.transfersToThirdCountries implies a.details->exists(Information_Type::TRANSFERS_THIRD_COUNTRIES)
            and a.areErasureTimeLimitsKnown implies a.details->exists(Information_Type::TIME_LIMITS_FOR_ERASURE)
            and a.areSecurityMeasuresKnown implies 
                a.details->exists(Information_Type::SECURITY_MEASURES)    
        )
    )
    and processors(dp)->forAll(a|
        let processor = a.oclAsType(Processor),
        proc_activities = processor.processing_activity,
        proc_details = proc_activities->collect(details) in
        proc_details->includes 
        (Bag{Information_Type::PROCESSOR_NAME,
            Information_Type::PROCESSORS_DETAILS,
            Information_Type::DPO_DETAILS
        })
        and processor.actingAsControllers->forAll(ac|
            proc_details->includes(ac.processing_activity->collect(details))
        )
        and let rep_activities = processor.representative.processing_activity in
        proc_activities->includes(rep_activities->collect(details))
        and proc_activities->forAll(a|
            a.transfersToThirdCountries implies a.details->exists(Information_Type::TRANSFERS_THIRD_COUNTRIES)
            and a.areSecurityMeasuresKnown implies 
                a.details->exists(Information_Type::SECURITY_MEASURES)    
        )
    )
)
\end{lstlisting}
\bigbreak
\begin{lstlisting}[linewidth=\columnwidth,xleftmargin=1em, breaklines=true]
inv C24:
processing->forAll(dp|
 dp.actors->select(a|a.oclIsKindOf(Processor) 
    or a.oclIsKindOf(Controller) 
    or a.oclIsKindOf(Representative)
    )->forAll(a:Actor|a.legal_authority->notEmpty())
)
\end{lstlisting}
\bigbreak
\begin{lstlisting}[linewidth=\columnwidth,xleftmargin=1em, breaklines=true]
def:main_intervenants (dp:Data_Processing) : Set(Actor) = dp.actors->select(a|a.oclIsTypeOf(Main_Intervenant))
inv C25:
processing->forAll(dp|
    let organizational_measures = dp.security_measure->select(sm|sm.oclIsTypeOf(Organizational)),
    technical_measures= dp.security_measure->select(sm|sm.oclIsTypeOf(Technical)) in
    main_intervenants(dp)->forAll(a|
        let appropriate_measures = a.oclAsType(Main_Intervenant).getAppropriateSecurityMeasures(dp) in
        dp.security_measure->collect(type)->includes(appropriate_measures)
        and 
        technical_measures->collect(type)->includes(Bag{Technical_Measure_Type::PSEUDOANONYMIZATION,
        Technical_Measure_Type::ENCRYPTION})
    )
     and main_intervenants(dp)->forAll(a|
         let intervenant =oclAsType(Main_Intervenant) in
         intervenant.ensuredPrinciples(dp)->includes(Bag{Integrity_And_Confidentiality,Availability, Resilience})
         and 
         intervenant.canRestorePersonalData
         and 
         intervenant.evaluatedMeasuresRegularly
     )
)
\end{lstlisting}
\bigbreak
\begin{lstlisting}[linewidth=\columnwidth,xleftmargin=1em, breaklines=true]
inv C26:
    processing->forAll(dp|dp.breach->
        forAll(b|
        let detector:Main_Intervenant = b.detectedBy in 
        detector.oclIsTypeOf(Controller)
        implies (let controller=detector.oclAsType(Controller) in
            Bag{Risk_Severity::LOW,Risk_Severity::MEDIUM,Risk_Severity::HIGH}->includes(b.severity)
                implies controller.recordBreach(b)
            and Bag{Risk_Severity::MEDIUM,Risk_Severity::HIGH}->includes(b.severity)
                implies controller.notifyBreachToSA(b,controller.getConcernedSA(dp))<72
            and Risk_Severity::HIGH = b.severity 
                implies personal_data.data_subject->flatten()->asSet()->forAll(ds|controller.notifyBreachToDS(b,ds)))
        and detector.oclIsTypeOf(Processor) 
        implies detector.oclAsType(Processor).notifyBreachToAllControllers(b,dp)
        )
    )
\end{lstlisting}
\bigbreak
\begin{lstlisting}[linewidth=\columnwidth,xleftmargin=1em, breaklines=true]
inv C27:
processing->forAll(dp|
    let dpias:Set(Data_Protection_Impact_Assessment) = dp.data_protection_impact_assessment in 
    dp.risk->exists(r|r.severity=Risk_Severity::HIGH 
        or r.impactsOnDS<>""
        or dpias->exists(dpia:Data_Protection_Impact_Assessment|dpia.dpia_motivation->exists
            (m|m.motivation=Enumerations::DPIA_Motivation::HIGH_RISK)
        )) implies 
        (controllers(dp)->forAll(c|c.oclAsType(Controller).assessProcessingImpact(dp)) 
        and dpias->notEmpty())
    and 
    (dp.operation=Operation_Type::PROFILING 
        or dp.purposes->includes(Processing_Context::BEHAVIOUR_MONITORING_OR_PROFILING)
        or dpias->exists(dpia|dpia.dpia_motivation->exists(dm|dm.motivation=Enumerations::DPIA_Motivation::PROFILING))
    )
    implies dpias->forAll(isMandatory) 
    and 
    dp.purposes->includes(Processing_Context::TAKE_AUTOMATED_DECISIONS)
    and 
    (dp.targetsSpecialDataCategories 
        and dpias->exists(dpia|dpia.dpia_motivation->exists(dm|dm.motivation=Enumerations::DPIA_Motivation::LARGE_SCALE_PROCESSING))
    )
    implies dpias->forAll(isMandatory) 
     and 
     dpias->exists(dpia|dpia.dpia_motivation->exists(dm|dm.motivation=Enumerations::DPIA_Motivation::SYSTEMATIC_MONITORING))
     implies dpias->forAll(isMandatory)
     and
     let mustHaveTypes:Set(Processing_Context)= mustHaveDPIA(dp.getConcernedSupervisoryAuthority().oclAsType(Supervisory_Authority)) in
     dp.purposes->exists(p|mustHaveTypes->includes(p._context))
     and 
     dpias->forAll(areInvolvedProcessingSimilar implies dpias->forAll(not isMandatory))
     and 
     if dpias->notEmpty()
     then 
     dpias->forAll(dpia|dpia.data_processing->notEmpty() and dpia.data_processing->forAll(not dp.type.oclIsUndefined()))
     and dpias->forAll(dpia|dpia.dpia_information->exists(info|info.type=DPIA_Information_Type::PROPORTIONALITY_ASSESSEMNT))
     and dpias->forAll(dpia|dpia.dpia_information->exists(info|info.type=DPIA_Information_Type::RISK_ASSESSMENT))
     and dpias->forAll(dpia|dpia.dpia_information->exists(info|info.type=DPIA_Information_Type::MEASURES_DESCRIPTION))
     else false
     endif
)
\end{lstlisting}
\bigbreak
\begin{lstlisting}[linewidth=\columnwidth,xleftmargin=1em, breaklines=true]
inv C28:
processing->forAll(dp|
dp.data_protection_impact_assessment->forAll(dpia:Data_Protection_Impact_Assessment|
    dpia.dpia_information->forAll(di:DPIA_Information|
        if (di.type=DPIA_Information_Type::RISK_ASSESSMENT
            and let tokens = di.summary.toLowerCase().tokenize(" ") in 
            tokens->includes("high")
            )
        then dpia.Consultation->forAll(isMandatory)
        else false
        endif
    )
    and 
    if dpia.Consultation<>oclIsUndefined()
        then (
            dpia.Consultation->forAll(feedback<>oclIsUndefined())
            and
            (dpia.Consultation->forAll(date<dpia.startDate+8))
            and
            controllers(dp)->forAll(a:Actor|let controller = a.oclAsType(Controller) in 
                controller.communicateDPIAToSA(controller.getConcernedSA(dp),dpia)
            )
        )
    else false
    endif
    )
)
\end{lstlisting}
\bigbreak
\begin{lstlisting}[linewidth=\columnwidth,xleftmargin=1em, breaklines=true]
def: controllers(dp: Data_Processing): Set(Actor) = dp.actors - > select(a | a.oclIsKindOf(Controller))
inv C29:
processing->forAll(dp|
let controllers:Set(Actor) = controllers(dp),
processors:Set(Actor) = processors(dp) in
if (controllers->union(processors)->exists(a|
    a.type=Actor_Type::PUBLIC_ORGANIZATION 
    and (a.oclIsKindOf(Legal_Authority)
    or a.oclIsKindOf(Accreditation_Body)
    or a.oclIsKindOf(Certification_Body))
    and not a.oclIsKindOf(Court))
    or 
    (dp.purposes->exists(p|
        p._context=Processing_Context::BEHAVIOUR_MONITORING_OR_PROFILING
        or dp.data_protection_impact_assessment->exists(dpia| 
            let dpia_motivations = dpia.dpia_motivation in
            dpia_motivations->includes(Bag{LARGE_SCALE_PROCESSING, SYSTEMATIC_MONITORING})
            or
            dpia_motivations->includes(Bag{LARGE_SCALE_PROCESSING,INVOLVES_SPECIAL_OR_CRIMINAL_DATA}) 
            )))
)
then controllers->union(processors)->forAll(designates->size()=1)
else controllers->union(processors)->forAll(designates->size()>=0)
endif)
\end{lstlisting}
\bigbreak
\begin{lstlisting}[linewidth=\columnwidth,xleftmargin=1em, breaklines=true]
inv C30:
    Certification.allInstances()->forAll(cert: Certification |
        cert.isVoluntary and cert.isAvailableViaTransparentProcess and not cert.legal_authority - > 
        isEmpty() and cert.validity = 3
    )
\end{lstlisting}
\bigbreak
\begin{lstlisting}[linewidth=\columnwidth,xleftmargin=1em, breaklines=true]
inv C31:
Supervisory_Authority.allInstances()->forAll(
sa:Supervisory_Authority|sa.approved_legal_basis.data_transfer->forAll(
dt:Data_Transfer|
if 
dt.IsWithinEU
or dt.adequacy_decision->notEmpty()
or dt.legal_basis->exists(lb|
    lb.areApproved  
    and (lb.oclIsTypeOf(Binding_Corporate_Rules)
    or lb.oclIsTypeOf(Standard_Contractual_Clauses)
    or lb.oclIsTypeOf(Administrative_Arrangement)
    or lb.oclIsTypeOf(Certification)))
    or dt.code_of_conduct.isApproved=true
and not (
    let transfer_derrogations = dt.legal_basis->select(oclIsTypeOf(Transfer_Derogation)) in 
        Bag{
            Transfer_Derogation_Types::SUPPORTED_BY_CONSENT,
            Transfer_Derogation_Types::NECESSARY_FOR_CONTRACT,
            Transfer_Derogation_Types::PROTECT_DS_VITAL_INTERESTS,
            Transfer_Derogation_Types::PUBLIC_INTEREST,
            Transfer_Derogation_Types::EXERCISE_OR_DEFENCE_OF_LEGAL_CLAIMS,
            Transfer_Derogation_Types::PUBLIC_CONSULTATION
        }->includes(transfer_derrogations)
        or transfer_derrogations->exists(let td = oclAsType(Transfer_Derogation) in 
        not td.isTransfertRepetitive implies td.limitedNumberOfConcernedDS)
)
then dt.isForbidden = false
else dt.isForbidden = true
endif
)
)
\end{lstlisting}
\bigbreak
\begin{lstlisting}[linewidth=\columnwidth,xleftmargin=1em, breaklines=true]
inv C32:
    Data_Transfer.allInstances()->forAll(dt |
        let additional_requirements: Set(String) = 
        dt.adequacy_decision.additionalRequirements->flatten()->asSet() in
            not dt.adequacy_decision.oclIsUndefined() and additional_requirements->forAll(
            requirement: String | requirement <> "")
        implies dt.importer.provideEvidence(additional_requirements)->exists(dt.Adequacy_Evidence)
    )
    
\end{lstlisting}
\bigbreak
\begin{lstlisting}[linewidth=\columnwidth,xleftmargin=1em, breaklines=true]
inv C33:
Data_Transfer.allInstances()->forAll(dt|
    if dt.adequacy_decision.basedOnInternationalAgreement
    and dt.adequacy_decision.country->exists(not isEUMemberState)
    and dt.adequacy_decision.requiresDiscloseOfPersonalData 
    then dt.adequacy_decision.mayBeRecognised = true
    and dt.adequacy_decision.mayBeEnforceable = true
    else dt.adequacy_decision.mayBeRecognised = false
    and dt.adequacy_decision.mayBeEnforceable = false
    endif)
\end{lstlisting}
\bigbreak
\begin{lstlisting}[linewidth=\columnwidth,xleftmargin=1em, breaklines=true]
inv C34:
    Supervisory_Authority.allInstances().approved_legal_basis->collect(oclIsKindOf(Binding_Corporate_Rules))->
    forAll(lb |
        let bcr = lb.oclAsType(Binding_Corporate_Rules) in
            bcr.areLegallyBinding
        and bcr.areEnforcedToEveryMember and bcr.applyToEveryMember and 
        bcr.conferEmforceableRightsOnDS and bcr.contract_information.type->includes(
            Bag {
                Transfer_Contract_Information::UNDERTAKINGS_STRUCTURE, 
                Transfer_Contract_Information::CONTACT_DETAILS
            }) and bcr.contract_information.type->includes(
            Bag {
                Transfer_Contract_Information::DATA_CATEGORIES,
                Transfer_Contract_Information::TYPE_PROCESSING_AFTER_TRANSFER,
                Transfer_Contract_Information::PURPOSES_PROCESSING_AFTER_TRANSFER,
                Transfer_Contract_Information::TYPE_DS_AFFECTED
            }) and bcr.contract_information.type->includes(
            Bag {
                Transfer_Contract_Information::TARGET_COUNTRIES,
                Transfer_Contract_Information::INTERNAL_COUNTRIES_BINDING_LAWS,
                Transfer_Contract_Information::EXTERNAL_COUNTRIES_BINDING_LAWS
            }) and bcr.contract_information.type-> 
            includes(Transfer_Contract_Information::APPLIED_GDPR_PRINCIPLES) and 
            bcr.data_transfer.right->select(r | r.oclIsTypeOf(Right_To_Be_Informed))->notEmpty()
            and bcr.data_transfer.breach->notEmpty() implies bcr.contract_information.type->
            includes(Transfer_Contract_Information::LIABILITY_SHARING) and 
            bcr.contract_information.type->includes(Bag {
                Transfer_Contract_Information::HOW_DS_INFORMED,
                Transfer_Contract_Information::DPO_TASKS,
                Transfer_Contract_Information::COMPLIANCE_PROCEDURES,
                Transfer_Contract_Information::REPORTING_MECHANISMS,
                Transfer_Contract_Information::REPORTING_MECHANISMS,
                Transfer_Contract_Information::PERSONAL_TRAINING
            })
    )
\end{lstlisting}
\bigbreak
\begin{lstlisting}[linewidth=\columnwidth,xleftmargin=1em, breaklines=true]
inv C35:
    Corrective_Action.allInstances()->forAll(ca: Corrective_Action |
        (
            ca.type = SA_Corrective_Action_Type::ADMINISTRATIVE_FINES 
            and(ca.main_intervenant.obligation -> exists(obligation: Obligation | not 
            obligation.isFulfilled) or ca.infringement.type -> 
            exists(Infringement_Type::CHILD_CONSENT_VIOLATION) or ca.infringement.type -> 
            exists(Infringement_Type::OBLIGATION_VIOLATION) or ca.infringement.type -> 
            exists(Infringement_Type::CERTIFICATION_OBLIGATION_VIOLATION) or ca.infringement.type -> 
            exists(Infringement_Type::CODE_OF_CONDUCTS_MISS_APPLICATION))
        ) implies ca.main_intervenant -> forAll(mi |
            ca.TotalFines = 0.02 * mi.annualRevenue or ca.TotalFines <= 10000000) and(
            ca.type = SA_Corrective_Action_Type::ADMINISTRATIVE_FINES and(
                ca.infringement.type -> exists(Infringement_Type::PRINCIPLE_VIOLATION) or 
                ca.infringement.type -> exists(Infringement_Type::DS_RIGHT_VIOLATION) or 
                ca.infringement.type -> exists(Infringement_Type::CROSS_BORDER_TRANSFER_VIOLATION) or
                ca.main_intervenant.getAdditionalObligations() -> exists(ob | not ob.isFulfilled) or 
                ca.infringement.type -> exists(Infringement_Type::OTHER_LOCAL_LAW_VIOLATION)
            )
        ) implies ca.main_intervenant -> forAll(mi |
            ca.TotalFines = 0.04 * mi.annualRevenue or ca.TotalFines <= 20000000)
    )
\end{lstlisting}

\clearpage

\newpage
\section*{{Appendix~C}} \label{appendixC}
In this section, we present the resolutions (including new OCL constraints, if needed) of the 20 variation points discussed in Section \ref{variabilitytable} of Appendix A.
\bigbreak
\noindent\textbf{V1 resolution}:  Add V1 and implement V\_getMinimumAgeForDS
\begin{lstlisting}[linewidth=\columnwidth,xleftmargin=1em, breaklines=true, caption=]
context Data_Subject inv V1:
    let minDSAge: Integer = Variability.V_getMinimumAgeForDS(self) in
        if (self.oclIsTypeOf(Child_Data_Subject))
        then self.getAge() < minDSAge
        else self.getAge() >= minDSAge endif
\end{lstlisting}
\bigbreak
\noindent\textbf{V2 resolution}: Add V2 and implement V\_checkParentDocuments
\begin{lstlisting}[linewidth=\columnwidth,xleftmargin=1em, breaklines=true]
context Natural_Person inv V2: 
    self.children->forAll(c:Child_Data_Subject|self.V_checkParentDocuments(c)
    )
\end{lstlisting}
\bigbreak
\noindent\textbf{V3 resolution}: Implement V\_prohibitionCanBeLiftedByConsent used in C6
\bigbreak
\noindent\textbf{V4 resolution}: Add V4, implement V\_verifyFurtherConditionsAndLimit
\begin{lstlisting}[linewidth=\columnwidth,xleftmargin=1em, breaklines=true]
inv V4:
    processing->forAll(dp: Data_Processing | dp.personal_data->forAll(pd |
        Bag {
            Data_Category::HEALTH, Data_Category::GENETIC, Data_Category::BIOMETRIC
        } -> includes(pd.category) implies Variability.V_verifyFurtherConditionsAndLimit(dp) = true)
    )
\end{lstlisting}
\bigbreak
\noindent\textbf{V5 resolution}: (1) Adapt C2, and (2) Adapt from C9 to C22.
\bigbreak
\noindent\textbf{V6 resolution}: Implement V\_processWithoutControllerInstructions, used in C22
\bigbreak
\noindent\textbf{V7 resolution}: Add V7 and implement V\_processWithoutControllerInstructions
\begin{lstlisting}[linewidth=\columnwidth,xleftmargin=1em, breaklines=true]
inv V7:
    processing->forAll(dp: Data_Processing |
        dp.actors->select(a | a.type = Actor_Type::NATURAL_PERSON and a.actsUnderAuthorityOf()-> 
        notEmpty())->forAll(np |
            Variability.V_processWithoutControllerInstructions(dp, np) implies 
            np.actedOnControllerDocumentedInstructions(dp) = false)
    )
\end{lstlisting}
\bigbreak
\noindent\textbf{V8 resolution}: Add V8, and RECONCILIATION\_ASSESSMENT enumeration literal in DPIA\_Information\_Type, and implement V\_ReconcileByLaw
\begin{lstlisting}[linewidth=\columnwidth,xleftmargin=1em, breaklines=true]
inv V8:
    processing->forAll(dp |
        Variability.V_ReconcileByLaw(dp) implies dp.data_protection_impact_assessment.dpia_information->
        collect(type)->includes(DPIA_Information_Type::RECONCILIATION_ASSESSMENT)
    )
\end{lstlisting}
\bigbreak
\noindent\textbf{V9 resolution}: Implement V\_bodiesMustDesignateDPO used in C29
\bigbreak
\noindent\textbf{V10 resolution}: Add V10 and implement V\_verifyTransferLimits
\begin{lstlisting}[linewidth=\columnwidth,xleftmargin=1em, breaklines=true]
inv V10:
Supervisory_Authority.allInstances()->forAll(sa : Supervisory_Authority |
    sa.approved_legal_basis.data_transfer->forAll(dt : Data_Transfer |
    (Variability.V_limitTransfers() and dt.adequacy_decision->isEmpty() and dt.purposes->exists(p |
            p._context = PUBLIC_INTEREST)) implies Variability.V_verifyTransferLimits(dt) = true))
\end{lstlisting}
\bigbreak
\noindent\textbf{V11 resolution}: Add V11
\begin{lstlisting}[linewidth=\columnwidth,xleftmargin=1em, breaklines=true]
inv V11:
    processing->forAll(dp: Data_Processing |
        dp.actors->select(a | a.type = Actor_Type::NON_PROFIT_ORGANIZATION)->forAll(a | 
        a.canLodgeComplaints = true)
    )
\end{lstlisting}
\bigbreak
\noindent\textbf{V12 resolution}: --V12 resolution: Replace C35 by V12\_1, and add V12\_2
\begin{lstlisting}[linewidth=\columnwidth,xleftmargin=1em, breaklines=true]
inv V12_1:
    Corrective_Action.allInstances()->forAll(ca: Corrective_Action |
        (ca.type = SA_Corrective_Action_Type::ADMINISTRATIVE_FINES and(ca.main_intervenant.obligation -> 
        exists(obligation: Obligation | not obligation.isFulfilled) or ca.infringement.type -> 
        exists(Infringement_Type::CHILD_CONSENT_VIOLATION) or ca.infringement.type -> 
        exists(Infringement_Type::OBLIGATION_VIOLATION) or ca.infringement.type -> 
        exists(Infringement_Type::CERTIFICATION_OBLIGATION_VIOLATION) or ca.infringement.type -> 
        exists(Infringement_Type::CODE_OF_CONDUCTS_MISS_APPLICATION))) implies ca.main_intervenant -> forAll(mi |
            mi.type < > Actor_Type::PUBLIC_ORGANIZATION implies ca.TotalFines = 0.02 * 
            mi.annualRevenue or ca.TotalFines <= 10000000
        ) and(ca.type = SA_Corrective_Action_Type::ADMINISTRATIVE_FINES and(ca.infringement.type -> 
        exists(Infringement_Type::PRINCIPLE_VIOLATION) or ca.infringement.type -> 
        exists(Infringement_Type::DS_RIGHT_VIOLATION) or ca.infringement.type -> 
        exists(Infringement_Type::CROSS_BORDER_TRANSFER_VIOLATION) or 
        ca.main_intervenant.getAdditionalObligations() -> exists(ob | not ob.isFulfilled) or 
        ca.infringement.type -> exists(Infringement_Type::OTHER_LOCAL_LAW_VIOLATION))) implies 
        ca.main_intervenant -> forAll(mi |
            mi.type <> Actor_Type::PUBLIC_ORGANIZATION implies ca.TotalFines = 0.04 * mi.annualRevenue or 
            ca.TotalFines <= 20000000)
    )

inv V12_2:
    Corrective_Action.allInstances()->forAll(ca: Corrective_Action |
        (ca.type = SA_Corrective_Action_Type::ADMINISTRATIVE_FINES implies ca.main_intervenant -> forAll(mi |
            mi.type <> Actor_Type::PUBLIC_ORGANIZATION implies Variability.V_setAdministrativeFines(ca)
            )
        )
    )
\end{lstlisting}
\bigbreak
\noindent\textbf{V13 resolution}: Add V13
\begin{lstlisting}[linewidth=\columnwidth,xleftmargin=1em, breaklines=true]
inv V13:
    Corrective_Action.allInstances() -> forAll(ca: Corrective_Action |
        ca -> forAll(Variability.V_setInfrimegementsPenalties(ca)
        )
    )
\end{lstlisting}
\bigbreak
\noindent\textbf{V14 resolution}: Add 14
\begin{lstlisting}[linewidth=\columnwidth,xleftmargin=1em, breaklines=true]
inv V14:
    processing->forAll(dp | dp.purposes->forAll(p |
        Bag {
            Processing_Context::PUBLIC_INTEREST,
                Processing_Context::PUBLIC_HEALTH
        } -> includes(p._context) implies dp.data_protection_impact_assessment ->
        forAll(dpia | dpia.Consultation <> oclIsUndefined())
    ))
\end{lstlisting}
\bigbreak
\noindent\textbf{V15 resolution}: Add V15, implement V\_checkedIDProcessing and add the IDENTIFICATION literal to the Data\_Category enumeration 
\begin{lstlisting}[linewidth=\columnwidth,xleftmargin=1em, breaklines=true]
inv V15:
    processing->forAll(dp |
        dp.personal_data->forAll(pd |
            pd.category = Data_Category::IDENTIFICATION implies Variability.V_checkIDProcessing(dp) = true
        )
    )
\end{lstlisting}
\bigbreak
\noindent\textbf{V16 resolution}: Add the EMPLOYMENT\_ASSESSMENT literal to the DPIA\_Information\_Type enumeration
\begin{lstlisting}[linewidth=\columnwidth,xleftmargin=1em, breaklines=true]
inv V16:
    processing->forAll(dp |
        dp.purposes->forAll(p |
            p._context = Processing_Context::EMPLOYMENT implies dp.data_protection_impact_assessment.
            dpia_information->collect(type)->includes(DPIA_Information_Type::EMPLOYMENT_ASSESSMENT)
        )
    )
\end{lstlisting}
\bigbreak
\noindent\textbf{V17 resolution}: Add V17, and implement V\_checkDerrogationsFromRights. 
\begin{lstlisting}[linewidth=\columnwidth,xleftmargin=1em, breaklines=true]
inv V17:
    processing->forAll(dp |
        dp.purposes->forAll(p |
            Bag {
                Processing_Context::RESEARCH, Processing_Context::STATISTICAL_PURPOSES
            } -> includes(p._context) implies Variability.V_checkDerrogationsFromRights(dp) = true
        )
    )
\end{lstlisting}
\bigbreak
\noindent\textbf{V18 resolution}: Add V18, and implement V\_checkedDerrogationsFromRights
\begin{lstlisting}[linewidth=\columnwidth,xleftmargin=1em, breaklines=true]
inv V18:
processing->forAll(dp|
    dp.purposes->forAll(p|
        Bag{Processing_Context::PUBLIC_INTEREST, Processing_Context::ARCHIVING_PURPOSES}->includes(p._context)
        implies Variability.V_checkDerrogationsFromRights(dp)=true
    )
)
\end{lstlisting}
\bigbreak
\noindent\textbf{V19 resolution}: Add V19, and implement V\_checkDerrogationsFromRights
\begin{lstlisting}[linewidth=\columnwidth,xleftmargin=1em, breaklines=true]
inv V19:
    processing->forAll(dp |
        dp.personal_data -> exists(pd | dp.isCoveredByObligationOfSecrecy(pd)) implies 
        Variability.V_checkSARules(dp)
    )
\end{lstlisting}
\bigbreak
\noindent\textbf{V20 resolution}: Add V20, and the CHURCH\_OR\_RELIGIOUS\_ORGANIZATION literal to the Actor\_Type enumeration  
\begin{lstlisting}[linewidth=\columnwidth,xleftmargin=1em, breaklines=true]
inv V20:
    processing->forAll(dp|
        dp.actors->exists(type=Actor_Type::CHURCH_OR_RELIGIOUS_ORGANIZATION) implies 
        Variability.V_checkReligiousRules(dp)
    )
\end{lstlisting}

\twocolumn

\bibliographystyle{spmpsci}
\bibliography{paper}


\end{document}